\documentclass{article}

\usepackage{arxiv}
\usepackage{subfig}
\pdfoutput=1 
\usepackage[utf8]{inputenc} 
\usepackage[T1]{fontenc}    
\usepackage{hyperref}       
\usepackage{url}            
\usepackage{booktabs}       
\usepackage{amsfonts}       
\usepackage{nicefrac}       
\usepackage[pdftex]{graphicx}
\usepackage{bm}
\usepackage{float}
\usepackage{amsmath}
\usepackage{enumitem}
\usepackage[english]{babel}
\usepackage{placeins}
\usepackage[boxruled, lined, vlined]{algorithm2e}
\usepackage{caption}
\usepackage{amsthm}
\usepackage{xcolor}

\usepackage{algorithmic}  

\usepackage{multirow}

\DeclareMathSizes{10}{8}{5}{2}

\allowdisplaybreaks

\theoremstyle{theorem}
\newtheorem{theorem}{Theorem}

\theoremstyle{lemma}
\newtheorem{lemma}{Lemma}[section]

\theoremstyle{definition}
\newtheorem{definition}{Definition}[section]

\title{Cyber Loss Model Risk Translates to Premium Mispricing and Risk Sensitivity}

\author{
Gareth W. Peters \\
Statistics \& Applied Probability\\
University of California Santa Barbara\\
  \texttt{garethpeters@ucsb.com} \\
     \And
Matteo Malavasi \\
Actuarial Studies and Business Analytics,\\
Macquarie University, Australia\\
     \And
Pavel V. Shevchenko \\
Actuarial Studies and Business Analytics,\\
Macquarie University, Australia\\
     \And
Georgy Sofronov \\
Mathematical and Physical Sciences, \\
Macquarie University, Australia\\
     \And
Stefan Tr\"uck \\
Actuarial Studies and Business Analytics, \\
Macquarie University, Australia\\
     \And
Jiwook Jang \\
Actuarial Studies and Business Analytics,\\
Macquarie University, Australia\\
}

\begin{document}
\maketitle

\begin{abstract}
We focus on model risk and risk sensitivity when addressing the insurability of cyber risk. The standard statistical approaches to assessment of insurability and potential mispricing are enhanced in several aspects involving consideration of model risk. Model risk can arise from model uncertainty, and parameters uncertainty. We demonstrate how to quantify the effect of model risk in this analysis by incorporating various robust estimators for key model parameter estimates that apply in both marginal and joint cyber risk loss process modelling. We contrast these robust techniques with standard methods previously used in studying insurabilty of cyber risk. This allows us to accurately assess the critical impact that robust estimation can have on tail index estimation for heavy tailed loss models, as well as the effect of robust dependence analysis when quantifying joint loss models and insurance portfolio diversification. We argue that the choice of such methods is akin to a form of model risk and we study the risk sensitivity that arise from choices relating to the class of robust estimation adopted and the impact of the settings associated with such methods on key actuarial tasks such as premium calculation in cyber insurance. Through this analysis we are able to address the question that, to the best of our knowledge, no other study has investigated in the context of cyber risk: is model risk present in cyber risk data, and how does is it translate into premium mispricing? We believe our findings should complement existing studies seeking to explore insurability of cyber losses. In order to ensure our findings are based on realistic industry informed loss data, we have utilised one of the leading industry cyber loss datasets obtained from Advisen, which represents a comprehensive data set on cyber monetary losses, from which we form our analysis and conclusions.
\end{abstract}


\section{Introduction}
Cyber risk continues to gain relevance in our society, as companies and enterprises increasingly rely on information systems, see a detailed overview of the state of cyber risk understanding in insurance contexts in \cite{eling2020cyber}. A successful cyber attack can cause major damage to a public or private institution. It can directly affect the budgetary bottom line, in addition to a business' standing and consumer trust. Cyber attack security breaches can be categorised basically into three core components in which losses can be realised after a cyber attack: financial, reputational damage and legal, see detailed discussions in \cite{eling2016we} and \cite{peters2018understanding}.

Furthermore, cyber attacks can carry a significant direct economic and financial cost, see discussions in \cite{romanosky2017content} and empirical analysis in \cite{biener2015insurability,edwards2016hype,shevchenko2021quantification}. These costs can manifest for instance as losses due to: theft of corporate information; or theft of financial information such as customer records; direct theft of money or assets; business disruption for critical systems such as trading or the inability to process transactions; and the loss of business or contracts, to name but a few of the types of economic and financial loss types that may result from a serious cyber attack, see cluster analysis of cyber event types for insurance and risk contexts in \cite{peters2018statistical}. Furthermore, there are also often significant losses arising from incurred costs associated with repairing affected systems, networks and devices. This is often required after major events in order to meet regulatory standards or satisfy investors or clients of the risk reduction changes made post significant cyber events, see an overview on discussions on Basel banking regulation requirements for operational risk cyber loss in \cite{cruz2015}. 

In addition, after a cyber attack there are also a variety of indirect costs that often arise which may  due to reputational damage borne from news of the attack reaching the public or customers affected by data breach of their records, see discussions on data breach fines in cyber risk in \cite{ceross2017use}. Cyber attacks can damage a business' reputation and erode the trust of the customer, leading to customer attrition. The effect of reputational damage can even impact on an institutions suppliers, or affect relationships with partners, investors and other third parties vested in a business. Other impacts from cyber events can include legal and regulatory consequences. In many jurisdictions, both private and public entities are required to provide certain guarantees on data privacy under data protection and privacy laws which require firms to manage the security of all personal data being held on staff and customers. If this data is accidentally or deliberately compromised, and the firm in question can be deemed to have failed to deploy appropriate security measures, after which they may face fines and regulatory sanctions from multiple jurisdictions from which affected customers may be derived. 

As such, it is increasingly becoming apparent that mitigation of cyber risk and cyber losses alone will not suffice to protect both public and private institutions from the potential for catastrophic monetary losses arising from cyber attacks. Therefore, upon the realisation that cyber attacks can never truly be completely mitigated, especially with the increasing pace of technology adoption and growth, then there is a growing need to find effective risk transfer strategies. One such strategy, not to mitigate a cyber loss but rather to ensure that the affected institution or firm is able to recover and fund any required losses is through insurance and re-insurance markets. The cyber risk insurance markets and available products are still very much in their infancy, see discussions in \cite{eling2016we,eling2018cyber} and a recent US case study in \cite{xie2020cyber}. As the interest in the effects of cyber risk grows, so does the number of papers in the actuarial studies tackling several important questions on cyber risk \cite{eling2017,eling2019,edwards2016, jung2021extreme}. 

In this manuscript, the focus is on aspects of the perception of the insurability or not of cyber risk. What makes this work distinct from previous studies of a related nature arises in two key factors of our study focus: the first is the fact that we have used one of the industry gold standards for cyber loss data, Advisen Cyber Loss Data (\url{https://www.advisenltd.com/data/cyber-loss-data}), which represents a comprehensive data set on cyber monetary losses, from which we form our analysis and conclusions; the second, perhaps more important aspect is that we question the standard statistical approaches to assessment of this question of insurability in several important aspects. In particular, one may conclude that we assess the question of insurability of cyber loss taking into account a previously unaccounted for dimension related to model risk. We seek to answer a question that, to the best of our knowledge no other study has investigated in the context of cyber risk: is model risk present in cyber risk data, and how does is it translate into premium mispricing?

Model risk can arise from two different factors: model uncertainty, and parameter uncertainty. While model uncertainty generally refers to the assumptions that one makes in developing a statistical model representation, parameter uncertainty revolves around the idea of predictive inference \cite{frohlich2018}. In this paper we focus on both aspects of model structure uncertainty as well as parameters uncertainty, and investigate two main channels of transmissions, using the Advisen cyber loss dataset. In this regard we focussed on two core components related to assessing the question of insurability for both an individual cyber risk threat type or a portfolio of multiple cyber risk threat types. The first component is the marginal tail behaviour of a cyber risk loss process and how assumptions regarding the validity of core details in reported losses and completeness of such records, obtained from Freedom of Information (FOI) requests, may influence the outcome of determinations of the insurability of cyber losses. We achieve this by considering a variety of parametric and non-parametric estimators for the tail index of the loss processes under study and we add to this analysis the dimension of robust estimation, which involves the ability to question the validity, completeness, quantisation, providence, accuracy and veracity of losses by trimming or weighting exceptionally large losses that directly influence the marginal question of insurability. Secondly, we also study the effect of uncertainty in the model structure and estimation of dependence in cyber risk losses. To achieve this we employ novel methods for the quantification of copula dependence structures, robust estimation techniques for correlation analysis and tail dependence estimation. This aspect of model risk allows us to assess the impact on diversification of cyber risk that may or may not be present in an insurers portfolio, should they offer cyber risk to clients across a range of different cyber risk loss types or over a range of different industry sectors.

\subsection{Contributions and Outline}
Our studies are undertaken in two parts and offer a variety of contributions to the understanding of model risk, parameter uncertainty and its translation to premium mispricing in cyber risk settings. Ultimately, we argue this provides a new dimension to understanding the insurability of cyber risk as quantified by required insurance premiums.

In part 1 we study model and parameter uncertainty risk as it relates to key idea of tail index estimation. In particular we demonstrate significant challenges with working with cyber loss data when estimating tail indexes and we demonstrate the variation obtained in the different parametric and non-parametric tail index estimators under varying assumptions on the data quality and accuracy as reflected in modifications to our estimators. As such, we first consider how parameter uncertainty impacts insurance premium calculations. Several studies have shown that cyber risk severity follows a heavy tailed distribution \cite{eling2019, eling2017, edwards2016}. We focus our analysis on tail index of the cyber event severity distribution and recall some well known facts about various estimators proposed by the literature. Then, we  aggregate cyber related losses by business sectors, adopting the North America Industry Classification (NAIC), and compare the tail index estimates using various estimators. We observe significant variations among tail index estimates, indicating the presence of parameter uncertainty. This acts as a motivation for robust and trimmed tail index estimators. In this process, we further explain how and why cyber loss data may need to be trimmed and then show the effect this has on the tail index estimations. Trimmed tail index estimators are valid alternative in the context of cyber risk, where many extreme loss events are often of very high monetary amounts and one could question whether they do turn into realised losses, see \cite{brazauskas2000robust, zou2020extreme, goegebeur2014robust, peng2001robust}. Using the trimmed estimator in \cite{bhattacharya2017trimming}, for different trimming values, we confirm once more presence of parameter uncertainty given the great variability in the values of the tail index estimates. Ultimately, based on this estimation analysis, we then demonstrate for each tail index estimation method and set of assumptions, using both trimmed and non-trimmed estimators, how a utility based pricing of an insurance premium, under an indifference pricing framework, produces variation/sensitivity for a single insurance line by NAIC. Ultimately, we demonstrate, using the zero utility principle, how parameter uncertainty translates into insurance premium mispricing risk, which could jeopardise the perspective of the insurability of cyber risk with regard to required premiums to be incurred. We note that at present, premiums for such products are generally deemed prohibitive unless contracts are specifically designed to be bespoke and restricted in scope of coverage.

In part 2 of our studies, we assess model and parameter uncertainty from a multivariate perspective, and determine how such model risk factors affect diversification of an insurers portfolio of insured cyber risk lines, when multiple lines or multiple industry sectors are being offered coverage. In particular, we explore copula model uncertainty as well as robust versus non-robust estimators of correlation dependence. To achieve this we adopt a novel approach to estimation of the incurred premium for a portfolio of loss types or business lines. In particular, we assess model risk as it relates to key idea of diversification of risk from risk pooling. To achieve this we study the linear correlation for all NAICs, then we focus on top 5 NAICs and study robust correlation analysis using 3 robust estimators: median based Sum-of-Squares (SSD), Quadrant (sign) correlation coefficient, Minimum Covariance Determination (MCD). We are able to then demonstrate how the risk based diversification coefficient will vary for differing robust estimators and the role this sensitivity will have on portfolio net coverage, thereby demonstrating sensitivity to premium calculations in marginal versus conditional risk profile perspectives. Lastly, we also fit a variety of pairwise copulas on quarterly data and show sensitivity to these fits and resulting sensitivity of premiums calculated.

The remainder of the manuscript is structured as follows. In Section \ref{sec:Data} a detailed data description is provided for the Advisen cyber loss data set, followed by several sets of empirical data analysis of the cyber risk data according to Advisen risk types and according to industry sectors determined by the North American Industry Classification System (NAIC) codes, which act as the standard used by US Federal statistical agencies in classifying business establishments for the purpose of collecting, analysing, and publishing statistical data related to the U.S. business economy. Section \ref{sec:NonParametricTails} addresses the quantification of heavy tails in cyber risk losses. A variety of methods are explored including smoothed Hill plots, extremogram estimators and analysis of tail index estimators from a variety of different statistical perspectives, including empirical characteristic function asymptotic regression methods, Hill type estimation methods and variations. These methods are briefly mathematically outlined and then applied to study cyber risk loss data from Advisen. Section \ref{subsec:RobustHill} outlines the challenges associated with working with real world cyber loss data that include: inaccuracies, rounding, truncation, partially settlement and unreliable massive reported cyber total losses. To address this and determine how it can manifest as a form of model risk via  model uncertainty, and parameters uncertainty, Sections \ref{subsec:RobustHill} and \ref{sec:NonParametricTails} introduced robust methods. This includes overviews of relevant classes of robust trimmed Hill estimators that are applied subsequently to the cyber loss data. Section \label{sec:NonParametricTails} presents a comprehensive analysis of the Advisen Loss Data using the various proposed robust tail index estimation methods as well as a study of robust dependence. Subsequently, a detailed insurance pricing example is performed for various cyber insurance lines of business to show how model risk in tail index estimation transfers to potential for miss pricing in cyber risk insurance, ranging from uninsurable due to exorbitant costs through to affordable, depending on the modelling approach adopted. Furthermore, a detailed analysis of dependence between different cyber risk loss processes is studied via robust dependence estimation methods, copula estimation methods and a little known Monte Carlo based simulation method is detailed in order to perform insurance pricing and portfolio diversification assessment in standard and robust contexts to further assess aspects of multivariate model risk in insurance pricing contexts. The paper concludes with Section \ref{sec:Conclusions}.

\section{Data Description and Attributes} 
\label{sec:Data}
There is an ongoing exploration on the various ways to classify and taxonomize cyber risk loss events, see discussions in \cite{shevchenko2021quantification,rea2017systematic} and \cite{elnagdy2016understanding}. In this study the focus has been on the US cyber risk experience as it is generally the environment where the largest commercial cyber loss data collection effort has been instigated, both in terms of breadth of industry and loss type as well as in terms of duration of collection and reporting. In this regard, the paper will focus on the  Advisen Cyber Loss Data. This data set provides a historical view of more than 132,126 cyber events from 2008 to  2020, affecting 49,495 organisations across the world. Advisen is a US-based for-profit organisation which collects and processes cyber reports form reliable and publicly verifiable sources such as news media, governmental and regulatory sources, state data breach notification sites, and third-party vendors. Given that the interest in cyber risk is on the rise, many recent studies on cyber risk have made use of Advisen Cyber Loss Data \cite{romanosky2016, cyentia2020}. The understanding and classification of the array of cyber loss event and risk types is diverse and can differ by sectors and industry as well as varying over time. 

More than the 80\% of the events recorded affect organisations residing in the USA and for each event accident timeline, i.e. first notice date, accident date, loss start date, and loss end date, a detailed explanation are reported. One of the key advantages with respect other commonly used data sets, such as the Chronology of Data Breaches” provided by the Privacy Rights Clearinghouse (PRC), is that Advisen dataset gives direct information of monetary losses linked to cyber risk event, providing an empirical measurement of financial losses that can be then used for modelling purposes. 

Following \cite{eling2017} and \cite{edwards2016} we remove all the observations that do not give information on the monetary losses, and restrict the analysis to the observation for which complete information on company specific characteristics, such as yearly revenue and number of employees are available, leaving the total number of observations considered in this study to 3,792, corresponding to roughly the 2.6\% of the total. A detailed analysis of the basic attributes, non-statistical in nature is provided in the industry white paper \cite{shevchenko2021quantification} and an overview of how events are classified according to Advisen's own classification, based on the type of cyber threat, are provided in a detailed overview in \cite[Section 3]{malavasi2021cyber}.

Other possible classifications of cyber risk events are available in the literature, see \cite{eling2019} who suggests to divide cyber risk events into categories, according to operational risk classification: Actions by People, System and Technical Failure, Failed Internal Process, External Events. On the other hand, \cite{romanosky2016} provides cyber risk driven categories, such as Data Breach, Security Incident, Privacy Violation, Phishing Skimming, and Other. 

In this work we will instead work with industry related partitions based on the United States  NAIC Sector decomposition's widely used in insurance practice. The organisation of NAICS at a sector level produces 24 unique sector/subsector combinations in which to partition the loss data according to a wide variety of industry types. Such a cyber risk analysis has not previously been undertaken and we believe this will shed some interesting insight into how different sectors are coping with cyber threats that they all increasingly face in doing business in a digital environment.

\subsection{Basic Empirical Data Description and Attributes} 
\label{subsec:EmpData}
In this section we first provide a basic summary of the Advisen data under two decomposition's: by risk type and then secondly by NAIC sectors, for the US based company cyber risk loss histories.  The NAIC sector codes are provided at \url{https://www.naics.com/search/} and at the level 1 categorisation, those used by Advisen, correspond to: 11 `Agriculture, Forestry, Fishing and Hunting'; 21 `Mining, Quarrying, and Oil and Gas Extraction'; 22 `Utilities'; 23 `Construction'; 31 `Manufacturing Part A'; 32 `Manufacturing Part B'; 33 `Manufacturing Part C'; 42 `Wholesale Trade'; 44 `Retail Trade Part A'; 45 `Retail Trade Part B'; 48 `Transportation and Warehousing Part A'; 49 `Transportation and Warehousing Part B'; 51 `Information'; 52 `Finance and Insurance'; 53 `Real Estate and Rental and Leasing'; 54 `Professional, Scientific, and Technical Services'; 55 `Management of Companies and Enterprises'; 56 `Administration/Support/Waste Management/Remediation Services'; 61 `Educational Services'; 62 `Health Care and Social Assistance'; 71 `Arts, Entertainment, and Recreation'; 72 `Accommodation and Food Services'; 81 `Other Services (except Public Administration)'; and 92 `Public Administration'.

To ensure that records are related to the recent history of cyber risk data, we have excluded from analysis Advisen data that goes back to the 1950's as we are not confident on its accuracy or on its ability to reflect realistic cyber threat environments faced by modern corporations. In this regard we have selected a time window in which we take the earliest reported accident date as 01/01/1990 00:00 through to the most recent accident date recorded being 20/09/2020 00:00. Furthermore, we focus on the analysis of loss records that satisfy that the total loss amount was positive, ignoring many records that register empty or zero cells due to in-completion of the claim or non-settlement or payout. Below in Table \ref{tab:my-table} and Table \ref{table:descriptive} we show the summary statistics of the data used under each decomposition.

\begin{table}[]
\centering
\resizebox{\textwidth}{!}{%
\begin{tabular}{|c|l|l|l|l|l|l|l|l|l|l|}
\hline
NAIC Sector &
  N &
  Min &
  Max &
  Q1 &
  Median &
  Q3 &
  Mean &
  St. Dev. &
  Skew &
  Kurt \\ \hline
11 &
  2 &
  10,000 &
  2.70M &
  NA &
  NA &
  NA &
  1.36M &
  NA &
  NA &
  NA \\ \hline
21 &
  7 &
  16,600 &
  7.00M &
  39,185 &
  299,600 &
  814,000 &
  1.29M &
  2.55M &
  1.54 &
  0.64 \\ \hline
22 &
  45 &
  30 &
  10.03M &
  3,000 &
  42,337 &
  2.40M &
  1.51M &
  2.63M &
  1.90 &
  2.85 \\ \hline
23 &
  62 &
  300 &
  22.41M &
  6,162 &
  32,919 &
  718,663 &
  1.37M &
  3.90M &
  3.83 &
  14.81 \\ \hline
31 &
  31 &
  6,817 &
  84.00M &
  62,199 &
  310,000 &
  1.80M &
  4.48M &
  15.10M &
  4.70 &
  21.83 \\ \hline
32 &
  \multicolumn{1}{l|}{43} &
  500 &
  410.00M &
  25,488 &
  231,768 &
  4.20M &
  20.10M &
  78.00M &
  4.21 &
  16.61 \\ \hline
33 &
  117 &
  200 &
  2.00B &
  29,000 &
  487,000 &
  3.30M &
  27.61M &
  190.76M &
  9.57 &
  94.95 \\ \hline
42 &
  126 &
  44 &
  2.0B &
  20,000 &
  237,500 &
  1.8M &
  0.18M &
  178M &
  10.94 &
  118.86 \\ \hline
44 &
  \multicolumn{1}{l|}{280} &
  14 &
  298.00M &
  10,850 &
  265,887 &
  2.29M &
  6.98M &
  31.45M &
  7.19 &
  54.48 \\ \hline
45 &
  126 &
  60 &
  291M &
  5,987 &
  196,250 &
  1.68M &
  6.57M &
  28.95M &
  8.09 &
  73.11 \\ \hline
48 &
  51 &
  202 &
  177.00M &
  55,000 &
  300,000 &
  4.73M &
  13.19M &
  37.91M &
  3.38 &
  10.32 \\ \hline
49 &
  12 &
  1,632 &
  400.00M &
  8,642 &
  29,101 &
  193,981 &
  33.76M &
  115.34M &
  2.65 &
  5.48 \\ \hline
51 &
  668 &
  1 &
  5.00B &
  5,290 &
  37,740 &
  1.00M &
  12.05M &
  195M &
  25.05 &
  636.62 \\ \hline
52 &
  1027 &
  10 &
  460M &
  11,608 &
  98,005 &
  1.39M &
  43.44M &
  20.95M &
  13.67 &
  249.46 \\ \hline
53 &
  66 &
  1,295 &
  28.40M &
  15,250 &
  132,249 &
  1.20M &
  132,249 &
  4.33M &
  4.12 &
  20.20 \\ \hline
54 &
  442 &
  89 &
  4.00B &
  6,305 &
  50,180 &
  840,500 &
  17.01M &
  199.04M &
  18.45 &
  361.45 \\ \hline
55 &
  18 &
  4,057 &
  24.69M &
  15,416 &
  28,736 &
  313,070 &
  2.20M &
  5.90M &
  3.06 &
  8.74 \\ \hline
56 &
  737 &
  69 &
  1.35B &
  4,500 &
  23,500 &
  1.04M &
  6.36M &
  59.50M &
  17.59 &
  363.81 \\ \hline
61 &
  146 &
  800 &
  26.00M &
  21,250 &
  135,050 &
  726,153 &
  1.03M &
  2.99M &
  6.08 &
  42.74 \\ \hline
    62 &
  289 &
  100 &
  190.00M &
  26,469 &
  200,000 &
  875,000 &
  2.48M &
  13.42M &
  11.51 &
  145.43 \\ \hline
  71 &
  40 &
  600 &
  30.95M &
  51,980 &
  933,750 &
  3.80M &
  4.26M &
  7.72M &
  2.10 &
  3.36 \\ \hline
  72 &
  126 &
  27 &
  226.00M &
  20,000 &
  160,948 &
  2.28M &
  6.36M &
  24.16M &
  7.03 &
  56.03 \\ \hline
    81 &
  86 &
  48 &
  90.00M &
  9,183 &
  93,975 &
  688,023 &
  2.88M &
  11.49M &
  5.93 &
  38.43 \\ \hline
  92 &
  384 &
  200 &
  1.0B &
  20,850 &
  139,300 &
  633,600 &
  5.35M &
  57.16M &
  15.41 &
  250.38 \\ \hline
\end{tabular}%
}
\caption{NAIC Sectors and Descriptions used by Advisen data to partition loss data into industrial sectors that allow for the study of sector specificity in cyber event data. The columns represent: N - number of loss events, Mean- average total loss event, Median - typical total loss event, St. Dev. - dispersion of the total loss events, Skew and Kurt. - excess skewness and kurtosis relative to a Gaussian for total loss events.}
\label{tab:my-table}
\end{table}

\FloatBarrier
It will also be insightful to see equivalent summary statistics for the total losses also partitioned according to the Advisen risk type classifications, as shown in Table \ref{table:descriptive}.

\begin{table}[ht!]
	\centering
	\caption[Descriptive Statistics]{This table reports some descriptive statistics of cyber risk related losses aggregated by categories, expressed in million dollars. The losses exhibit great variability in terms of median and first four moments across the risk types. IT  –  Configuration/Implementation  Error, Privacy – Unauthorised Data Collection, and Industrial Controls have the highest average loss amongst all the cyber risk categories.}
	\label{table:descriptive}
	\begin{tabular}{l|cccccc}
		Risk Type                        &	N	& Mean	&  Median &  St. Dev. &  Skew	 & Kurt \\
		\hline
        Privacy - Unauthorised Contact or Disclosure &2237 & 3.698 &0.033 &25.844  &25.197 &799.476\\
        Privacy - Unauthorised Data Collection       &157  &40.283 &0.84  &401.093 &12.171 &150.848\\
        Data - Physically Lost or Stolen             &93   &24.974 &0.212 &207.490 &9.424  &90.201\\
        Identity - Fraudulent Use/Account Access     &914  &1.035  &0.028 &6.146   &10.562 &131.617\\
        Data - Malicious Breach                      &768  &20.975 &0.5   &176.715 &17.591 &361.688\\
        Phishing, Spoofing, Social Engineering       &161  &9.219  &0.516 &59.273 &10.726  &124.611\\
        IT - Configuration/Implementation Errors     &44   &6.065  &0.804 &22.852 &5.890   &37.472\\
        Data - Unintentional Disclosure              &131  &2.612  &0.250 &11.601 &9.147   &94.133\\
        Cyber Extortion                              &105  &0.634  &0.010 &3.177  &5.998   &39.287\\
        Network/Website Disruption                   &207  &7.933  &0.090 &46.196 &7.731   &63.533\\
        Skimming, Physical Tampering                 &87   &1.471  &0.051 &5.930  &6.855   &54.070\\
        IT - Processing Errors                       &33   &48.872 &0.925 &120.700 &2.826  &9.733\\
        Industrial Controls \& Operations            &5    &2.247  &0.040 &4.359   &1.457  &3.187\\
        Undetermined/Other                           &17   &1.890  &1.500 &2.647   &3.014  &11.627\\
	    \hline
		
	\end{tabular}
\end{table}
\FloatBarrier

Having explored the basic empirical statistics to summarise the NAIC sector data, we will also now provide some empirical statistical analysis of the data based around three interesting statistical quantities: the smoothed Hill plots of \cite{resnick1997smoothing}, power law Pareto based Quantile-Quantile plots, and an extremogram time series analysis of \cite{davis2009extreme}.

For an ordered independent and identically distributed (i.i.d.) sequence of $n$ losses, we will define the increasing sequence of order statistics by \(0 < X_{(1,n)}\leq X_{(2,n)}\leq \cdots\leq X_{(n,n)} \) and the decreasing sequence of order statistics by \(X^{(1,n)}\geq X^{(2,n)}\geq \cdots\geq X^{(n,n)}>0 \), such that $X_{(1,n)} = X^{(n,n)}$ and $X_{(n,n)} = X^{(1,n)}$, the Hill (1975) estimator, discussed in Section \ref{sec:NonParametricTails}, using \(k\) order statistics is given by 
\begin{equation}\label{Eqn:kthOrderHill}
H_{k,n}=\frac{1}{k}\sum_{i=1}^{k} \ln\left(\frac{X^{(i,n)}}{X^{(k,n)}}\right),
\end{equation}
which is the pseudo-likelihood estimator ($\widehat{\xi}=H^{(k,n)}$) of reciprocal of the tail index \(\xi=1/\alpha>0\) for regularly varying tails (e.g. Pareto distribution). 
Note, when $k$ is too small, only a few observations influence $\widehat{\alpha}$ and the variance of the estimator, given assymptotically by $\alpha^2/k$ is too large. When $k$ is too large, the assumption underlying the derivation of the estimator typically degrades and bias increases. 

Recall a few basic facts, where a positive measurable function $f$ is called regularly varying (at infinity) with index $\alpha \in \mathbb{R}$ if it is defined on some neighbourhood $[x_0,\infty)$ of infinity and
\[
\lim_{x \rightarrow \infty}\frac{f(tx)}{f(x)} = t^{\alpha}, \;\; \forall t > 0.
\]
An example of such a distributional law that satisfies the resulting power-law tail behaviour is the Pareto distribution which is given by 
\[
\mathbb{P}\left[X \leq x\right] = 1 - \left(\frac{x_m}{x}\right)^{\alpha}
\]
and which admits a heavy tailed power law tail in which the tail index $\alpha$ determines the degree of heavy tailedness. See a detailed discussion on heavy tailed loss models in \cite{peters2015advances} and the references therein.

The Hill estimator is defined on orders \(k>2\), as when \(k=1\) the $H^{(1,n)}=0$. Once a sufficiently low order statistic is reached the Hill estimator will be constant, up to sample uncertainty, for regularly varying tails. The Hill plot is a plot of $H^{(k,n)}$ against the \(k\). Symmetric asymptotic normal confidence intervals assuming Pareto tails are provided. To avoid well known challenges with interpreting the Hill plot, we have opted for the log scale smoothed Hill plot of \cite{resnick1997smoothing}: 
\begin{equation}
\widetilde{H}_{k,n}=\frac{1}{(r-1)k}\sum_{j=k+1}^{rk} H_{j,n},
\end{equation}
where $r$ is the smoothing factor and the order is also on a log scale which is equivalent to plotting the points \((\theta, H^{\lceil n^\theta\rceil, n})\) for \(0\le \theta \le 1\). 

\begin{figure}[h] 
\includegraphics[scale=0.3]{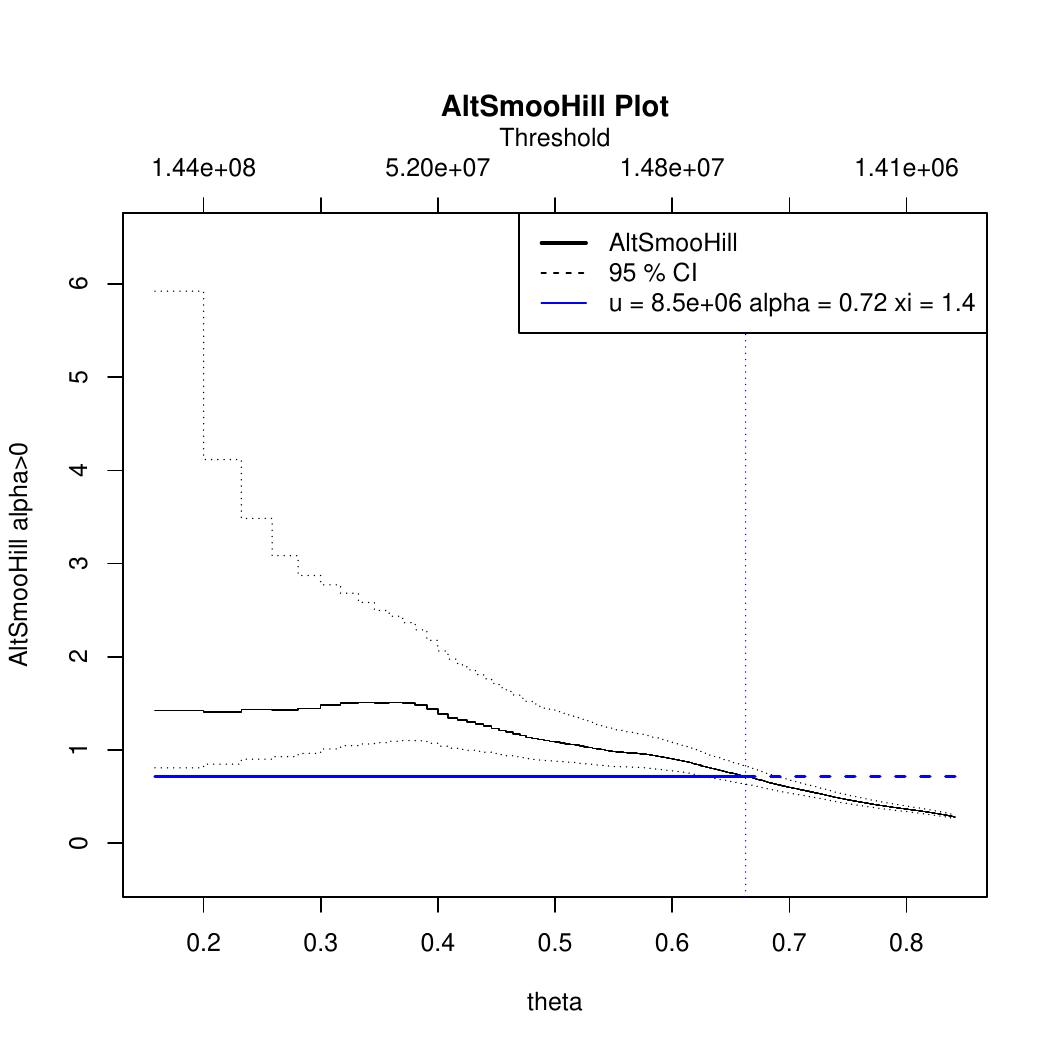}
\includegraphics[scale=0.3]{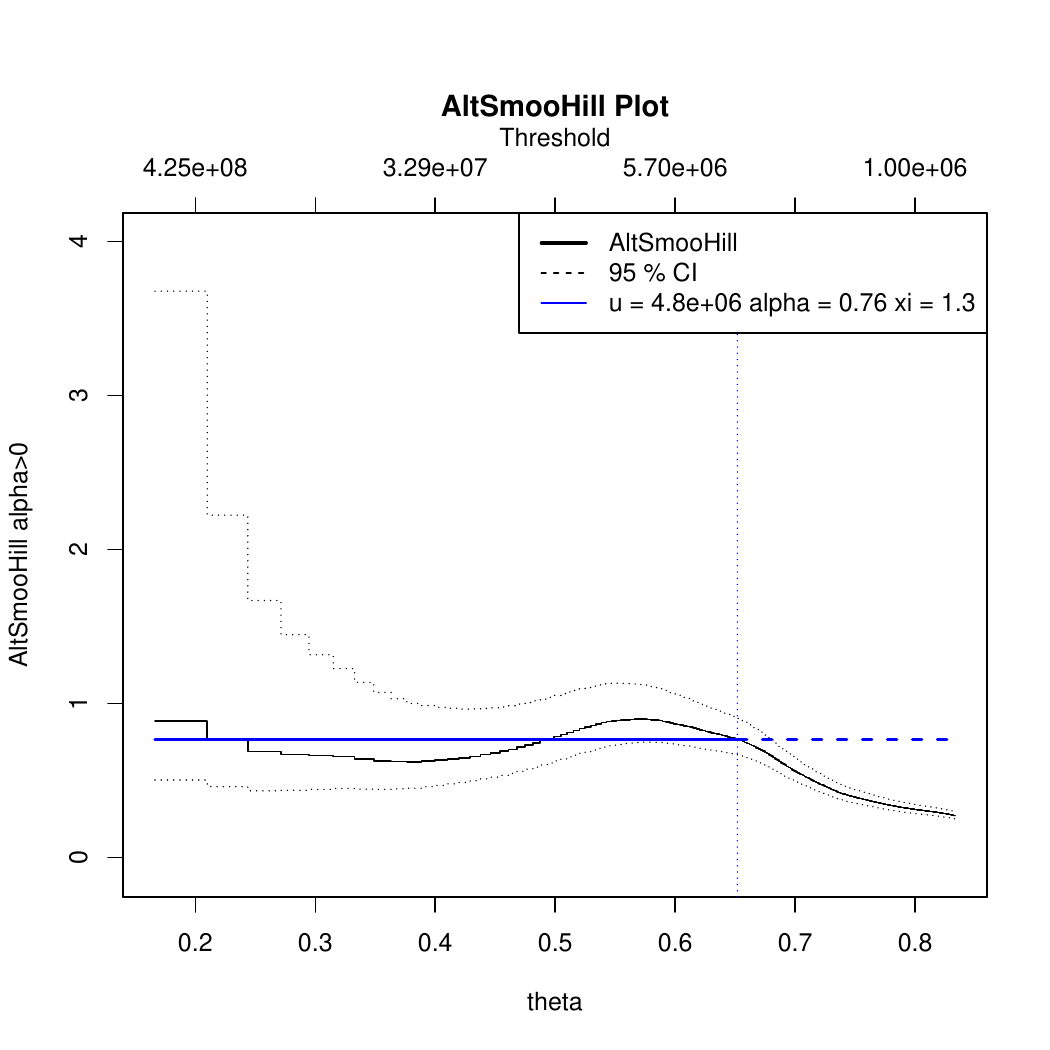}
\includegraphics[scale=0.3]{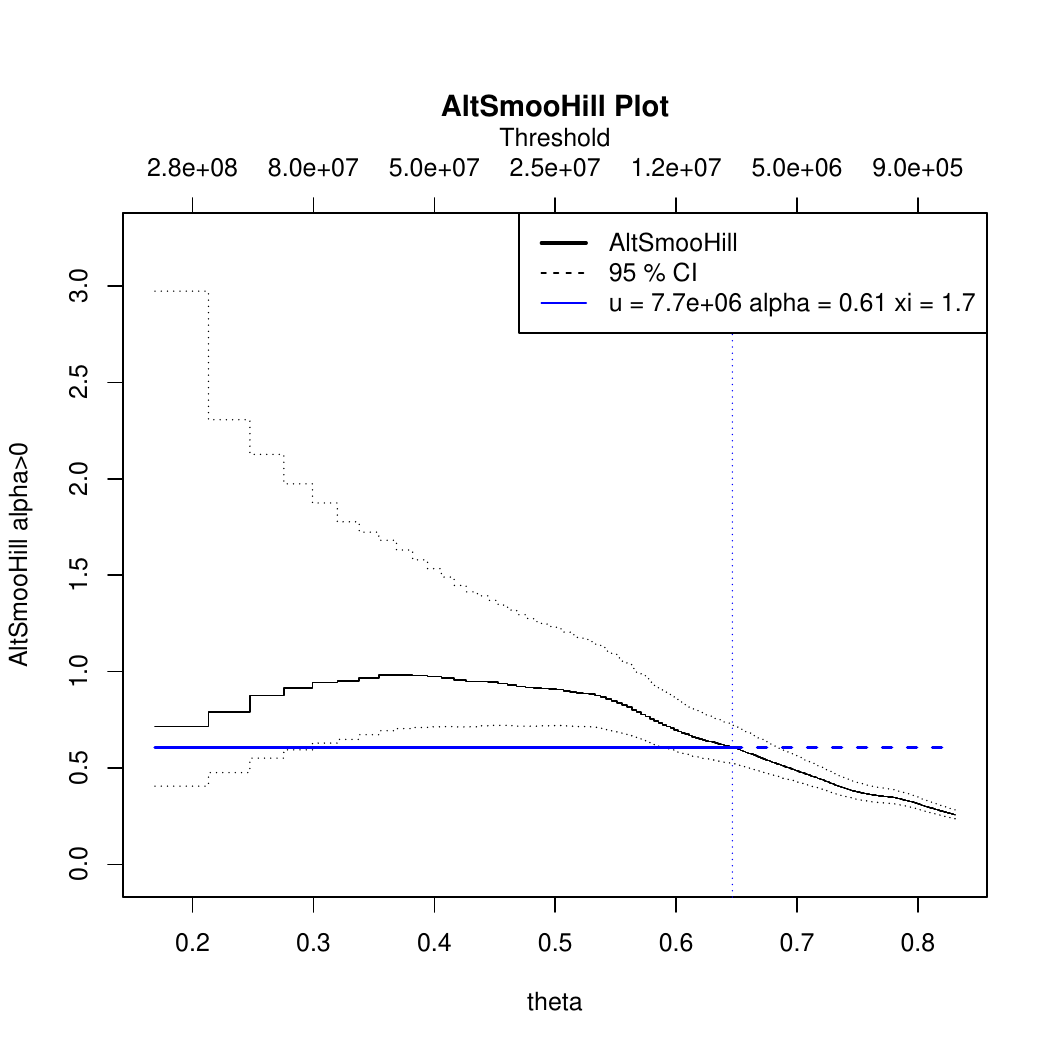}
\includegraphics[scale=0.3]{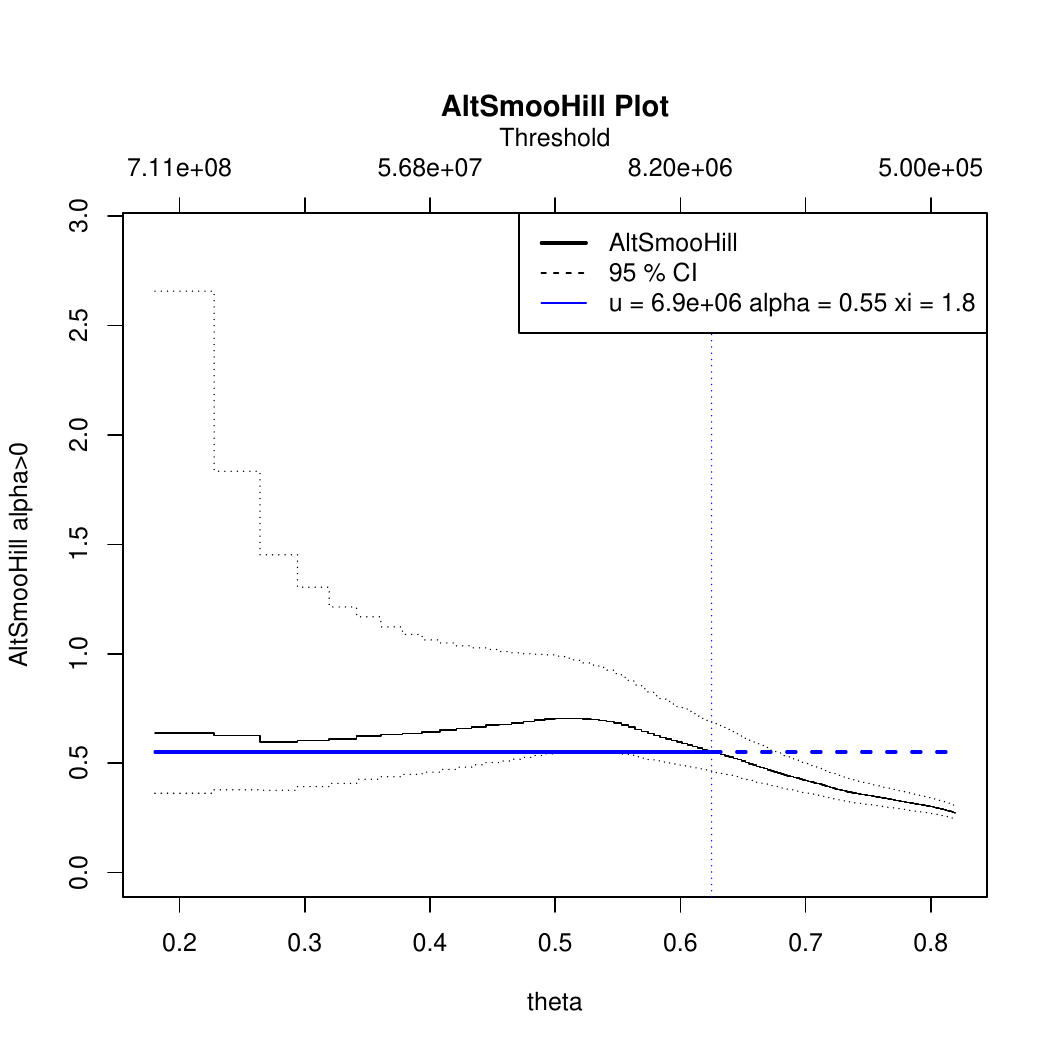}
\includegraphics[scale=0.3]{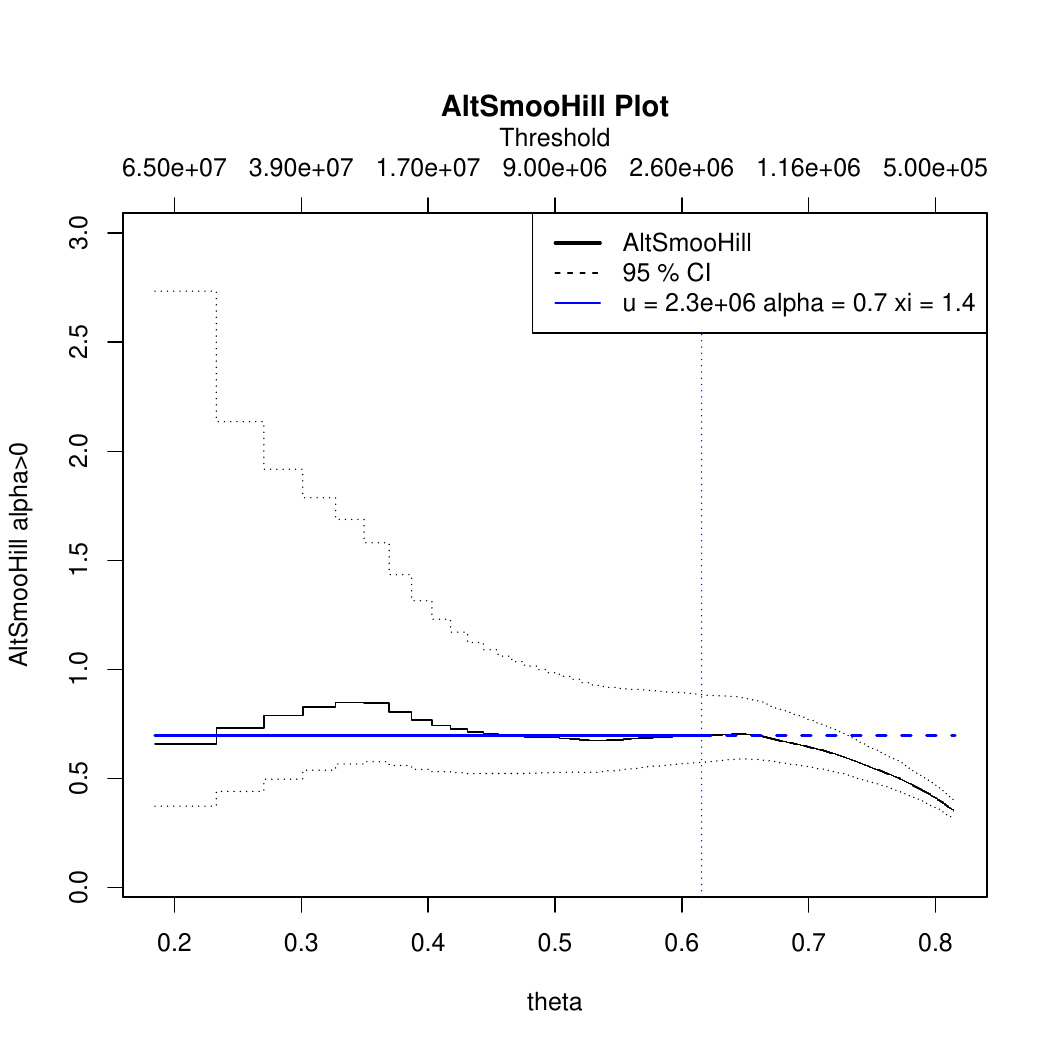}
\centering
\caption{Smoothed Hill Plots. Top Left Subplot: NAIC Sector 52. Top Middle Subplot: NAIC Sector 56
Top Right Subplot: NAIC Sector 51. Bottom Left Subplot: NAIC Sector 54. Bottom Right Subplot: NAIC Sector 92}
\label{Fig:UniExtremogramTop5}
\end{figure}

In Figure \ref{Fig:UniExtremogramTop5} we present results for the NAIC sectors with the top 5 largest loss count records corresponding, in order from highest to lowest: NAIC$=52$ `Finance and Insurance'; NAIC$=56$ `Administrative and Support, and Waste Management and Remediation Services'; NAIC$=51$ `Information'; NAIC$=54$ `Professional, Scientific and Technical Services'; and NAIC$=92$ `Public Administration'.

\FloatBarrier

\begin{figure}[h] 
\includegraphics[scale=0.3]{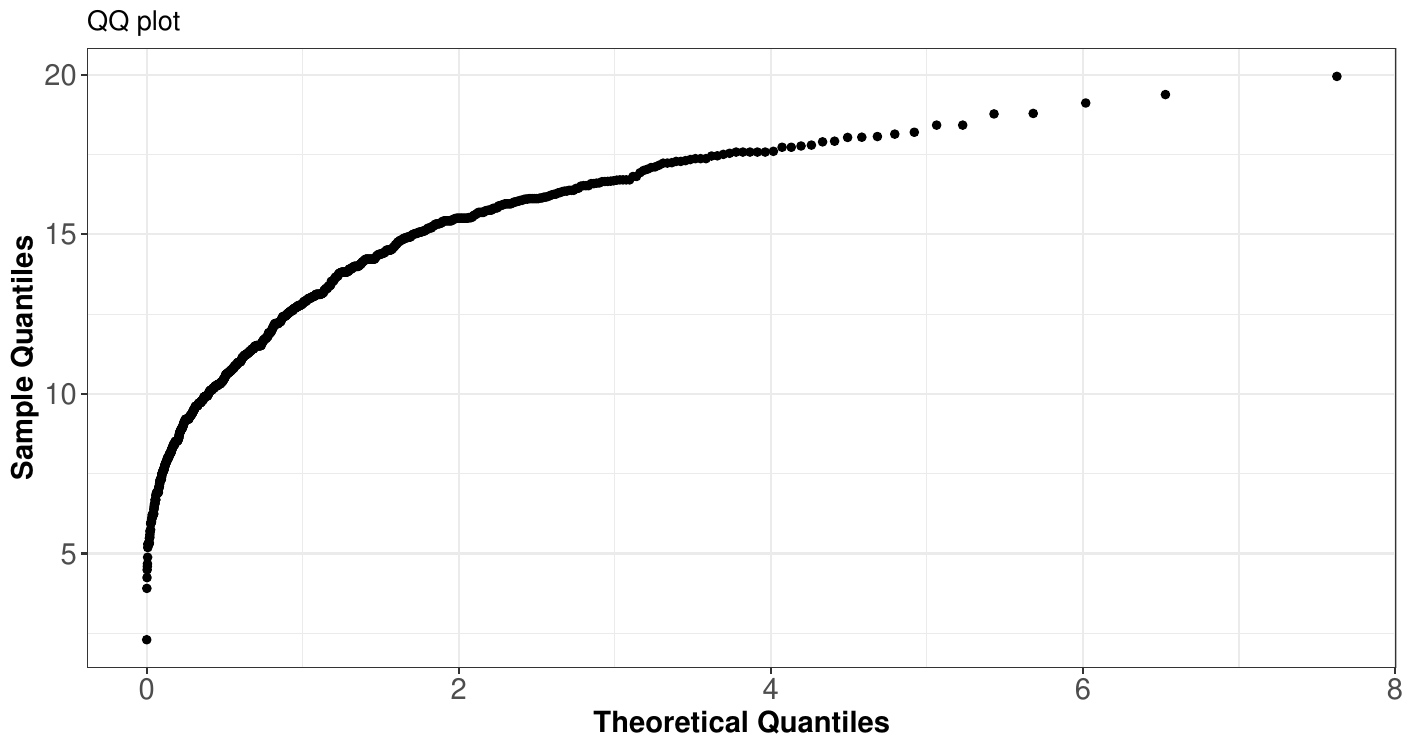}
\includegraphics[scale=0.3]{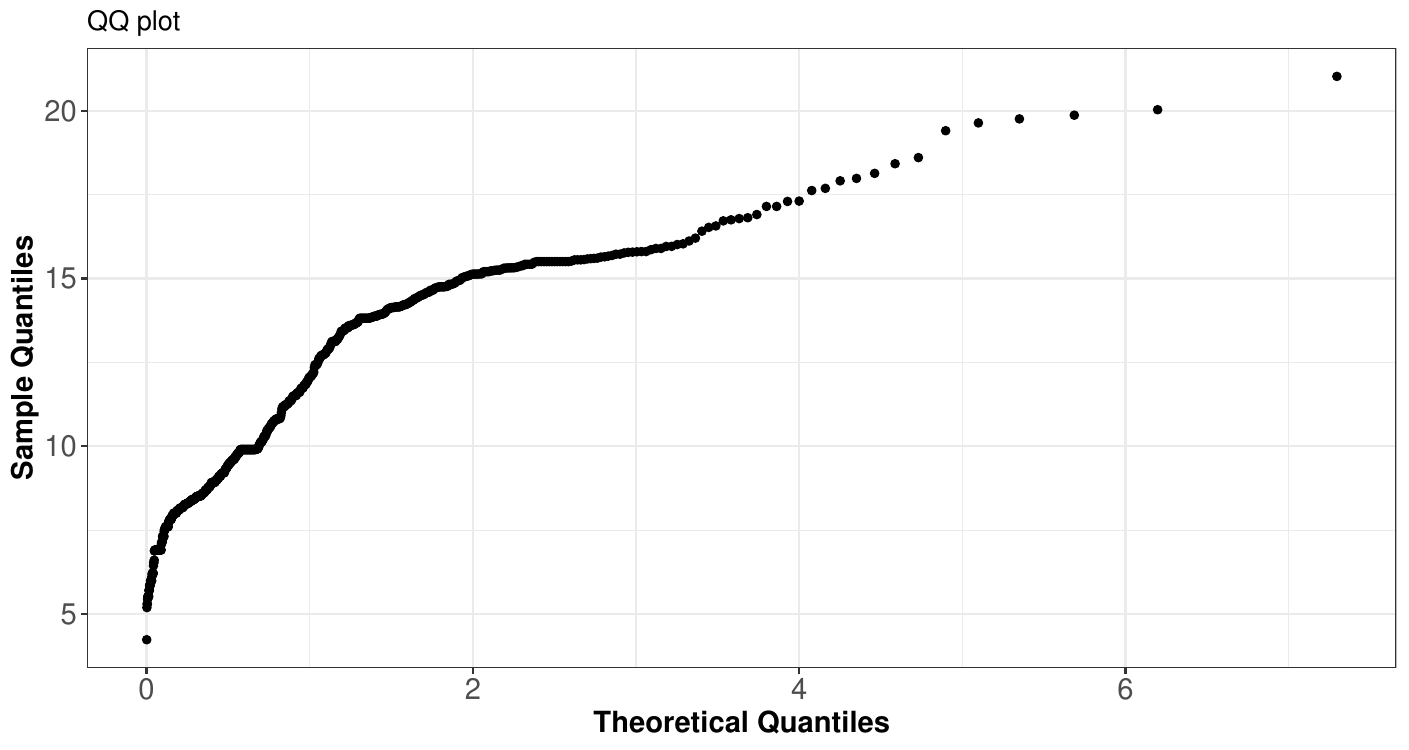}
\includegraphics[scale=0.3]{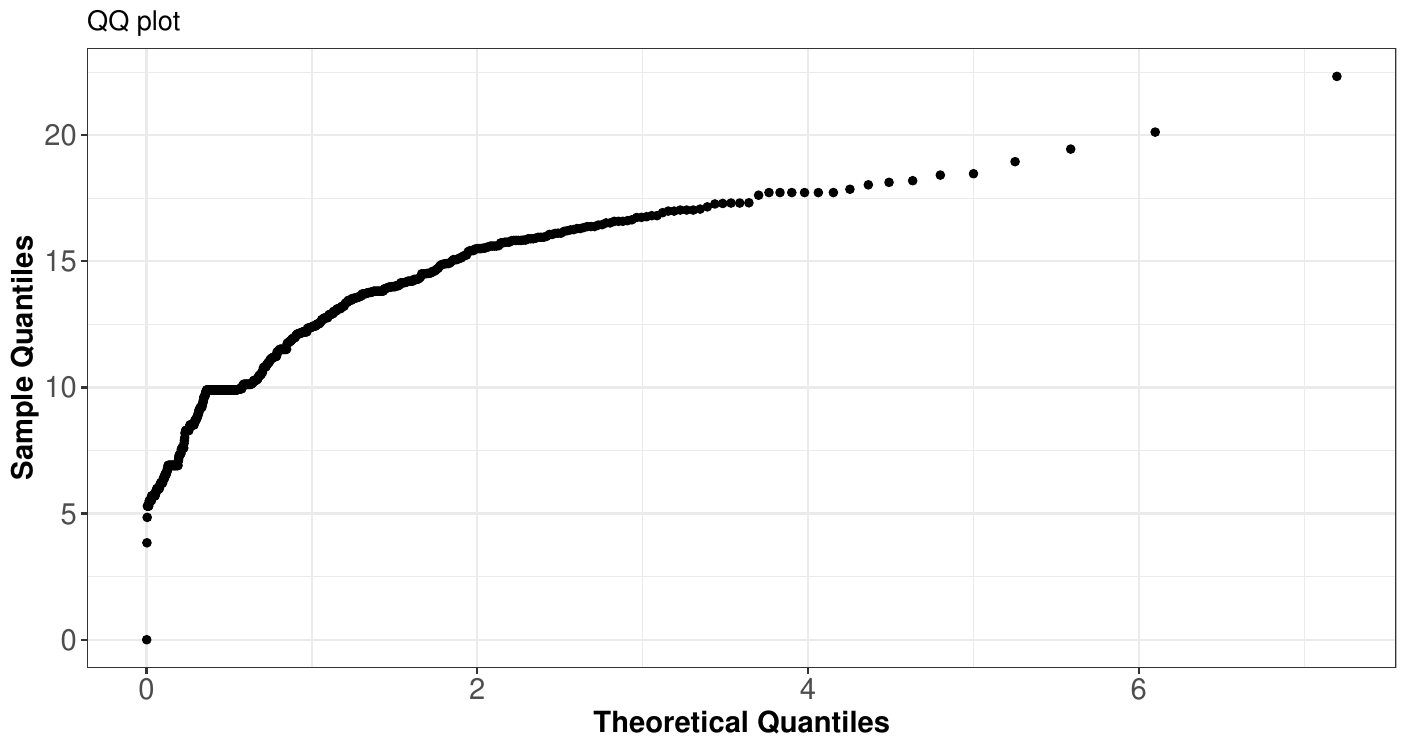}
\includegraphics[scale=0.3]{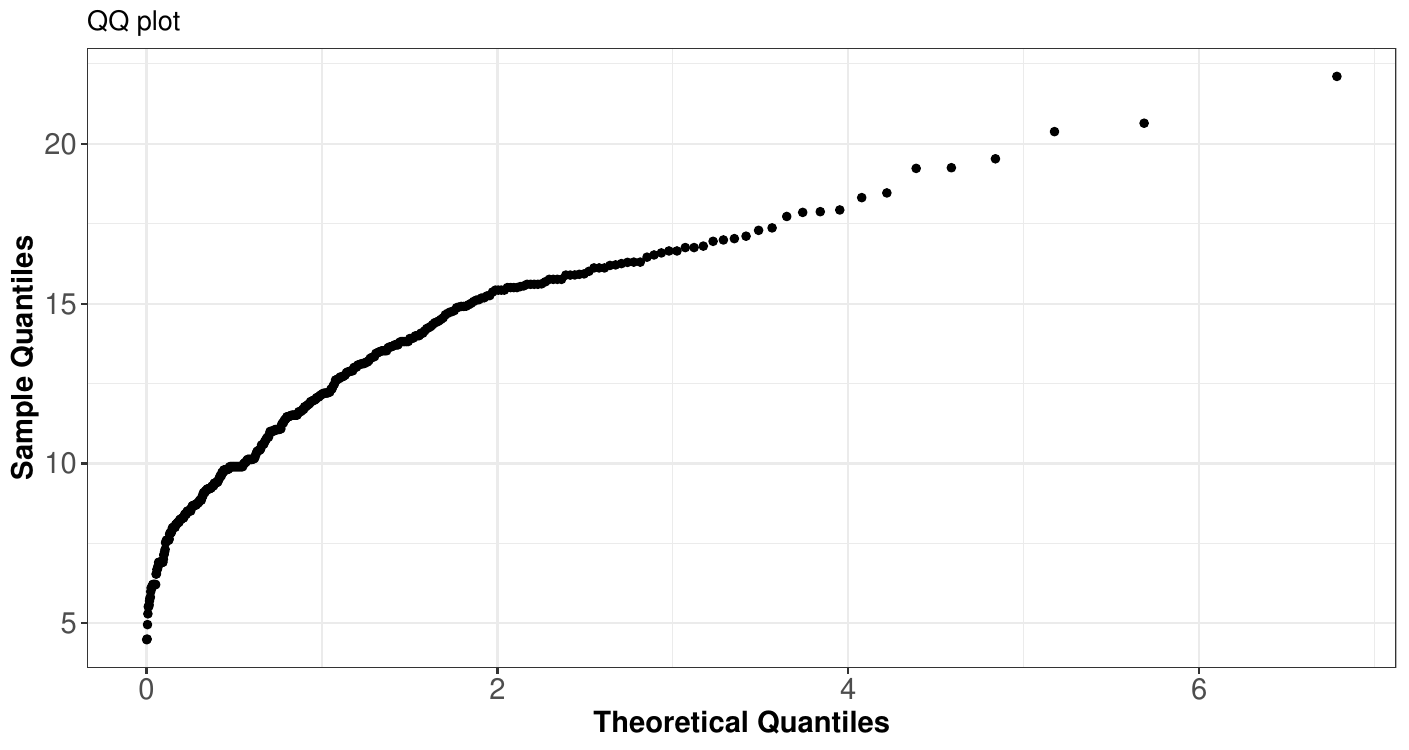}
\includegraphics[scale=0.3]{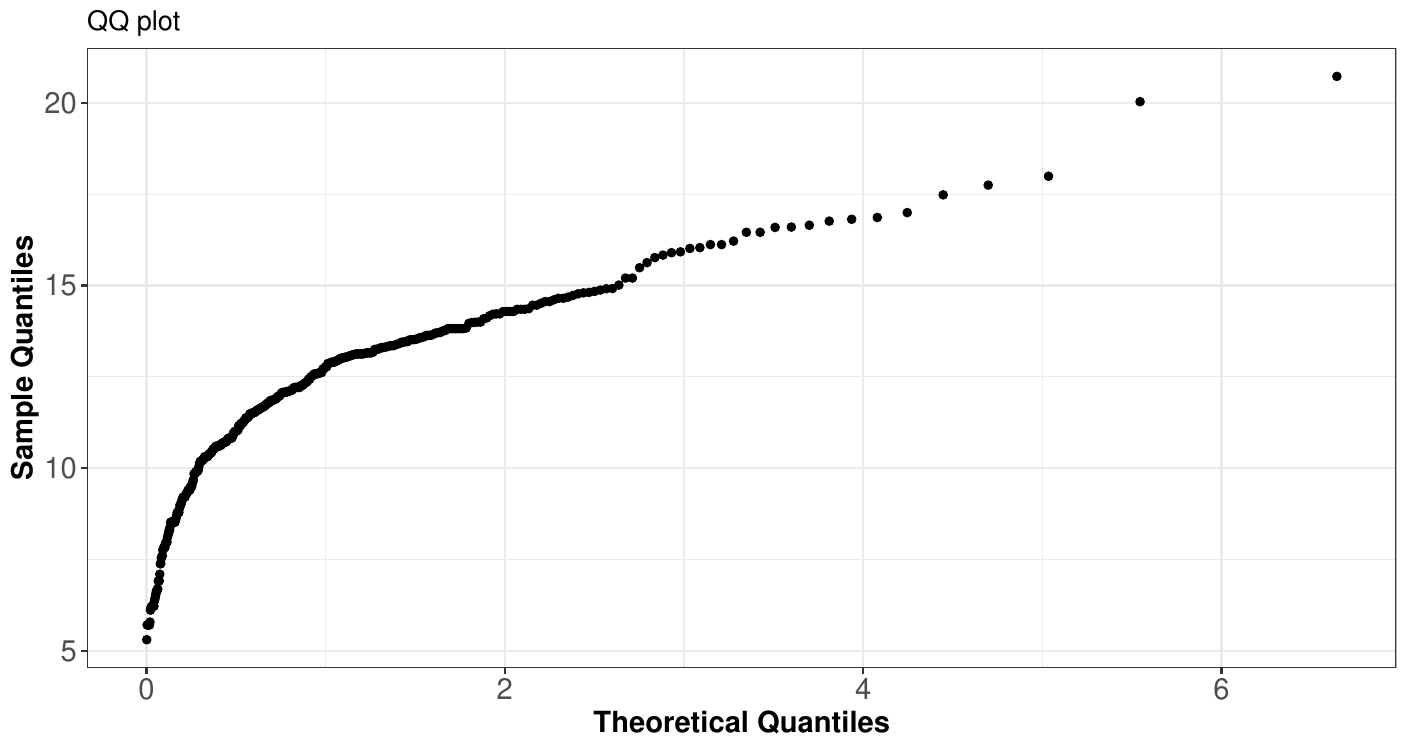}
\centering
\caption{Pareto Quantile-Quantile Plots. Top Left Subplot: NAIC Sector 51. Top Right Subplot: NAIC Sector 54. Bottom Subplot: NAIC Sector 92}
\label{Fig:QQPareto}
\end{figure}

The tail estimation methods proceeding this section will be based on statistical assumptions that relate to heavy tailed estimation of tail index based on a power law, a regular variation assumption or explicitly a Pareto law or asymptotic Pareto tail behaviour assumptions. Therefore, we also analyse the cyber loss data based on NAIC sectors for Pareto law type behaviour. We will do this in two ways, via Pareto Quantile-Quantile plots as shown in Figure \ref{Fig:QQPareto} and secondly in following analysis via a hypothesis test. 
 
To conclude this empirical analysis of the leading NAIC categories by total loss amounts, we will explore the extremal correlations as captured by the extremogram. This empirical estimator allows us to capture inter-temporal characteristics of the cyber risk loss data categorised by NAIC sectors and aggregated over quarterly periods from 1990 to 2021. The extremogram can be considered as a correlogram for extreme events and was introduced originally in works \cite{davis2009extreme} as a tool to measure the extremal dependence in $\mathbb{R}^d$-valued time series $(X_t)$. The extremogram is defined as a
limiting sequence given by
\begin{equation}\label{eq:extrem}
\gamma_{AB}(h)=\lim_{n \rightarrow \infty}\,{\rm cov}\bigl(I_{\{a_n^{-1} X_0\in A\}} ,I_{\{a_n^{-1} X_h\in B\}}\bigr)\,,\quad
  h\ge 0\,,
\end{equation}
with sequences $(a_n)$ suitably chosen as normalisation sequence and $A,B$ are two fixed sets bounded away from zero. A popular choice for the intervals $A$ and $B$ is to set $A=B=[q_{\alpha},\infty)$ with $q_{\alpha}$ being the $\alpha$-percentile of $(X_t)$. An example of selection for $A,B$ that is familiar to the actuarial audience will be to select $A=B=(1,\infty)$ which will reproduce the so-called upper tail dependence coefficient of the vector $(X_0,X_h)$ given as the limit
\begin{equation}\label{eq:rho}
\rho(h)=\lim_{x \to \infty} \mathbb{P}(X_h>x\mid X_0>x)\,.
\end{equation}

In the context used in this work to study cyber risk we will plot a sequence of extremograms marginally for each NAIC sector. Since this method requires regular time series and not an event driven time record set, we have aggregated the losses into quarterly time series of total losses between 1990 to 2020. For each NAIC sector we produce results for univariate extremograms. We will focus on indexed choices of $A=B \in \left\{q_{0.01},\ldots,q_{0.99}\right\}$ corresponding to quantile levels $\alpha \in \left\{0.01,\dots,0.99 \right\}$. Under this construction, as $\alpha \uparrow 1$ the events $\{X_0\in a_n A\}$ and  $\{X_h\in a_n B\}$ are increasingly considered as extreme ones and $\gamma_{AB}(h)$ measures the influence of the time zero extremal event $\{X_0\in a_n\,A\}$ on the extremal event $\{X_h\in a_n\,B\}$, $h$ lags apart, i.e. $h$ quarters later. This will result in construction of a matrix of values denoted by $\Gamma$, whose $i,j$-th element is given by $\Gamma_{ij}:=\gamma_{A_i=B_i=q_{\alpha_i}}(j)$ for $j$-th quarter from 1990 such that $j \in \left\{1,2,\ldots,124\right\}$ and for quantile thresholds $q_i \in \left\{0.01,0.02,\ldots,0.99\right\}$. 

\begin{figure}[h] 
\includegraphics[scale=0.3]{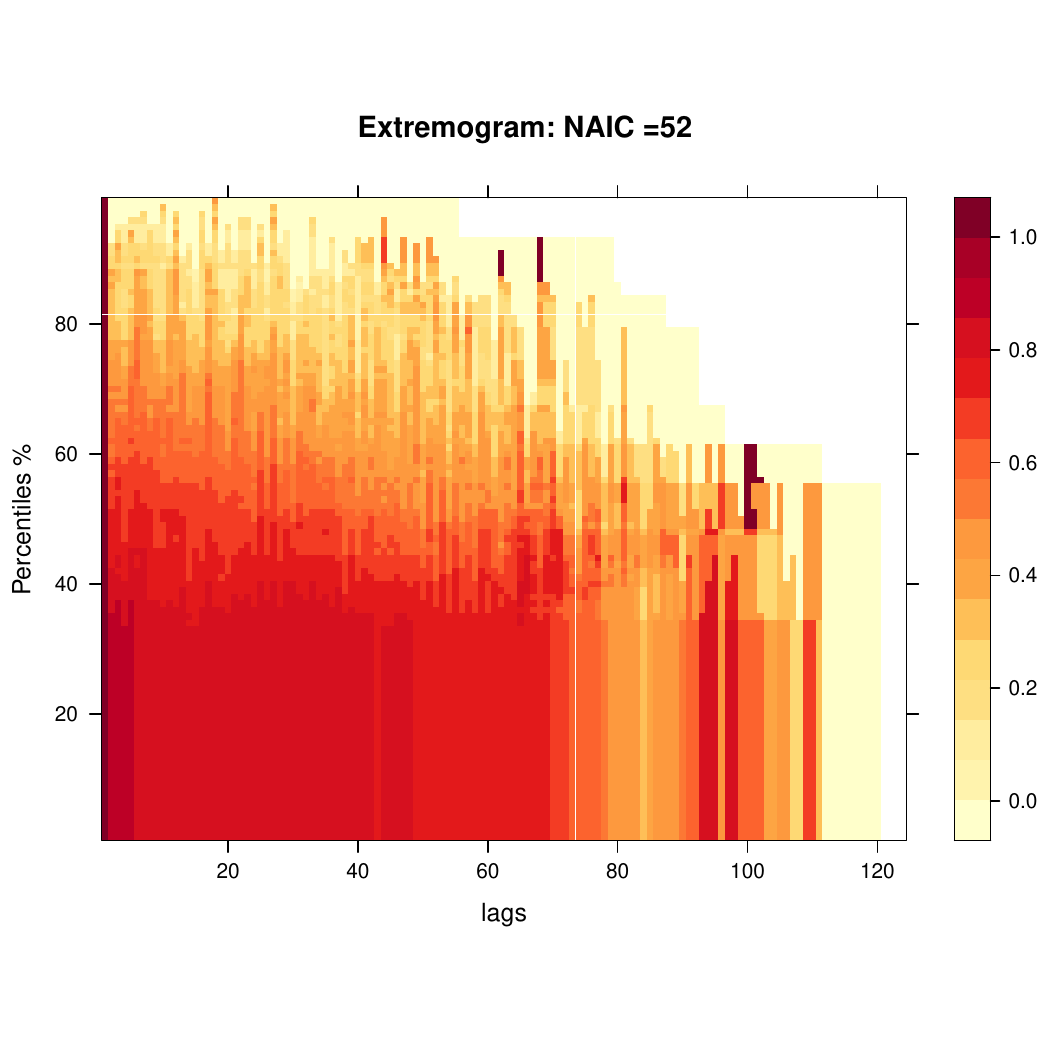}
\includegraphics[scale=0.3]{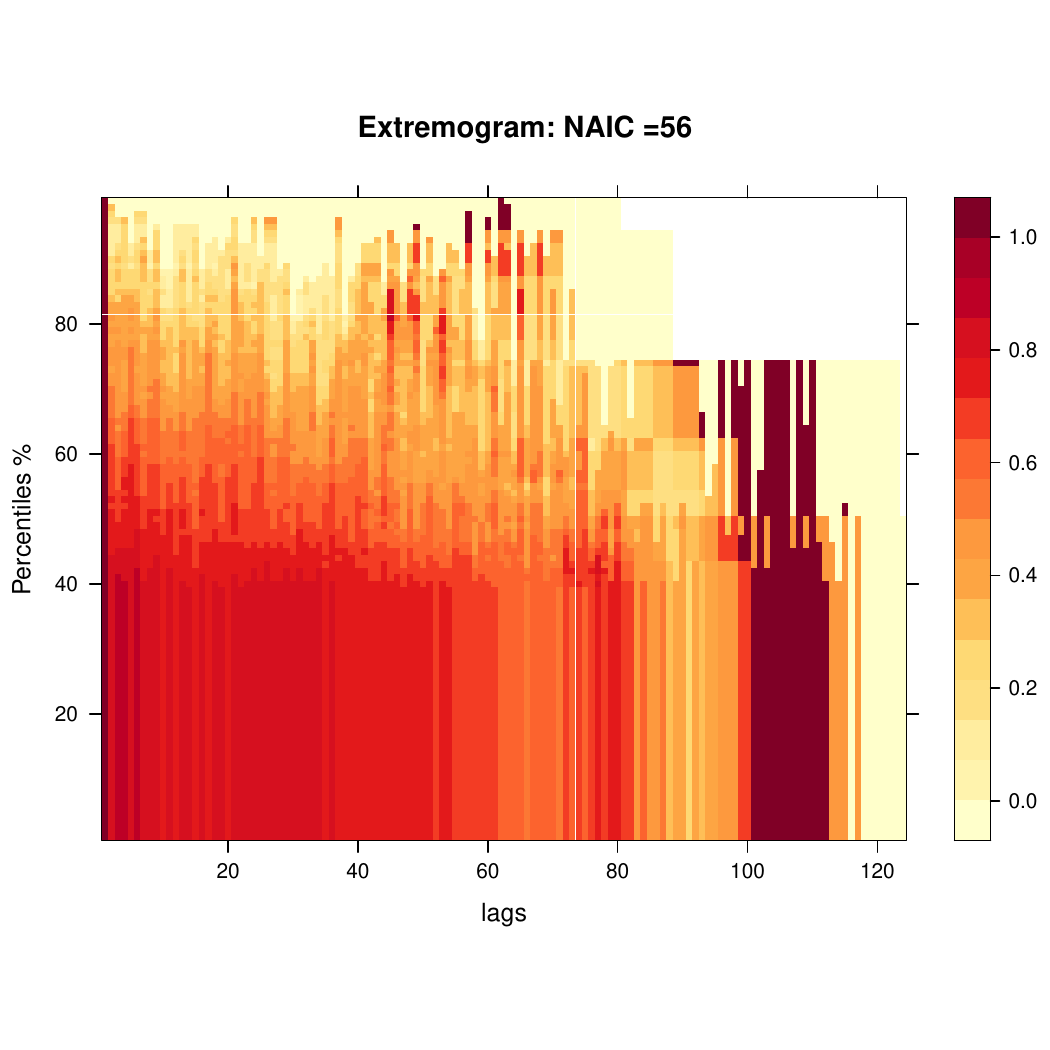}
\includegraphics[scale=0.3]{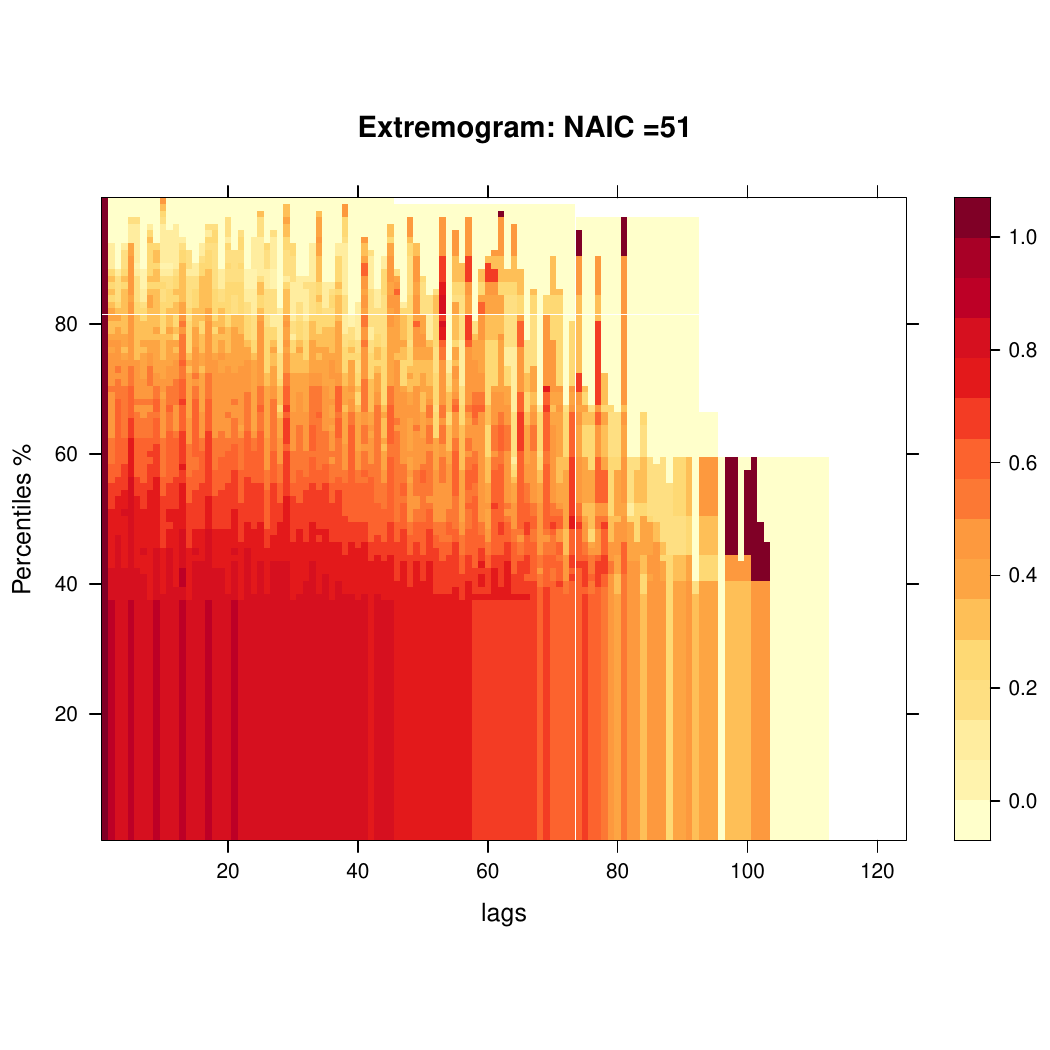}
\includegraphics[scale=0.3]{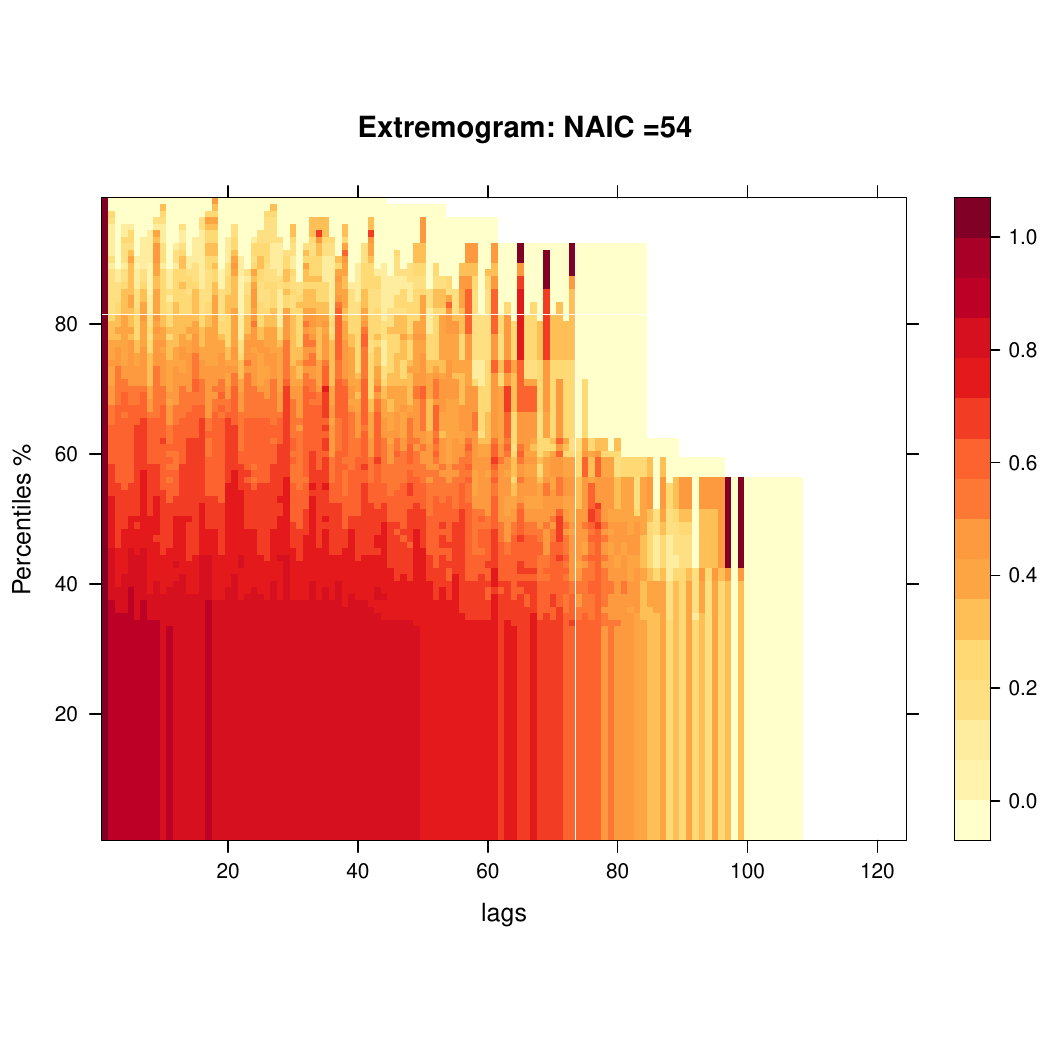}
\includegraphics[scale=0.3]{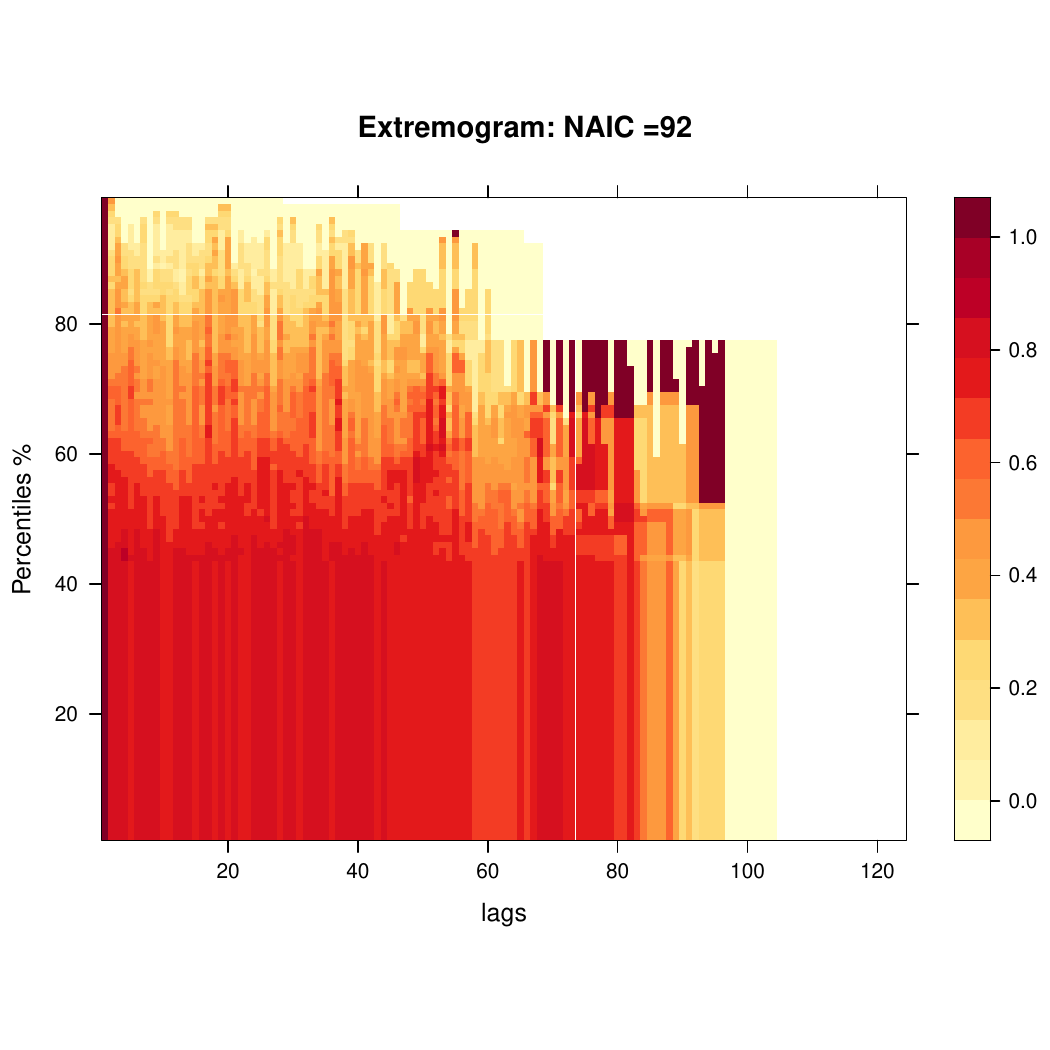}
\centering
\caption{Extremogram Plots. 
Top Left Subplot: NAIC Sector 52. Top Middle Subplot: NAIC Sector 56
Top Right Subplot: NAIC Sector 51. Bottom Left Subplot: NAIC Sector 54. Bottom Right Subplot: NAIC Sector 92}
\label{Fig:Extremograms}
\end{figure}

We note that in presenting these extremograms in matrix $\Gamma$, whilst the finite quantile sequences for each row $i$ always exist for finite $\alpha_i$, the limit as $\alpha_i \uparrow 1$ need not exist. As studied in \cite{davis2009extreme}, it is sufficient for existence of the limits $\gamma_{AB}(h)$ to assume a regularly varying sequence of quarterly loss random variables $(X_t)$, which has power law tails for every lagged vector $(X_1,\ldots,X_h)$, $h\ge 1$. This assumption whilst not required to study finite sample extremogram profiles, is required if one wanted to look for extremal asymptotic tail dependence within a NAIC sectors losses using this methodology. The results of the extremogram analysis for the current selection of the NAIC sectors, selected according to the top 5 with largest loss counts, is provided in Figure \ref{Fig:Extremograms}

\section{Quantifying Heavy Tails in Cyber Risk Loss Models} 
\label{sec:NonParametricTails}
In this work we seek to study loss processes which admit heavy tailed annual loss distribution profiles in the context of cyber risk losses. We will be interested in a variety of ways of classifying cyber risk losses annually by risk type and business sector. It will be valuable to first explain some basic background on how we will seek to quantify heavy tailed loss data both non-parametrically and parametrically with a severity model. We will assume throughout that losses will take a positive support and as such the right tail of the loss distribution is of interest when quantifying heavy tailed loss behaviour.

There is no unique way to characterise universally the notion of a heavy tailed distribution, and as such numerous definitions and characterisations have been proposed. We will explore a few key characterisations in this section. Starting with a widely used concept that a heavy tailed loss model is characterised via the existence of moments. Under this characterisation, a heavy tailed loss model $F$ will not have finite moments of some order, and the heavier the tail, the fewer moments will exist. A simple condition that shows this relationship is stated in the following Lemma \ref{Lemma:TailFunctionMoments}.

\begin{lemma} \label{Lemma:TailFunctionMoments}
The distribution $F$ possesses an absolute moment of order $\alpha > 0$ if and only if (iff) $|x|^{\alpha -1}\left[1 - F(x) + F(-x)\right]$ is integrable over $(0,\infty)$.
\end{lemma}

Another way to characterise heavy tailed models that is also often explored in risk modelling theory, is to state that heavy-tailed distributions are probability distributions whose tails are not exponentially bounded: that is, they have heavier tails than the exponential distribution.  Under this characterisation, one considers those distributions for which the moment-generating function does not exist on the positive real line such that
\begin{equation*}
\int e^{sx}dF(x) = \infty, \; \forall s > 0.
\end{equation*}
In other words, take the standard Markov's inequality with $\psi$ a monotonically increasing nonnegative function for the nonnegative reals, $X$ is a random variable, $a \geq 0$, and $\psi(a) > 0$, then applying
$$
\mathbb{P}(|X|\geq a)\leq \frac{\mathbb{E}(\psi(|X|))}{\psi(a)}
$$
for the exponentially decaying `light' tail behaviour of a loss distribution 
\begin{equation*}
\overline{F}(x) \leq \exp(-sx)\mathbb{E}\left[\exp(sX)\right], \;\; \forall x > 0
\end{equation*}
does not apply.

In addition to multiple characterisations, there are also numerous ways to represent and study a heavy tailed loss distribution, beyond just the distribution function, that will be useful to briefly recall.

\begin{definition}[Hazard Function and Hazard Rate]\label{DefnHazardFnRate} 
For a loss distribution $F$ on $\mathbb{R}^+$, the hazard function is given by \begin{equation*}
R(x) = -\ln \overline{F}(x).
\end{equation*}
If the loss distribution $F$ has a loss density $f$, then such a distribution's tail behaviour can be characterised also by the hazard rate, given by 
\begin{equation*}
r(x) := \frac{dR(x)}{dx} = \frac{f(x)}{1-F(x)} = \frac{f(x)}{\overline{F}(x)} = -\frac{\overline{F}_X(x)'}{\overline{F}_X(x)}.
\end{equation*}
\end{definition}

\begin{lemma}[Hazard Rate of a Loss Distribution]\index{Hazard Rate}
The following three right limiting possibilities of the hazard rate function of a loss distribution characterise its tail behaviour:
\begin{enumerate}
\item[1]{ If $\lim_{x \rightarrow \infty} r(x) = 0$, then the loss distribution $F$ will be a heavy-tailed distribution function.}
\item[2]{If $\lim_{x \rightarrow \infty} r(x) > 0$, then the loss distribution $F$ is not heavy tailed and the exponential moments will exist up to $\lim \inf_{x \rightarrow \infty} r(x) > \lambda > 0$}
\item[3]{If $\lim_{x\rightarrow \infty} r(x)$ does not exist but one has $\lim \inf_{x \rightarrow \infty} r(x) = 0$, then the distribution $F$ can be either heavy or light tailed, and one needs further information to determine the characteristics.}
\end{enumerate}
\end{lemma}

One can summarise the relationship between different representations of a heavy tailed loss distribution as follows \cite[Theorem 2.6]{foss2011introduction}: 
\begin{theorem}
For any distribution $F$, the following assertions are equivalent: 
\begin{itemize}
\item $F$ is a heavy-tailed distribution.
\item $\overline{F}$ survival function is heavy-tailed.
\item Corresponding hazard function $R$ satisfies $\liminf_{x \rightarrow \infty} R(x)/x = 0$. 
\item For some fixed $T > 0$, the function $F$ on interval $(x,x+T]$ is heavy-tailed. 
\item If distribution $F$ is absolutely continuous with density function $f$ then if $F$ is heavy-tailed, the density $f$ is also heavy-tailed.
\end{itemize}
\end{theorem}

There are numerous more refined categorisations of heavy-tailed distributions, we will recall an important class of heavy tailed loss models, those that correspond to the regularly varying tail behaviour characterisation.

\begin{definition}[Regularly Varying Tail]
A probability distribution $F$ has regularly varying tails, $\overline{F} \in \mathcal{R}$ iff for some $\alpha \geq 0$ and any $y > 0$, it holds that
$$
\lim_{x\rightarrow \infty}\frac{\overline{F}(xy)}{\overline{F}(x)} = y^{-\alpha}.
$$
\end{definition}

We say a function $L(x)$ is slowly varying if $L \in \mathcal{R}$ and $\alpha = 0$. 

If we consider the characteristic function of loss random variable $X$, or of $F$, is $\varphi$, defined for all $t$ by 
$$
\varphi(t) = \int_{-\infty}^{\infty} \exp\left(itx\right)dF(x),
$$
then another way to characterise the heavy tailed nature of the loss random variable is through the relationship between the value of $F(x)$ for large $x$ and the value of $\varphi(t)$ for small $t$ in the neighbourhood of the origin. We will briefly rewrite the characteristic function as follows:
$$
\varphi(t)=\int_{-\infty}^{\infty}\cos(tx)dF(x) + i \int_{-\infty}^{\infty}\sin(tx)dF(x) = U(t) + i V(t). 
$$
Now, we will consider the distribution function re-expressed in terms of the tail sum $H(x) = 1 - F(x) + F(-x)$ which in the case of loss distribution models in which the support is positive, this reduces to $H(x)=\overline{F}(x)$. 

To proceed, we will assume $\overline{F}(x)$ is regularly varying at infinity, i.e.
$$
\overline{F}(x) = x^{-\alpha}L(x), \;\; \text{as } x \rightarrow \infty,
$$
where $\alpha > 0$ and $L(x)$ is slowly varying at infinity. Now consider the characteristic function of $X$ for all real $t$ split into real and imaginary components and integrate by parts to obtain
$$
1 - U(t) = t\int_{0}^{\infty}\sin(tx)\overline{F}(x)dx.
$$
which means that the behaviour of the tail sum depends only on $U(t)$ the real part of the characteristic function. To see this proceed as follows:
\begin{equation*}
\begin{split}
\varphi(t) &= \int_{-\infty}^0 e^{itx}dF(x) + \int_{0}^{\infty} e^{itx} d\left[F(x)-1\right]\\
&= \int_{-\infty}^0 \cos(tx)dF(x) + \int_{0}^{\infty}\cos(tx)d[F(x)-1] \\
& \;\; + i\left\{\int_{-\infty}^0 \sin(tx)dF(x) + \int_{0}^{\infty}\sin(tx)d[F(x)-1]\right\}.
\end{split}
\end{equation*}
Integrating by parts gives
\begin{equation*}
\begin{split}
\varphi(t) &= F(0) + t \int_{-\infty}^0 \sin(tx)F(x)dx \\
& - (F(0)-1) + t\int_{0}^{\infty}[F(x)-1]\sin(tx)dx \\
& - it \left\{\int_{-\infty}^0 \cos(tx)F(x)dx + \int_{0}^{\infty}\cos(tx)[F(x)-1]dx\right\}.
\end{split}
\end{equation*}
Which then allows one to obtain for the real component of the characteristic function $\varphi(t)$ the identity
\begin{equation*}
\begin{split}
U(t) - 1 &= t \int_{-\infty}^0 \sin(tx)F(x)dx +  t \int_{0}^{\infty} [F(x)-1]\sin(tx)dx\\
&= -t \int_{0}^{\infty} \sin(tx)F(-x)dx + t\int_{0}^{\infty} [F(x)-1]\sin(tx)dx\\
&= -t \int_{0}^{\infty} \sin(tx)H(x)dx.
\end{split}
\end{equation*}

This result was a critical part of the work of \cite{pitman1968behaviour} who went on to show that for infinite variance loss random variables with tail sum function $H(x)$, of index of regular variation $0<\alpha<2$, that as $x \rightarrow \infty$, one has the following relationship between the real part of the characteristic function near the origin and the regularly varying tail function
$$
1 - U(t) \sim s(\alpha)H(1/t) = s(\alpha)L(1/t)t^{\alpha}, \;\; \text{ as } t\downarrow 0,
$$
with 
$$
s(u) = \begin{cases}
\frac{\pi/2}{\Gamma(u)\sin(u\pi/2)}, & \text{if}\;\; u >0 \\
1, & \text{if} \;\; u=0,
\end{cases}
$$
with $s(u)$ finite for any $u$ not an even positive integer. Other cases for $\alpha =2$ and $\alpha > 2$ were studied but are not directly of relevance to the work in this paper where we concentrate of heavy tailed models with non-finite variance or non-finite mean.

Hence in summary, as $t \downarrow 0$, depending on value of tail index parameter $\alpha$ one can obtain the relationship
\begin{equation}\label{EqnHurstTheory}
\ln\left(1 - U(t)\right) = \begin{cases}
\ln\left(s(\alpha) L(1/t)\right) + \alpha \ln t, & \text{ if } 0<\alpha<2\\
\ln\left(\int_0^{1/t}xH(x)dx\right) + 2 \ln t, & \text{ if } \alpha=2\\
\ln\left(\frac{\mu_2}{2}\right), & \text{ if } \alpha > 2.
\end{cases}
\end{equation}
This identity is the precursor to tail estimators such as Hill estimators, which we will use to study in the non-parametric analysis of cyber risk loss data in Section \ref{sec:NonParametricTails}. To understand this, consider taking loss data and estimating the real part of the empirical characteristic function using the observed cyber risk loss samples $\left\{X_1, \ldots, X_n \right\}$ assumed i.i.d. from $F$ to produce
$$
\varphi_n(t) = \frac{1}{n} \sum_{j=1}^n \exp\left(i t X_j\right)
$$
and therefore the empirical estimator for the real component is $U_n(t) = \frac{1}{n} \sum_{j=1}^n \cos\left(t X_j\right)$. From this one can then define the empirical quantity for a grid of values $t_1, \ldots, t_N$ around the origin which produce tuples $\left\{\left(t_i,\ln\left[1-U_n(t_i)\right]\right)\right\}$ which can be regressed given a general assumption about the slowly varying tail function $L$ to produce an estimator for the tail index $\widehat{\alpha}_n$ based on linear regression obtained in the general form
\begin{equation} \label{Eqn:RegCharFn}
\ln\left(1-U_n(t_j)\right) \sim \ln C  + \alpha\ln t_j + \ln \frac{1- U_n(t_j)}{1- U(t_j)} , \;\;\;\; j = 1, 2,\ldots, m,
\end{equation}
where one can assume that this looks like a simple linear regression with 
$y_i = \ln\left(1-U_n(t_j)\right)$, $Z_j = \ln t_j$ and error $\epsilon_j = \ln \frac{1- U_n(t_j)}{1- U(t_j)}$ which then admits a least squares estimator for $\alpha$. This method is interesting as it is basically non-parametric and relies upon an estimator formed from the empirical characteristic function around the origin. Note, that in such estimators for the tail index $\alpha$, the bias in the estimation will be directly related to the assumption regarding the model for $L(x)$ which clearly enters through $U$.

Other methods of similar nature make more specific assumption about $H(x)$ and $L(x)$ or at least about the asymptotic functional form of these quantities as $x \rightarrow \infty$. This naturally then leads to classes of estimators such as the Hill estimator \cite{hii1975simple} and generalisations such as those discussed in an excellent work exploring aspects of tail index estimation by Peter Hall, see \cite{hall1982some}. This basic observation of using an asymptotic relationship and a simple linear regression to estimate the tail index can be studied in a plethora of other related approaches, we will not review all of these in this work. Our focus in this work will be on two classes of general estimator for tail index consistent with a regular variation assumption, those based on characteristic function asymptotics and those based on maximum likelihood principle. Within these two categories there are in fact more than 100 currently known estimators for the tail index based various different assumptions about $H(x)$ and its tail behaviour, see a comprehensive account many of these estimators in \cite{fedotenkov2020review}. This aforementioned review is a very comprehensive and well written review of univariate Pareto-type tail index estimators for i.i.d. non-truncated data. With regard to the maximum likelihood based approaches, the classic approach most widely used by practitioners is known as the Hill estimator. We discuss in this paper a few different variants of this estimator that can treat bias in the cyber risk data collection and robustness considerations. 

In both contexts we will consider application of extensions that build upon the work of Peter Hall in \cite{hall1982some} where it was proposed to consider particular types of tail sum generic functional forms of the slowly varying function $L(x)$ that will parametrize $\overline{F}(x) = x^{-\alpha}L(x)$ as $x \rightarrow \infty$ in the regularly varying tail function class,
\begin{equation}\label{Eqn:Hall}
L(x) = C\left[ 1 + Dx^{-\beta} + o(x^{-\beta})\right]
\end{equation}
for $C>0$, $\alpha > 0$, $\beta > 0$ and $D$ is a non-zero real number, which gives tail sum function 
\begin{equation}\label{Eqn:HallL}
H(x) = Cx^{-\alpha}\left[ 1 + Dx^{-\beta} + o(x^{-\beta})\right] \text{ as } x\rightarrow \infty.
\end{equation}
Note this type of parametric form captures: 
\begin{itemize}
\item stable distributions with stability index $\alpha \in (1,2)$ by setting $\beta/2 < \alpha \leq \beta$, 
\item extreme value distributions with $F(x) = \exp\left(-x^{-\alpha}\right)$ for $x > 0$ and $\alpha = \beta$, 
\item powers of ``smooth'' distributions of loss $X$ where if $X = Y^{-1/\alpha}$ then $Y$'s distribution admits a Taylor series expansion of at least 3 terms about the origin.
\end{itemize}
In the following sections we will explore in detail working with the choice of regularly varying model assumptions consistent with the power-law type severity models such as stable Pareto-Levy or Pareto type heavy tailed loss models. To achieve this we will work with estimators of the tail index $\alpha$ for loss data based on empirical regression based estimators in one of two forms, either based on empirical characteristic function regressions near the origin or based on assumptions on the likelihood used to derive MLE based estimators like Hill estimators. Furthermore, we will also consider some robust versions of the Hill estimator recently developed that extend the classical Hill type estimators aforementioned to accommodate removal of potentially biased or misreported massive losses, subject to significant uncertainty, rounding error, non-actual payment, reporting error and more. 

\subsection{Tail Index Estimators for Cyber Risk based on Empirical Characteristic Function Asymptotic Regressions} 
\label{subsec:NonParametricTailsCharFn}
If one makes the assumption of Hall in Equation \ref{Eqn:Hall} then it was shown by 
\cite{pitman1968behaviour} and \cite{welsh1986use} that if Equation \ref{Eqn:HallL} is assumed then as $x \rightarrow \infty$ one can express the resulting asymptotic real component of the power-law distributions characteristic function as follows near the origin as $t \downarrow 0$:
\begin{equation}
    1 - U(t) = C s(\alpha) t^{\alpha} + D_1t^{\gamma} + o(t^{\gamma}),
\end{equation}
where $\alpha \in (0,2)$ and it satisfies for a non-negative integer $p$ the constraint $2p < \alpha + \beta < 2p + 2$, with $\gamma = \min\left\{\alpha + \beta, 2\right\}$ and constant $D_1 = CD s(\alpha + \beta)$ if $\alpha + \beta < 2$, see discussion in \cite{jia2014heavy} for more general characterisations that extend the representations for arbitrary real $\alpha$.

If one then substitutes this representation based on Equation \ref{Eqn:Hall} for $\alpha \in [0,2]$ then as $t \downarrow 0$ one obtains a resulting refinement to the asymptotic relationship in Equation \ref{Eqn:RegCharFn}, derived in \cite[Section 2.2]{jia2014heavy}  given, for $y = \ln(1-U(t))$ by:
\begin{equation}
\begin{split}
y &= C_{\alpha} + \alpha\ln t + \ln\frac{1-U_n(t)}{1-U(t)}, \\
C_{\alpha} &= \ln\left[C s(\alpha)\right] + \frac{D_1}{C s(\alpha)}t^{\gamma-\alpha} + o(t^{\gamma-\alpha}),
\end{split}
\end{equation}
where $C_{\alpha}$ is treated as the constant of the regression and the estimator of the tail index given by simple least squares produces:
\begin{equation} \label{Eqn:CharFnRegEst}
\begin{split}
    &\widehat{\alpha} = \frac{\sum_{j=1}^m a_j y_j}{S_{zz}},\\
    &\widehat{C}_{\alpha} = \overline{y} - \widehat{\alpha} Z,
\end{split}
\end{equation}
with $a_j = Z_j - \overline{Z} = \ln t_j - \frac{1}{m}\sum_{k=1}^m \ln t_k$ and $S_{zz} = \sum_{i=1}^m \left(Z_i - \overline{Z}\right)^2 = \sum_{i=1}^m a_i^2$. 
For further properties of this estimator including its mean squared error, guidance on optimal selection of points for the regression around the origin see details in \cite[Section 2.3]{jia2014heavy}. We will use these properties in forming analysis of Advisen data and we therefore provide briefly a few key properties for practitioners below.

\begin{theorem} \label{Thm:EmpCharRegEst}
Suppose that $H(x)$ the tail sum for the loss distribution characterising the cyber loss severity distribution satisfies the assumptions of Hall \cite{hall1982some}, such that
\begin{equation}\label{Eqn:HallL}
H(x) = Cx^{-\alpha}\left[ 1 + Dx^{-\beta} + o(x^{-\beta})\right] \text{ as } x\rightarrow \infty.
\end{equation}
For the heavy tailed case of $\alpha \in (0,2)$ and estimators given by Equation \ref{Eqn:CharFnRegEst}, with $t_j = j/\sqrt{n}$ for $j=1,\ldots,m = n^{\delta}$ with $\delta \in (0,1/2)$. Then as $n \rightarrow \infty$ the bias of this regression based tail index estimator is given by
\begin{equation}
\mathbb{E}\left[\widehat{\alpha}\right] - \alpha = \frac{D_1(\gamma-\alpha)}{Cs(\alpha)(\gamma-\alpha+1)^2}n^{(\delta-1/2)(\gamma-\alpha)}\left\{1 + o(1) \right\},
\end{equation}
and the variance of the estimator is given by
\begin{equation}
{\rm Var}\left(\widehat{\alpha}\right) = O\left(n^{\alpha/2 - 1 - \delta\alpha}\right).
\end{equation}
where the exponent power for the variance is always less than 0.
\end{theorem}
The proof for these results is provided in \cite{jia2014heavy}.

\subsection{Tail Index Estimators for Cyber Risk based on Hill Type Estimators} 
\label{subsec:NonParametricTailsHillType}
A second class of popular approach to tail index estimation that we will explore for cyber risk data, applicable when one is willing to assume an additive tail function $H(x)$ which is Pareto in law, is the class of Hill estimators. There are many variations on the Hill estimator, we will first show the basic form of the estimator and then discuss briefly important variations of this estimator to make the estimator more statistically robust. Numerous authors have contributed to this class of estimators development, see for instance influential works by \cite{hill1975simple,pickands1975statistical,hall1982limit,embrechts2013modelling}
and the references therein.
 
Suppose that $X_1,\dots,X_n$ is an i.i.d. sample from a heavy tailed distribution $F$. Namely,
\begin{equation}
\label{e:heavy-tail}
\mathbb{P}(X_1> x) \equiv 1-F(x) \sim \ell(x) x^{-1/\xi}\quad\mbox{ as }x\to\infty,
\end{equation}
for some $\xi>0$ and a slowly varying function $\ell:(0,\infty)\rightarrow (0,\infty)$, i.e.,  $\ell(\lambda x)/\ell(x)\to 1,\ x\to\infty,$ for all $\lambda>0$. 
The parameter $\xi$ is also often referred to as the {\em tail index} of $F$ and it is typically treated in this context as equivalent to $\alpha$ in previous sections being represented as $\alpha = 1/\xi$. It will be convenient in this cyber risk study to use both tail index notations: with $\alpha$ to refer to the regression type estimator based on the characteristic function, derived above; and $\xi$ to refer to the class of Hill estimators based on maximum likelihood estimation under an asymptotic Pareto power law assumption for the observed losses. At this point it will be informative to also recall the classic {\em Hill estimator} given by, which is the estimator of tail index $\xi$ which is the inverse of the tail exponent $\alpha$:
\begin{equation}
\label{e:hill}
\widehat \xi_{k,n}:=\widehat{H}_{k,n}=\frac{1}{k} \sum_{i=1}^{k} \ln \Bigg(\frac{x^{(i,n)}}{x^{(k,n)}} \Bigg)= \widehat{\alpha}^{-1}.
\end{equation}

Fortunately, \cite{munasinghetail} has produced a useful R package that implements a range of tail index estimators of Hill "type" which we will loosely refer to as the variety of related estimators based of the asymptotic Pareto power law assumptions for the cyber risk losses.  In all estimation methods based on extremal order statistics, one must determine a threshold for the order statistics to begin being used in the estimator since the tail index working under the assumption of Pareto distributed data either exactly or asymptotically. Therefore, to apply these methods for the general power law form, we would look to identify where tail behaviour starts, which is not a precise or easy task, the interested reader is referred to \cite{hill1975simple, hubert2013detecting, vandewalle2007robust} for further detail and \cite{fedotenkov2020review} for a catalogue of Pareto-tail index estimation techniques.

The different Hill type estimators we will consider to use will be:
\begin{enumerate}
\item \underline{Maximum Likelihood Estimation (MLE)}: the MLE formula gives estimator for inverse $\hat{\xi}$:
\begin{equation} \label{eq:MLE}
\hat{\alpha} = N \cdot \left[ \sum_{i=1}^n \frac{x_i}{\hat{x}_{\mathrm{min}}} \right]^{-1},
\end{equation}
where $x_i$ represents the data point for $i=1, \ldots, n$. The minimum value, $x_{\mathrm{min}} = x_{(1,n)}$, is estimated from the data set and hence denoted $\hat{x}_{\mathrm{min}}$. As noted in \cite{newman2005power} this leads to a biased estimator, however this estimate (\ref{eq:MLE}) can be converted to an unbiased version $\alpha^*$ as follows \cite{rizzo2009new}:
\begin{equation}
\alpha^* = \frac{n-2}{n} \cdot \hat{\alpha}.
\end{equation}
\item \underline{Weighted variants of Least Squares Estimation (WLS)}: This method is based on the order statistics, assumed sorted in increasing order. Then for each value $i$ (of $n$ data points) one calculates $y_i$ the number of points greater than the $i^{th}$ data point. This method seeks to minimise the sum of the squared errors between the rank plot and the logarithm of the cdf. The estimator is given by (see \cite{nair2013fundamentals}):
\begin{equation} \label{eq:LS1}
\hat{\alpha} = \frac{  \sum_{i=1}^n  \left( \hat{y_i} -   \frac{1 }{n} \sum_{i=1}^n \hat{y_i} \right) \left( \ln x_i  - \frac{1}{n} \sum_{i=1}^n \ln x_i \right) }{ \sum_{i=1}^n \left( \ln x_i  - \frac{1}{n} \sum_{i=1}^n \ln x_i \right)^2}.
\end{equation}
There is also a popular weighted variant where the sum of squared errors criterion from the LS method above is developed with a weight function. A common choice of weight function is given by
\begin{equation} \label{eq:WLSweight}
w_i = \left[ \ln \left( \frac{x_i}{\hat{x}_{\mathrm{min}}} \right). \right]^{-1}
\end{equation}
which gives a WLS solution closely related to the first estimator based on the MLE where each has the same asymptotic limiting results. Under this weight function the WLS tail index estimate is then given, assuming no ties in the sorted losses by
\begin{equation} \label{eq:WLS2}
\hat{\alpha} = \frac{ - \sum_{i=1}^n \ln \left[ (n+1-i) / n \right] }{ \sum_{i=1}^n \ln \left( x_i / \hat{x}_{\mathrm{min}} \right)}.
\end{equation}
\item \underline{Percentile Method (PM)}: this method develops an estimator for the tail index based on percentiles, typically based on a robust dispersion measure such as the inter-quartile range, producing estimators such as (see \cite{bhatti2018efficient}):
\begin{equation} \label{eq:PM}
\hat{\alpha} = \frac{ \ln 3}{ \ln (P^*_{75}) - \ln (P^*_{25})}.
\end{equation}
Here $P^*_{q}$ is the $q^{th}$ percentile of the data set. 
\end{enumerate}

\begin{figure}[h] 
\includegraphics[scale=0.6]{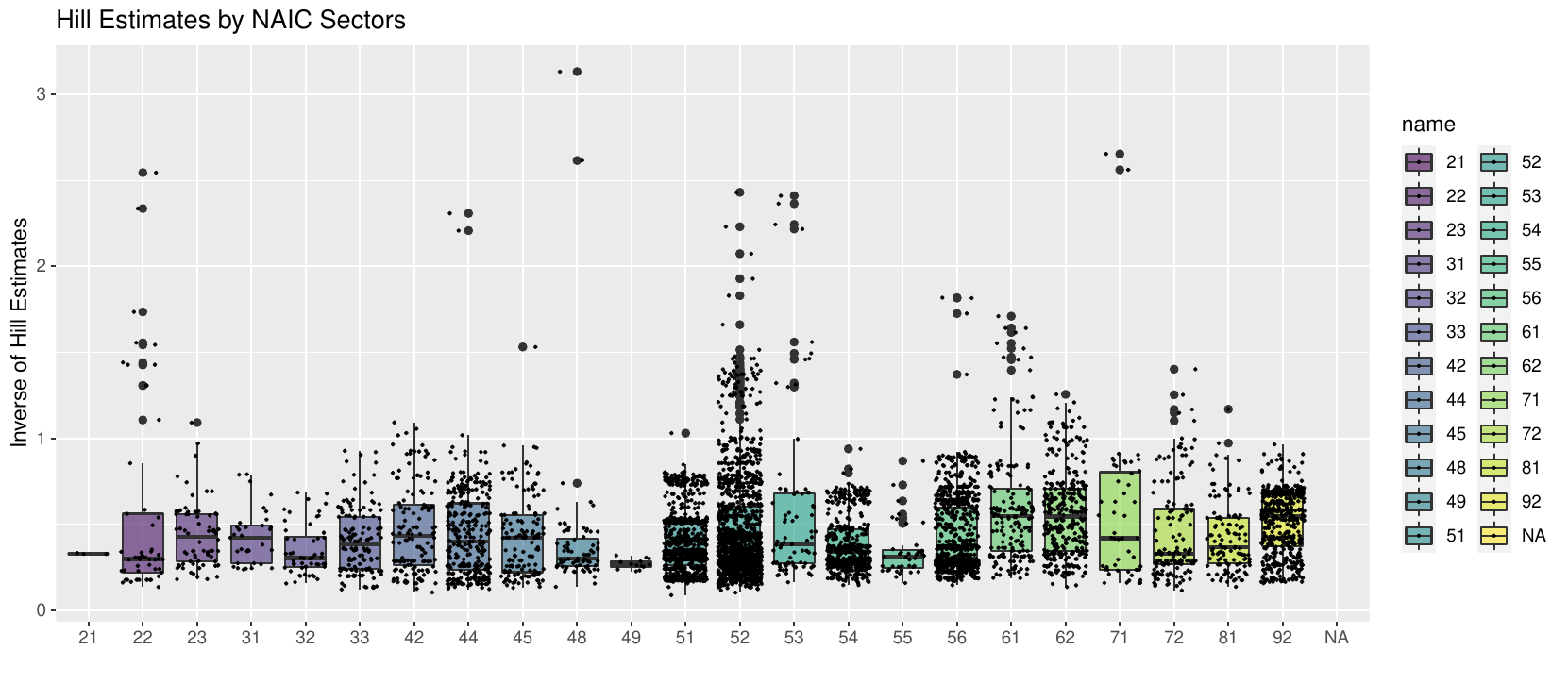}
\centering
\caption{Box plots of Hill Estimators obtained from a sequence of thresholds corresponding to percentile levels of the data from N-k to N }
\label{Fig:hill_naics}
\end{figure}

We begin the results analysis by looking at the basic Hill Estimator obtained for a sequence of order statistic thresholds $k$. The results are presented in Figure \ref{Fig:hill_naics} and Table \ref{tab:hill_estimates}.

\begin{table}[]

\centering
\resizebox{\textwidth}{!}{%
\begin{tabular}{|l|c|c|c|c|c|c|c|c|}
\hline
Sector NAIC &
  Emp.Char.Fn Reg. &
  MLE &
  LS &
  MoM &
  PM &
  MPM &
  GPM &
  WLS \\ \hline
11 &
  N.A. &
  (0.3572,10000) &
  (0.0724,10000) &
  N.A. &
  (0.1962,10000) &
  (1.7252,10000) &
  (0.1538,10000) &
  (0.1238,10000) \\ \hline
21 &
  N.A. &
  (0.3840,16600) &
  (0.2228,16600) &
  N.A. &
  (0.3004,16600) &
  (0.5329.16600) &
  (0.2430,16600) &
  (0.2796,16600) \\ \hline
22 &
  N.A. &
  (0.1332,30) &
  (0.1930,30) &
  N.A. &
  (0.1586,30) &
  (0.1692,30) &
  (0.1006,30) &
  (0.1216,30) \\ \hline
23 &
  N.A. &
  (0.1879,300) &
  (0.2663,300) &
  N.A. &
  (0.2210,300) &
  (0.2158,300) &
  (0.1494,300) &
  (0.1761,300) \\ \hline
31,32,33 &
  0.0612 $\pm$ 0.1574 &
  (0.1353,200) &
  (0.2638,200) &
  N.A. &
  (0.2292,200) &
  (0.2860,200) &
  (0.1623,200) &
  (0.1319,200) \\ \hline
42 &
  0.0013 $\pm$ 0.0263 &
  (0.1228,44.49) &
  (0.2423,44.49) &
  N.A. &
  (0.2433,44.49) &
  (0.3396,44.49) &
  (0.1557,44.49) &
  (0.1188,44.49) \\ \hline
44,45 &
  0.0083 $\pm$ 0.0192 &
  (0.1235,40.58) &
  (0.2320,40.58) &
  N.A. &
  (0.2028,40.58) &
  (0.3295,40.58) &
  (0.1417,40.58) &
  (0.1217,40.58) \\ \hline
48, 49 &
  N.A. &
  (0.1295,202) &
  (0.2411,202) &
  N.A. &
  (0.2380,202) &
  (0.2454,202) &
  (0.1604,202) &
  (0.1217,202) \\ \hline
51 &
  0.0457 $\pm$ 0.0326 &
  (0.0900,1) &
  (0.2551,1) &
  N.A. &
  (0.2081,1) &
  (0.2115,1) &
  (0.1444,1) &
  (0.0876,1) \\ \hline
52 &
  0.0077 $\pm$ 0.0128 &
  (0.1068,10) &
  (0.2783,10) &
  N.A. &
  (0.2288,10) &
  (0.2607,10) &
  (0.1550,10) &
  (0.1060,10) \\ \hline
53 &
  N.A. &
  (0.2140,1295) &
  (0.2869,1295) &
  N.A. &
  (0.2464,1295) &
  (0.3080,1295) &
  (0.1752,1295) &
  (0.2040,1295) \\ \hline
54 &
  0.0059 $\pm$ 0.0141 &
  (0.1478,88.98) &
  (0.2770,88.98) &
  N.A. &
  (0.2238,88.98) &
  (0.2453,88.98) &
  (0.1614,88.98) &
  (0.1457,88.98) \\ \hline
55 &
  N.A. &
  (0.3237,4056.84) &
  (0.2355,4056.84) &
  N.A. &
  (0.2565,4056.84) &
  (0.1976,4056.84) &
  (0.1625,4056.84) &
  (0.2811,4056.84) \\ \hline
56 &
  0.0243 $\pm$ 0.0183 &
  (0.1475,69.04) &
  (0.2826,69.04) &
  N.A. &
  (0.2017,69.04) &
  (0.1827,69.04) &
  (0.1358,69.04) &
  (0.1457,69.04) \\ \hline
61 &
  0.0197 $\pm$ 0.0210 &
  (0.1986,800) &
  (0.3470,800) &
  N.A. &
  (0.3040,800) &
  (0.4067,800) &
  (0.2147,800) &
  (0.1925,800) \\ \hline
62 &
  N.A. &
  (0.1341,100) &
  (0.3368,100) &
  N.A. &
  (0.3098,100) &
  (0.4634,100) &
  (0.2360,100) &
  (0.1312,100) \\ \hline
71 &
  N.A. &
  (0.1508,600) &
  (0.2335,600) &
  N.A. &
  (0.2395,600) &
  (0.4771,600) &
  (0.1778,600) &
  (0.1391,600) \\ \hline
72 &
  0.0078 $\pm$ 0.0275 &
  (0.1126,27) &
  (0.2391,27) &
  N.A. &
  (0.2289,27) &
  (0.2554,27) &
  (0.1534,27) &
  (0.1092,27) \\ \hline
81 &
  N.A. &
  (0.1367,47.95) &
  (0.2384,47.95) &
  N.A. &
  (0.2461,47.95) &
  (0.3386,47.95) &
  (0.1672,47.95) &
  (0.1310,47.95) \\ \hline
92 &
  0.0029 $\pm$ 0.0162 &
  (0.1586,200) &
  (0.3158,200) &
  N.A. &
  (0.3193,200) &
  (0.4543,200) &
  (0.2185,200) &
  (0.1564,200) \\ \hline
\end{tabular}%
}
\caption{N.A. means that the sample size was insufficient for a reliable estimator to be obtained. The different tail index estimators $\alpha$ are presented for each NAIC. \\
Emp.Char.Fn.Reg: Empirical Characteristic Function regression with point estimator $\alpha$ $\pm$ std. error. \\
MLE: Maximum Likelihood Estimators of Hill for tail index and scale parameters such that $(\alpha, scale)$\\
LS: Least Squares Estimator with no weighting for tail index and scale parameters such that $(\alpha, scale)$\\
MoM: Method of Moments - not relevant as all instances were too heavy tailed to admit reliable MoM estimators.\\
PM: Percentile Method estimator for tail index and scale parameters such that $(\alpha, scale)$\\
MPM: Modified Percentile Method estimator for tail index and scale parameters such that $(\alpha, scale)$\\
GPM: Generalised Percentile Method estimator for tail index and scale parameters such that $(\alpha, scale)$\\ 
WLS: Weighted Least Square estimator for tail index and scale parameters such that $(\alpha, scale)$.
}
\label{tab:hill_estimates}
\end{table}

We note that for the empirical characteristic function regression estimator, we utilised the results in Theorem \ref{Thm:EmpCharRegEst} to select the values for $t_1,\ldots,t_m$ such that $t_j = j/\sqrt{n}$ for $m = n^{\delta}$ with $\delta = 1/2$ which was adaptive for each NAIC sector as they have differing numbers of realised cyber losses. It is evident from this analysis firstly that there is significant variation in the estimators across the various methods of tail index estimation. The Method of Moments (MoM) failed in all cases due to the sample size requirements and the Empirical Characteristic Function Regression methods were also seemingly unable to produce reliable results in numerous NAIC examples and in those in which it did produce estimator results, the uncertainty associated with the estimators under this method in this cyber risk application were very large. The MLE, PM, MPM, GPM and WLS were seemingly better at capturing estimators that were more comparable with each other across each NAIC dataset. These estimators indicated the presence of very heavy tailed loss distributions for the majority of NAICs. This type of finding is consistent with other studies of tail behaviour in cyber risk loss data. In the following section, we will question and explore the validity of these findings, which if taken at face value would indicate a difficulty with insurability of such loss processes due to the heavy tailed nature of the loss processes that would result in exorbitant premiums. We will therefore explore it the situation is as bad as it looks to this point, by taking a more practical perspective on the analysis and by exploring assumptions regarding the loss data recording and the impact they may have on such conclusions regarding heavy tailed behaviour of cyber loss processes.

\section{Dealing with Real World Cyber Data: Inaccurate, Rounded, Truncated, Partially Settled Unreliable Massive Reported Cyber Total Losses}  \label{subsec:RobustHill}
In real cyber risk loss data, the total loss may be subject to a range of issues in the reporting. The attribution of all loss components to the total loss may be difficult when it concerns a combination of both direct and indirect loss aggregations. In general reporting of cyber risk, in the financial sector for instance, fall under two broad classification of loss type: Direct and Indirect losses.
\begin{itemize}
\item Direct losses: Resulting from the event itself, such as reparation, time lost, client compensation, regulatory fines, money lost in wrongful transactions...
\item Indirect losses: Resulting from the consequences of the event such as loss of customers resulting from damage to image or reputation, low morale amongst employees, regulatory scrutiny, increased insurance premiums Indirect losses are often linked to reputation damage!
\end{itemize}
In the Advisen data studied in this paper the total cyber loss per event is actually a composition of many direct and indirect loss components including: injury loss payouts awarded, loss of wages, loss of business income, loss of assets, property 1st party payouts, financial damages, loss of life expense payouts, defence costs for legal and regulatory, other expenses, punitive exemplary damages, other fines and penalties, pain and suffering awarded amounts, other costs, plaintiff legal fees and plaintiff fees. All these losses are difficult to accurately measure over multiple events that make up a total loss event for a given accident trigger. This makes it likely that when losses of 1 Bil. or more are recorded for total loss, they are often less specific and rounded compared to those losses observed in the 1,000's and millions range. In addition, it is also the case that not all total losses are settled - they may be awarded but in practice if they are in the billions they will likely not ever be completely settled. This causes uncertainty in the extreme losses used to estimate the Hill estimator, which are the most critical losses for accurate tail index estimation. In this section we will explore how to overcome this challenge. In practice, if the largest few order statistics may be {\em corrupted} or {\em unreliable} or {\em uncertain} as just discussed. This may lead to a severe bias in the estimation of $\xi$. In the worst case scenarios one may find that the computed estimate of $\xi$ may be completely constructed from a small number of corrupted observations. To reduce the effect of such observations that may corrupt the sample one may introduce a class of Hill estimators based on trimming and weighting that produces a class of robust tail index estimators.

\subsection{Robust Trimmed Hill Estimators for Cyber Losses} 
\label{subsec:RobustHill}
To reduce this bias several authors have looked at how to robustify the Hill estimator, see \cite{brazauskas2000robust,zou2020extreme,goegebeur2014robust,peng2001robust} and \cite{peng2001robust} who all studied various aspects of the general concept of robust Hill estimators. The objective of these methods was to robustify the estimator of $\xi$ by trimming or reducing the reliance on the potentially corrupted extreme losses, this can be done through a hard truncation, weighting or a soft truncation weighted trimming method. 

The trimmed Hill estimator, denoted $\mathcal{H}_{k_0,k,n}$ below, is based on a weighted version of the classical Hill estimator, for some selection of weights $\left\{w_{k_0,k}(i)\right\}$ for the order statistics between $i \in \left\{k_0,\ldots,n\right\}$. The selection of the weighting rule will clearly influence the statistical properties of the estimator obtained, generically given by:
\begin{equation}\label{e:xi-trimmed}
\widehat{\xi}^{\rm trim}_{k_0,k,n}:= \mathcal{H}_{k_0,k,n} = \sum_{i=k_0+1}^{k} w_{k_0,k}(i) \ln \Bigg(\frac{X_{(n-i+1,n)}}{X_{(n-k,n)}} \Bigg),\hspace{5mm}  0\le k_0<k<n-1.
\end{equation}
It is worthy to remark that when $k_0<k<n$ then one has set the weight contributions from the potentially corrupted higher order statistic losses through to zero contribution after adjusting for the intermediate order statistics from $k_0$ on-wards which are incorporated in the estimation with a weighting rule. The consequence of this approach is to improve the breakdown point of the robust estimator. However, the selection of $k_0$ is a delicate matter and cannot be readily determine apriori. Therefore it is proposed to consider utilising a trimmed Hill plot to help select the choice of the trimming parameter $k_0$ which is of key importance in practice. One can utilise a method of the \textit{trimmed Hill plot} to visually determine $k_0$. Then, by exploiting the elegant joint distribution structure of the optimal trimmed Hill estimators, \cite{bhattacharya2017trimming} devised an weighted sequential testing method for the identification of $k_0$. This leads to a new {\em adaptive trimmed Hill} estimator, which works well even if the degree of contamination in the top order statistics is largely unknown. 

Inference for the {\em truncated} Pareto model has been developed in the seminal work of \cite{aban2004generalized} and recently in \cite{beirlant2016tail}. In contrast to this, \cite{bhattacharya2017trimming} studied a soft truncation approach based on weighted trimming in which the class of weights that would be optimal in the sense of being the Best Unbiased Linear Estimator (BLUE) for the class of loss models given by a regularly varying Pareto law and were able to obtain a closed form representation for the weighting functions.

\cite{bhattacharya2017trimming} were able to demonstrate that in the ideal Pareto setting, it turns out that our trimmed Hill estimator is essentially finite--sample optimal among the class of all unbiased estimators of  $\xi$ with a fixed {\em strong upper break-down point}. Furthermore, they established the following properties of the estimator for $\xi$:
\begin{itemize}
\item asymptotic normality of the trimmed Hill estimator in the semi parametric regime \eqref{e:heavy-tail}, under second order conditions on the regularly varying function $\ell$ as in \cite{beirlant2004statistics}.
\item rate of convergence of the estimator being the same as the classic Hill as long as $k_0=o(k)$.
\end{itemize}

The optimal BLUE trimmed weights, $w_{k_0,k}(i)$ for which the estimator in \eqref{e:xi-trimmed} is unbiased for $\xi$ and also has the minimum variance produces the tail index estimator given by 
\begin{equation} 
\label{e:xi-opt}
\widehat{\xi}^{\rm trim\, opt}_{k_0,k,n}:= \mathcal{H}_{k_0,k,n}=\frac{k_0+1}{k-k_0} \ln \Bigg(\frac{X_{(n-k_0,n)}}{X_{(n-k,n)}} \Bigg)+\frac{1}{k-k_0} \sum_{i=k_0+2}^{k} \ln \Bigg(\frac{X_{(n-i+1,n)}}{X_{(n-k,n)}} \Bigg),  \hspace{5mm} 0\leq k_0<k<n-1.
\end{equation}
see \cite{bhattacharya2017trimming} for details. Furthermore, \cite{bhattacharya2017trimming} proposed a data driven parameter selection procedure for the threshold $k_0$'s selection.

The potential uncertainty regarding the validity of extreme losses reported, not only causes a bias in the estimation of the tail parameter $\xi$, it is also reasonable to assume that such uncertainty could translate into premium mispricing. Basing the insurance premium calculations on the trimmed Hill estimators, should reduce the impact of uncertainty and provide for more robust premium estimates. When the trimmed Hill estimates show some consistency, perhaps suggesting values of the tail parameter greater than 1, one could then conclude that the heavy tails of cyber risk are mainly caused by a few extreme losses that may be inaccurately reported or recorded or suffer from a great degree of uncertainty in their assessment, that could be corrupted or even never settled, and therefore there should be little to no variation in the corresponding insurance premiums when trimming these noisy records. On the other hand, if the values of the trimmed Hill estimates still showed no consistency, then one should conclude that the effect of uncertainty in the cyber related losses is not mitigated by the trimming procedure implying presence of further model risk in the case of cyber risk related to inconsistency of the cyber loss process with the statistical assumptions underlying the tail index estimators. To investigate this, we consider a simple insurance premium calculation, including in our analysis the trimmed Hill estimator. 

Consider a 1 year insurance policy protecting against each of $X_1,\dots,X_n$ random losses, with $n$ following a Poisson distribution, up to an aggregate top cover limit equivalent to a percentage $c$ of the total company wealth. According to the zero utility principle, the maximum premium $P$, a non satiable and risk averse decision maker, with total wealth $w$ is willing to pay, corresponds to the solution of the following non linear equation: 

\begin{equation}
\label{eq:decision_premium}
    \mathbb{E}\left[u\left(w-\sum_{i=1}^n X
    _i\right) \right] = \mathbb{E}\left[u\left( w-P -\sum_{i=1}^n X_i + \sum_{i=1}^n\min(X_i,cw)\right)\right].
\end{equation}

We consider a company with 1 billion dollars of total wealth, wishing to insure $10\%$ of its total capital. For $k = 1,\dots,20$, and $k_0 = 15,\dots,60$ with step size of $5$, we estimate the tail parameter using the trimmed Hill estimator in \ref{e:xi-opt}, and fit quarterly frequency on a Poisson distribution for each NAIC. Then, we proceed with insurance premium calculations using a simulation framework under this illustrative example.

We present the results for the leading NAIC by total cyber losses in Figure \ref{fig:trimmed_NAIC52}. The remaining 4 NAICS that are in focus have similar results which are presented in Appendix 1, see Figure~\ref{fig:trimmed_NAIC51}. These plots show the trimmed Hill estimates for various values of $k$ and $k_0$ (panel (a)) and the corresponding insurance premiums (panel (b)), of the top 5 NAIC sectors in terms of cyber event frequency. 

All the considered NAICs exhibit a variation in the estimates of the trimmed tail index, and we see that critically, this variation indeed then translate into variation in the insurance premiums required. This shows that depending on what model assumptions one is willing to make regarding the quality of the data used in tail index estimation, these assumptions have a consequential influence on the insurance pricing. This manifests as a form of model risk in dealing with cyber risk data, given that the effect of uncertainty cannot be filtered out by trimming procedures. 

Importantly, we see that for business sector NAIC52 and NAIC56 they both show trimmed inverse tail parameter estimates which now appear substantially lower than 1, suggesting that the extreme tail behaviour reported in Figure~\ref{Fig:hill_naics} and Table~\ref{tab:hill_estimates} is mainly driven by a few extreme losses. According to the assumptions underpinning the application of the trimming methodology, outlined previously, one can see that according to the application of this technique, it yields a switch from a heavy tailed model to a lighter tailed loss model. In other words, by changing the assumptions on the cyber risk loss data, one goes from a non-trimmed class of estimates which produces heavy tailed loss models for insurance pricing through to the trimmed estimates which produced lighter tailed loss models and the resulting consequences on the insurance premiums is substantial. We see for NAIC52 that this results in a difference in premiums in which the premium reduces by upto 86\% under the trimmed model assumptions compared to the non-trimmed. For NAIC56, results presented in Appendix 1, there is also a substantial premium change as trimming is applied, and in this case it also results in a shift from heavy tailed loss model to light tailed loss model that subsequently results in a reduction in premium of up to 93\%, a very substantial difference in premium pricing. We note that in the case of these two NAIC examples, we state that the model has shifted from heavy tailed to light tailed, since the results shift for any initial $k_0$ immediately from a heavy tailed to a light tailed model as soon as any trimming is applied and the inverse tail index continues to decrease as increasing trimming is applied. Importantly in both cases, it is seen that after a certain point of trimming, for a give lower threshold $k_0$, the trimmed estimates stabilise indicating that one can reliably fit a model to this data once problematic, noisy, inaccurate or corrupted data is removed. 

In the cases of the remaining three NAICs: NAIC51, NAIC54 and NAIC92, again results are presented in Appendix 1, and whilst these models still have heavy tailed loss models after trimming is applied, the resulting premium reductions from applying the trimmed results compared to the non-trimmed results is also substantial. The maximum premium reduction produced for NAIC51 was 99.3\%, for NAIC 54 it was 99.5\% and for NAIC92 it was a premium reduction of 99.6\%. These are so substantial that they clearly indicate the need to consider this source of model risk and the potential impact on pricing coming from the underlying modelling assumption and subsequent model risk.

Tail index estimates for business sectors NAIC51, NAIC54, and NAIC92, still suggest that the corresponding severity distributions are heavy tailed. Nevertheless, this is not consistent for every values of $k$ and $k_0$, implying that the estimates are highly sensitive to the choice of $k$ and $k_0$. The sensitivity of the tail index estimates directly translates into the insurance premiums. For business sectors NAIC52 and NAIC56, insurance premiums computed using a log-utility function appear to be lower than those for business sectors NAIC51, NAIC54, and NAIC92. However, all business sectors present great variability in insurance premiums, showing how the uncertainty in the tail index affects ultimately premium mispricing and cyber risk insurability. 

\begin{figure}
    \centering
    \subfloat[]{\includegraphics[width = \textwidth, height = 0.35\textwidth]{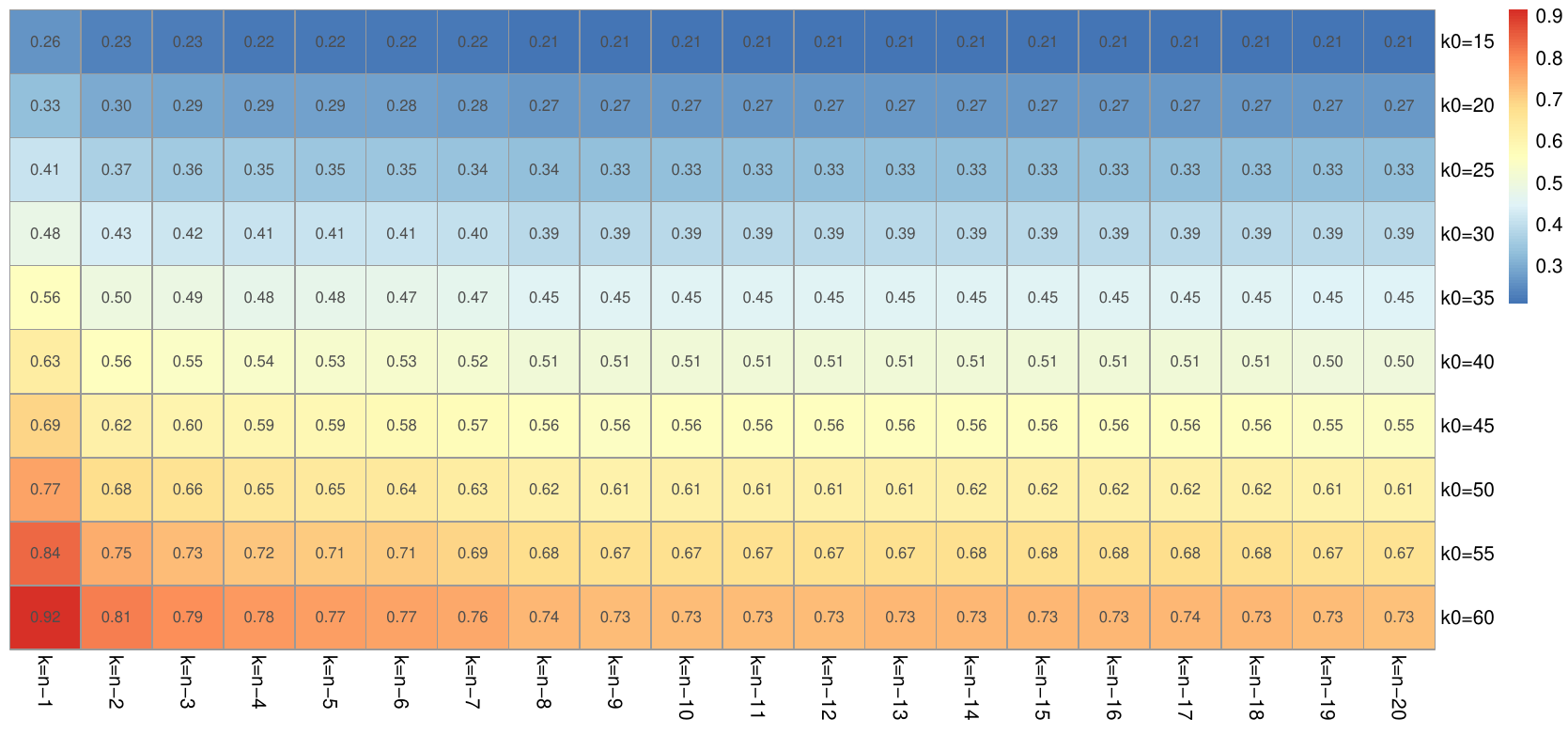}}\\
    \subfloat[]{\includegraphics[width = \textwidth, height = 0.35\textwidth]{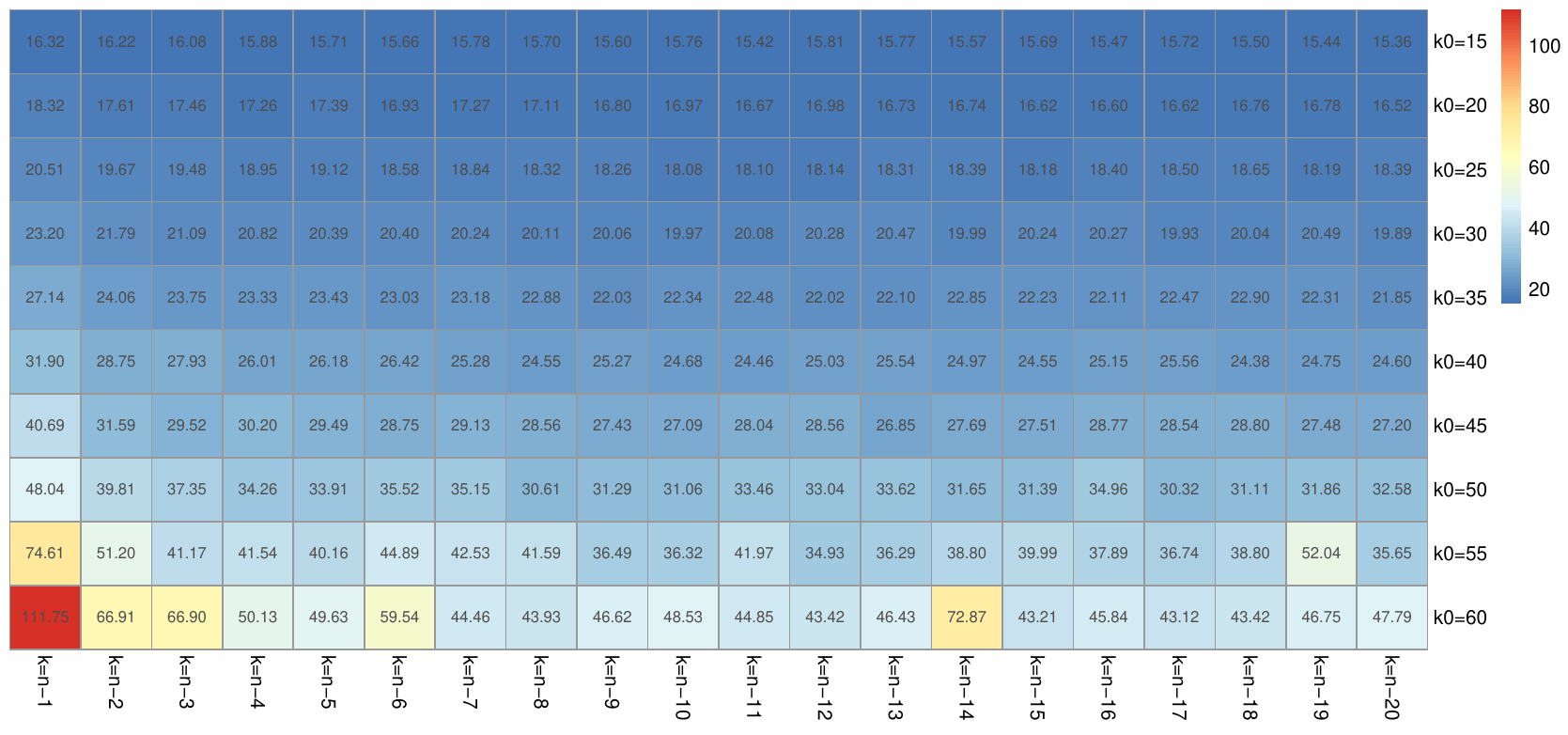}}

  \caption{Trimmed tail index estimator $\widehat{\xi}^{\rm trim\, opt}_{k_0,k,n}$ obtained by trimmed Hill estimators (on the top) and corresponding insurance premiums (on the bottom) for varying trimming parameters $(k_0,k)$ for NAIC 52. Premiums are computed on quarterly basis using 1000,000 Monte Carlo draws. The variation in the trimmed Hill estimates translates into significant variation in the resulting insurance premium calculations.}
    \label{fig:trimmed_NAIC52}
\end{figure}

\FloatBarrier

\section{Dependence and Tail Behaviour Estimation on Advisen NAIC Cyber Losses}  \label{sec:NonParametricTails}
In this section we will illustrate that in addition to model risk and parameter uncertainty on the marginal loss processes, manifesting in insurance pricing uncertainty. One can also find in the setting of cyber risk data significant model risk and parameter uncertainty in the joint dependence model between cyber loss processes, which again we will study across the NAIC industry sectors. 

Analogously to the analysis in the previous section that studied marginal tail behaviours under different data and model assumptions. In this section we will repeat this type of analysis but for the dependence structures between NAIC cyber loss processes. This will be performed in a sequence of stages, starting with a comparison between simple linear correlation estimates and various robust correlation estimators. Then we will develop this further to account for copula models under various assumptions and we will select optimal copula dependence structures. In terms of how these various studies of dependence manifest in an insurance context for cyber risk, we will explore the impact that parameter uncertainty, model misspecification and model risk in the dependence structures may have on risk diversification for insurers that may hold insurance portfolios for cyber risk across many industry sectors as captured by the NAIC codes.

\begin{figure}[h] 
\includegraphics[scale=0.5]{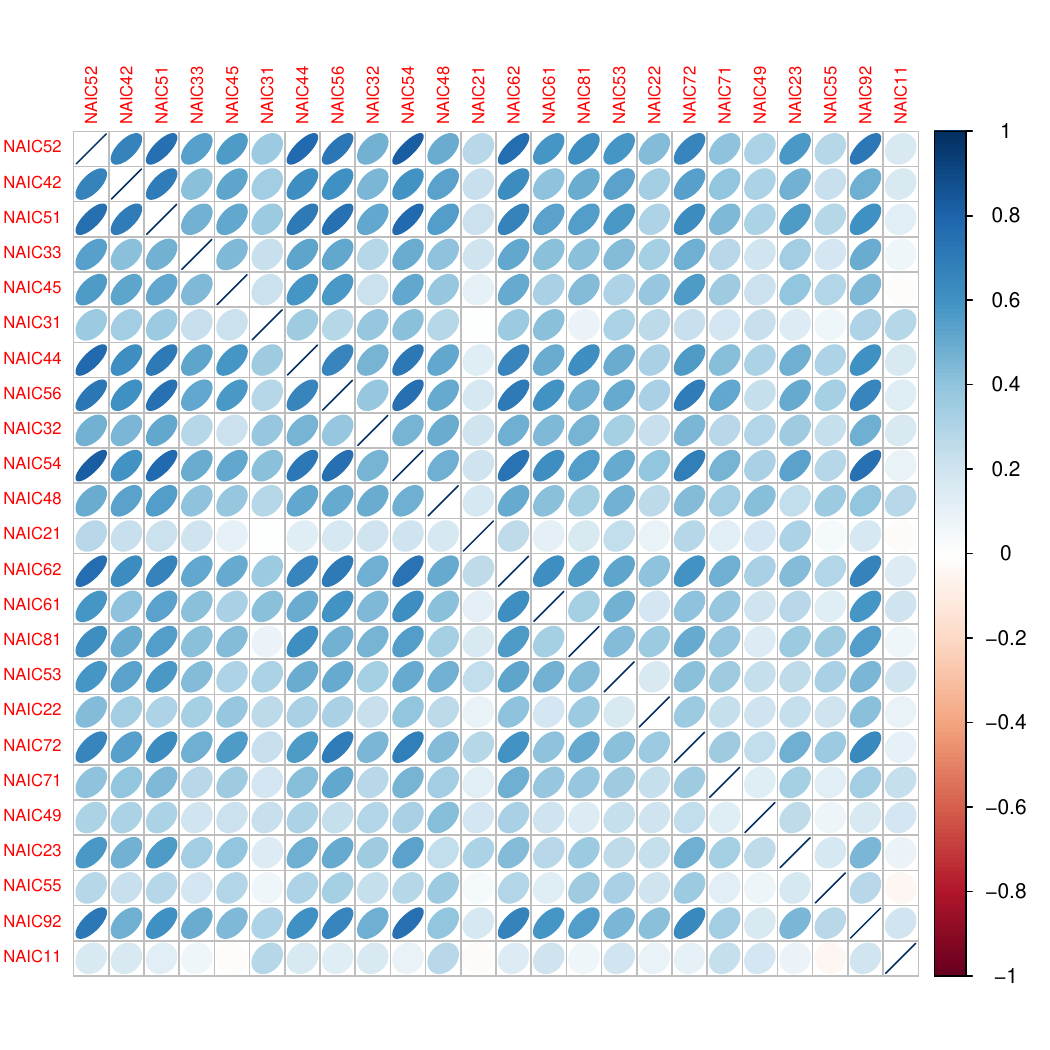}
\centering
\caption{Standard linear Pearson correlation coefficient (Eqn. \ref{eq:linear_corr}) estimates for quarterly aggregate historical Advisen cyber loss data across NAIC sectors in the US between 01/01/1990 and 20/09/2020.}
\label{Fig:TrimmedEstimator}
\end{figure}

We begin this section comparing standard linear correlation estimates between the loss data for each NAIC in the Advisen data. We will then go on to demonstrate how the correlation estimation may be effected by robust estimators that make different assumptions on the data when calculating the linear or rank correlations. In Figure \ref{Fig:TrimmedEstimator} we present the basic linear correlations between NAIC industry sectors. Note, throughout this section we will need to convert the loss data from event time series data where losses are have time stamps on the days of loss event, to a regular time series in order to compare dependence structures between NAICs. To achieve this we have decided after analysis of the data records that a reasonable time stratification is to perform a quarterly aggregation for the dependence analysis.

The Pearson linear correlation coefficient is given as follows in Equation \ref{eq:linear_corr},
\begin{equation}
\label{eq:linear_corr}
r = \frac{\sum(x_i - \overline{x})(y_i-\overline{y})}{\left[\sum (x_i-\overline{x})^2\sum (y_i - \overline{y})^2\right]^{1/2}},
\end{equation}
which in this analysis uses all loss events in the quarterly aggregates. This includes the indirect and direct loss events that for the extreme loss records were highly likely to suffer from different forms of inaccuracy, ranging from noisy reporting due to approximations, inaccurate records and rounding, misreporting or incomplete reporting, partially settled or unsettled and corrupted records which is particularly relevant for the extreme loss in the Advisen data, as discussed in Section \ref{subsec:RobustHill}

Therefore, to determine how such issues in the extreme losses may result in model uncertainty or model risk in this analysis, we will once again compare the standard linear Pearson correlation coefficient estimators ignoring these problems with the extreme loss records to estimators for dependence that are robust and can remove the influence of such problematic large loss records to various degrees, depending on the class of robust estimator. We will explore three robust methods of dependence estimation for correlation based on SSD Median, Quadrant (sign) correlation coefficient methods and MCD estimators each outlined below. In developing the analysis for the robust correlation estimation, we once again focus on the most important NAICS, studied in previous sections, that have the top five number of loss events reported: NAICs 52, 56, 51, 54, 92. 

As discussed in \cite{shevlyakov2011robust} one can robustify the sample correlation coefficient by replacing the linear procedures of averaging by the corresponding nonlinear robust counterparts according to
\begin{equation}
\label{eq:ssd_corr}
    r_{\alpha}(\Psi) = \frac{\sum_{\alpha}\Psi(x_i - \widehat{x})\Psi(y_i - \widehat{y})}{\left[\sum_{\alpha} \Psi^2(x_i-\widehat{x})\sum_{\alpha} \Psi^2(y_i - \widehat{y})\right]^{1/2}},
\end{equation}
where $\widehat{x}, \widehat{y}$ are robust estimators of location such as the median that are used to replace the mean. This is particular important in the case of infinite mean loss models. The function $\Psi(\cdot)$ is a monotonic function such as Huber's $\Psi$-function given by 
\begin{equation}
    \Psi(z,k) = \max\left\{-k, \min(z,k)\right\}
\end{equation}
and $\sum_{\alpha}$ is a robust version of the data summation that can trim values as follows
\begin{equation}
    \sum_{\alpha}z_i = nT_{\alpha}(z) = n(n-2r)^{-1}\sum_{i=r+1}^{n-r} z_{(i,n)}, \;\; 0\leq\alpha\leq 0.5, \;\; r=[\alpha(n-1)],
\end{equation}
with $[\cdot]$ the integer component. Note, when $\alpha = 0$ one recovers the standard summation and no trimming of order statistics is applied. If one wishes to recover the classical correlation median estimators of \cite{falk1998note} one can select $\alpha = 0.5, \widehat{x} = {\rm med}(x), \widehat{y} = {\rm med}(y)$ and $\Psi(z)=z$, where ${\rm med}(z)=z_{([n/2],n)}$ and we will study a version of this with trimming of extremes.

Furthermore, we will utilise a non-parametric measure for robust correlation of \cite{blomqvist1950measure} known as the quadrant (sign) correlation coefficient given by $\alpha=0$, $\widehat{x}={\rm med}(x),\widehat{y}={\rm med}(y)$ and $\Psi(z) = {\rm sgn}(z)$ to produce estimator 
\begin{equation}
\label{eq:quadrant_corr}
    r_{Q} = n^{-1}\sum {\rm sgn}(x_i - {\rm med}(x)){\rm sgn}(y_i - {\rm med}(y)).
\end{equation}

The final robust correlation estimator we will explore will be the Minimum Covariance Determination (MCD) estimator. This is obtained for a finite sample of observations $\left\{x_1 , \ldots, x_n\right\}$ in $\mathbb{R}^p$ by selecting that subset $\left\{x_{i_1}, \ldots, x_{i_h}\right\}$ of size $h$, with $1 \leq h \leq n$, which minimizes the generalized variance given by the determinant of the covariance matrix computed from the subset among all possible subsets of size $h$. The resulting robust location and scale estimators are then defined as
\begin{equation}
\label{eq:mcd_corr}
\begin{split}
\widehat{x} &= \frac{1}{h}\sum_{j=1}^h x_{ij},\\
\widehat{\Sigma} &= c_p \frac{1}{h}\sum_{j=1}^h (x_{ij}-\widehat{x})(x_{ij}-\widehat{x})^T,\\
\end{split}
\end{equation}
where $c_p$ is a consistency factor. The location estimator can also be replaced with a robust M-Estimator such as the median estimator for the trimmed sample. The choice $h = [(n+p+1)/2]$ is commonly preferred since it yields the highest possible breakdown point, see \cite{lopuhaa1991breakdown}. As these authors observed, setting it at-least as high as $h \approx n/2$ when the number of observations is much higher than the dimension, then breakdown point of the resulting multivariate scale estimator is defined as the smallest fraction of observations that you need to replace to arbitrary position before the estimated scatter explodes such that its largest eigenvalue tends to infinity or implodes such that its smallest eigenvalue tends to zero. 

\begin{figure}[h] 
\subfloat[Correlation matrix  of the top 5 NAICS in terms of occurrences, estimated using linear Pearson correlation coefficient estimator.]{\includegraphics[scale=0.35]{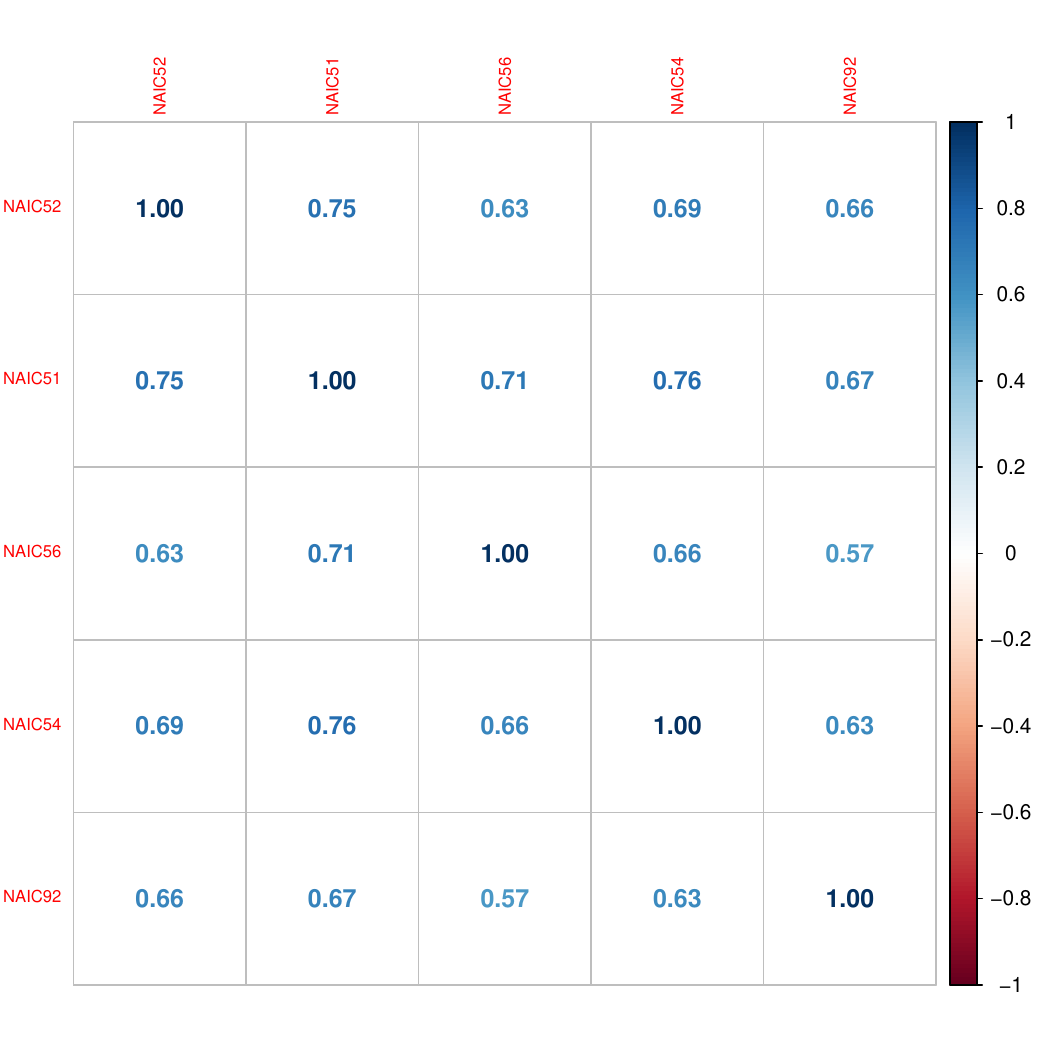}}\quad
\subfloat[Correlation matrix  of the top 5 NAICS in terms of occurrences, computed using SSD median estimator in Eqn. \ref{eq:ssd_corr}]{\includegraphics[scale=0.35]{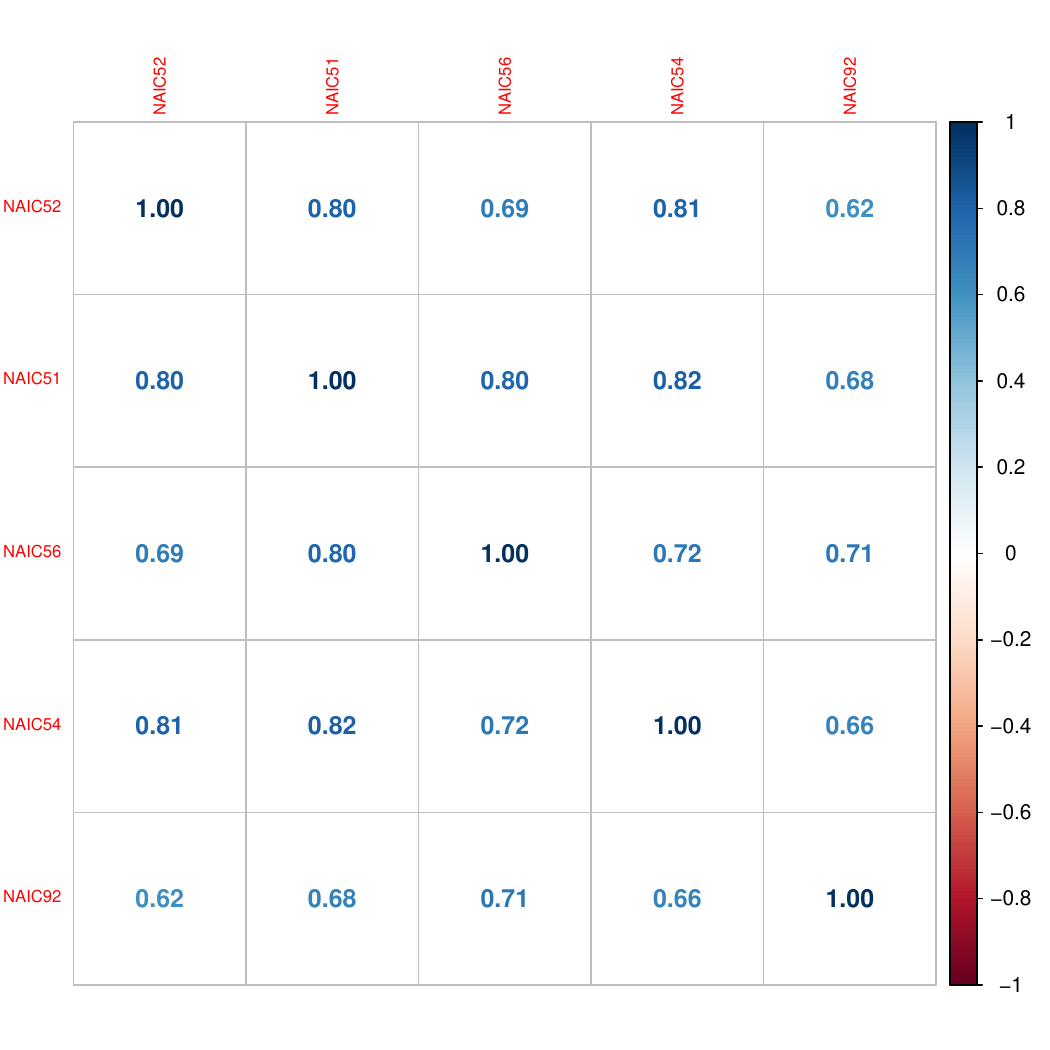}}\quad 
\subfloat[Correlation matrix  of the top 5 NAICS in terms of occurrences, computed using quadrant (sign) correlation estimator in Eqn. \ref{eq:quadrant_corr}]{\includegraphics[scale=0.35]{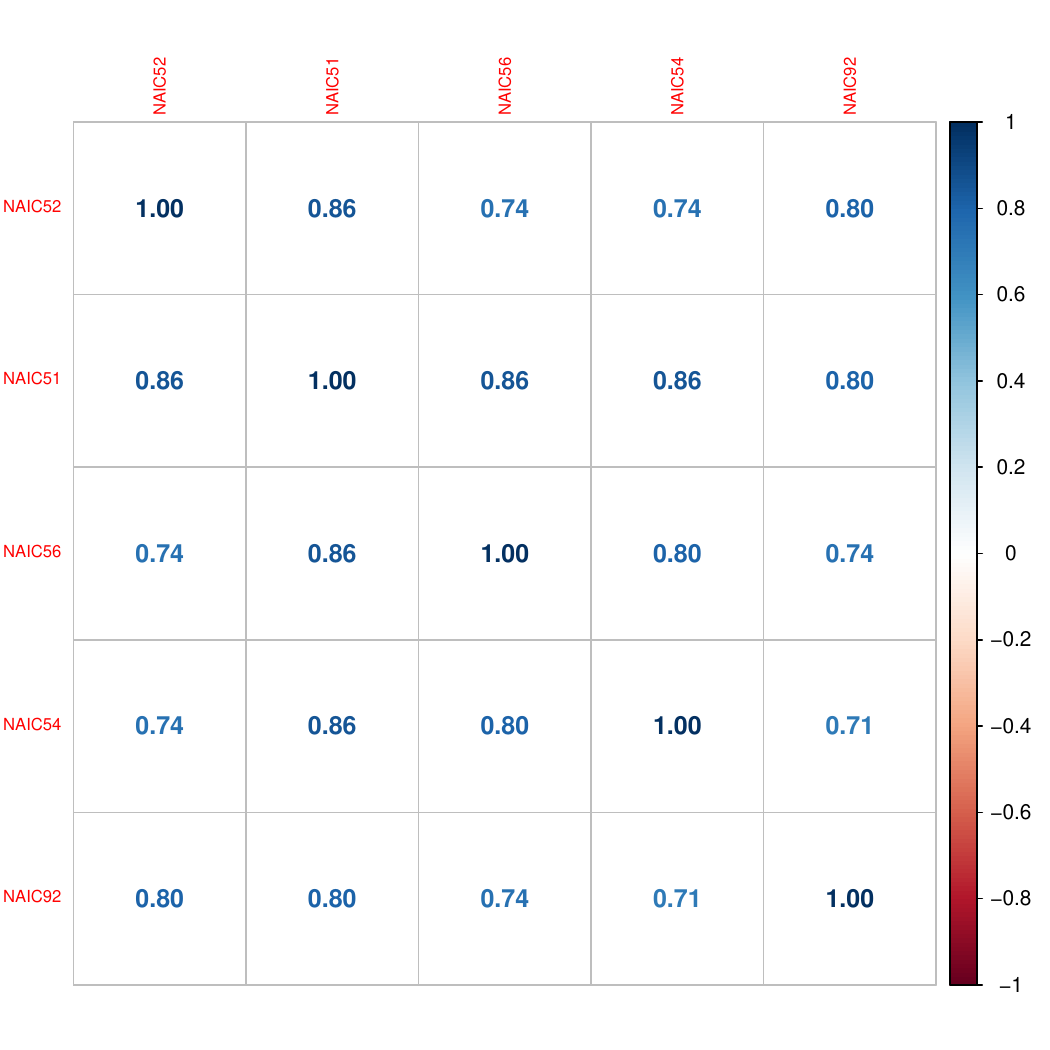}}\quad
\subfloat[Correlation matrix  of the top 5 NAICS in terms of occurrences, estimated using MCD estimator with the location and scale parameters given in Eqn. \ref{eq:mcd_corr}]{\includegraphics[scale=0.35]{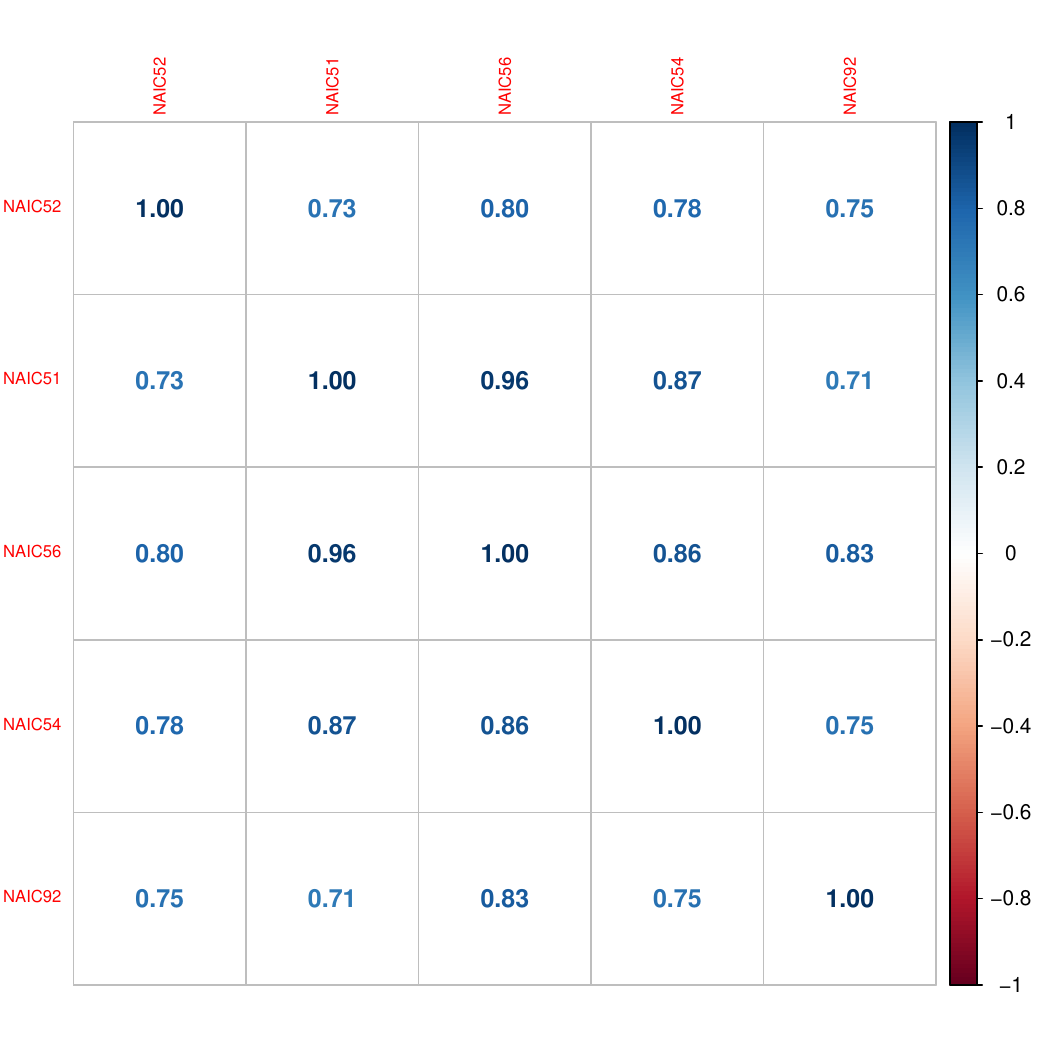}}
\centering
\caption{This figure shows the correlation matrices of NAIC52, NAIC51, NAIC56, NAIC54, and NAIC92. Estimates for correlation varies between the different estimators suggesting model risk also affects the dependence structure of cyber event severity. }
\label{Fig:robust_corr}
\end{figure}

Figure~\ref{Fig:robust_corr} shows the correlation matrices for cyber event severity occurring in the top 5 NAICs for number of events. Looking at the correlation estimates, the linear correlation case presents lower coefficients than the other robust estimators. This suggests that linear correlation might underestimate the strength of the dependence structure in cyber event severity. Moreover, the observed high degree of variation between the robust correlation estimates might suggest that cyber event severity dependence structure is also affected by parameter uncertainty.

We further investigate the impact of uncertainty in cyber event severity dependence parameters on insurance pricing, using a zero utility principle. In this case study, we consider a hypothetical insurance company with multiple lines of business for their cyber risk insurance policies. Each line of business corresponds to insurance policies the issue to companies in the US categorised under a given NAIC industry sector. This insurance company will then have a portfolio of insured cyber risks across various industry sectors. We will be interested in assessing in this section the influence on such an insurance portfolio of the model risk and parameter uncertainty associated with estimation of the dependence structure between the different cyber risk loss processes by line of business or NAIC. To continue the working illustration, we will focus on an insurance portfolio corresponding to the five NAICs studied in previous sections: NAIC52, NAIC 51, NAIC 56, NAIC54 and NAIC92. We will then modify the zero utility equation in \ref{eq:decision_premium} as follows to accommodate this insurance portfolio context, accounting for the dependence structures present between the NAICS, as shown in Equation \ref{eq:premium_copulas},
\begin{equation}
\label{eq:premium_copulas}
    \mathbb{E}\left[u\left(w-\sum_{i=1}^5\sum_{n=1}^{N_i}\omega_iX^i
    _n\right) \right] = \mathbb{E}\left[u\left( w-P -\sum_{i=1}^5\sum_{n=1}^{N_i}\omega_iX_n^i+\sum_{i=1}^5\sum_{n=1}^{N_i}\omega_i\min(X_n^i,cw)\right)\right],
\end{equation}
where $\omega_i$ corresponds to weight of each NAIC $i$. Note, we use notation $N_i$ here to denote that fact that it is a random variable for the number of losses in a given year. In order to numerically solve equation~\ref{eq:premium_copulas} one needs to know the joint distribution of cyber event frequency and severity occurring in the five considered NAICS.  While it's possible to employ copulas to approximate the multivariate compound process, in case of insurance premium calculations this might pose some challenges, given the presence of the top cover limit. Moreover, correlation estimates in Figure~\ref{Fig:robust_corr} refer to a quarterly aggregated losses. A possible solution is to implement an extension to more dimensions of the algorithm in \cite{cruz2015}, Chapter 12, where losses are drawn from a distribution of correlated aggregated losses. We will outline a summary of this approach as follows.

Consider a d-dimensional compound loss process $\bold{Z}=\left[Z^1,\dots,Z^d\right]$, with each component is a compound loss given by $Z^i=\sum_{n=1}^{N^i} X_n^i$, where $N^i\sim F_{N^i}$ and $X_n^i\sim F_{X^i}$, $n=1,\dots,N^i$ are random variables corresponding to frequency and severity of cyber event occurring in the NAIC $i$. The joint distribution of $\bold{Z}$, $F_{\bold{Z}}$ can then be uniquely expressed using a copula $C$ and the marginal distributions of each component, $F_{Z^i}$ as follows:
\begin{equation}
  F_{\bold{Z}}(z_1,\dots,z_d) = C\left(F_{Z^1}(z_1),\dots,F_{Z^d}(z_d)\right)\notag
\end{equation}

Then, the copula $C$ can be used to draw dependent variates for each compound loss $Z^{i*}$, and subsequently the insurance premiums can be computed on the corresponding random vector of losses $X^{i*} = \left[X_1^{i*},\dots,X_{N^{i*}}^{i*}\right]$. The steps of the algorithm in \cite{cruz2015} are summarised in the following psuedocode.

\begin{algorithm}
	\caption{Multivariate Compound Process}\label{algo:1}
	\begin{algorithmic}[0]
		\FOR{NAIC $i=1,2,\ldots d$}
			\FOR{Each MC draw $j=1,2,\ldots,J$}
				\STATE Simulate $N^{i,j}$ from $F_N$ and $X^{i,j}=\left[X_1^{i,j},\dots,X_{N^{i,j}}^{i,j}\right]$ from $F_X$
				\STATE Construct $Z^{i,j}=\sum_{n=1}^{N^{i,j}}X_n^{i,j}$
				\STATE Construct empirical cdf  $\widehat{F_{Z^i}(z)} = \frac{1}{J}\sum_{j=1}^{J}\mathcal{I}_{\left[ Z^{(i,j)}\leq z\right]}$ and the inverse cdf
				$\widehat{F_{Z^i}^{-1}(u)} = \inf\{j: Z^{(i,j)}\leq u \}$
			\ENDFOR
		\ENDFOR
		
	    \FOR{Each MC draw $s=1,\dots,S$}
	        \STATE Draw $u^{(i,s)}$ from the chosen copula
	        \STATE Find $j_s^*: Z^{(i,j_s^*)} = \widehat{F_{Z^i}^{-1}}(u^{i,s}) $ 
	        \STATE Create the "Multivariate" distribution with:  
	        $X^{i,j_s^*}=\left[X_1^{i,j_s^*},\dots,X_{N^{i,j_s^*}}^{i,j_s^*}\right]$
	    \ENDFOR
	\end{algorithmic} 
\end{algorithm}

We will now be particularly interested in the effect on diversification of this insurance portfolio coming from different dependence estimates arising from various assumptions about the extreme losses in the Advisen cyber risk data and how they manifest in parameter uncertainty and model risk that we conjecture also translates into impact on an actuaries perspective of the diversification of the insurance portfolio.

When the loss distributions exhibit extreme dependence, such as comonotonicty, and extreme tail behaviour, even well diversified positions fail to produce diversification benefit \cite{wang1998,dahen2010,ibragimov2011}. According to \cite{nevslehova2006}, in the case of infinite mean distribution, the Value-at-Risk, cannot be considered a coherent risk measure anymore, since subadditivity doesn't hold. In particular, when independent Pareto type heavy tailed risk sources are pooled together, the resulting value at risk becomes super additive:
\begin{equation}
    {\rm VaR}_{\alpha} \left(\sum_{i=1}^I Z^i\right) >\sum_{i=1}^I {\rm VaR}_{\alpha}(Z^i).
\end{equation}
To show how this affects the diversification benefit we consider the standard diversification measure given by the ratio of the value at risk of the position, and the weighted average of the value at risk of each NAIC:
\begin{equation}
\label{eq:diversification}
    D(Z) = \frac{{\rm VaR}_{\alpha} (\sum_{i=1}^I \omega_i Z^i) }{\sum_{i=1}^I \omega_i{\rm VaR}_{\alpha}(Z^i)}.
\end{equation}
In ideal conditions, $D(Z)$ is bounded in $[0,1]$, however we will show that this is not the case for cyber risk related losses in the Advisen data.

Table~\ref{table:net_ce} shows quarterly and yearly premium and diversification values for a company holding an equally weighted position in the five NAICs and a total wealth of 10 billion dollars. For each robust correlation estimator, we allow for a Gaussian copula as the dependence structure and generate the random losses using the algorithm in \ref{algo:1}.

\begin{table}[ht!]
	\centering
	\caption[Net Premiums]{This table shows the insurance premiums and the diversification measure for a equally weighted portfolio of NAIC52, NAIC51, NAIC56, NAIC54 and NAIC92. The estimator is reported along with the parameter standard error in brackets.}
	\label{table:net_ce}
	\begin{tabular}{l|cccc}
		Correlation Estimator & QuarterlyPremium & QuarterlyDiversification & YearlyPremium & YearlyDiversification\\
		\hline
Pearson correlation & 2157.723 & 1.754 & 7647.262 & 1.406 \\
                          & (59.014)  &  (0.191)       &  (43.951)        &  (0.268)     \\ 
SSD Median      & 2059.880 & 1.867 & 7482.660 & 1.269 \\
                          &   (59.540)     &  (0.295)   &    (44.215)     & (0.215)   \\ 
Quadrant (sign)      & 2018.442 & 1.547 & 7389.340 & 1.296 \\
                          &   (59.963)     &  (0.1883)   &    (43.069)     & (0.177)   \\ 
MCD & 1989.372 & 1.643 & 7313.716 & 1.419 \\
                          &   (59.662)     &  (0.1910)   &    (43.833)     & (0.194)   \\ 

        \hline
		
	\end{tabular}
\end{table}

Yearly and quarterly premiums in Table~\ref{table:net_ce} can be rearranged into a decreasing order starting with those corresponding to liner Pearson correlation and followed by SSD, quadrant (sign) correlation, and MCD correlation estimators. Combining these results with the correlation estimates in Figure~\ref{Fig:robust_corr}, where robust correlation estimators give higher estimates than the linear Pearson correlation, it can be inferred that under the combination of heavy tailed loss model marginals, combined with an eliptical copula with no tail dependence, such as the Gaussian copula model specification, then insurance premiums are negatively related with correlation estimates: ceteris paribas, as dependence strengthens, insurance premiums reduce. 

Moreover, the variability among the various robust correlation estimates also translates into insurance premium uncertainty that can be interpreted as potential mispricing if the data records for extreme losses that drive these model risk and parameter uncertainties identified are not adequately accounted for in the pricing calculations. In such cases the resulting premiums are shown to vary according to the underling assumptions about the data and model used in making an estimate of the dependence structure based on the correlations between the cyber risk loss processes. 

One can also observe that the diversification measure returns values greater than 1, both on a quarterly and a yearly basis. Given that the portfolio evaluated in this case is formed by heavy tailed risks, this is in agreement with the literature on diversification traps \cite{ibragimov2011}. In instances where the underlying risks are heavy tailed, the Value-at-Risk cannot be considered a subadditive risk measure anylonger and therefore, the resulting diversification measure is not bounded in the interval $[0,1]$. Moreover, the diversification measure defined in Equation \ref{eq:diversification} is in general not consistent with majorization orderings and in particular with the first order of stochastic dominance. Therefore, while insurance premiums are consistent with the riskiness of the position, this is not true in the case of the diversification measure. This is also confirmed by the values taken by the diversification measure in the various setting not following the ordering structure of the insurance premiums. Nevertheless we present these results for this measure of risk diversification as it is widely used in practice and so should be informative for practitioners.

Premiums in Table~\ref{table:net_ce} reflect how the uncertainty in correlation estimates affect the net exposure of a portfolio of five sources of cyber risk. To evaluate how the dependence structure affects the exposure of each individual NAIC we consider performing a sequence of conditional premium calculations. For each NAIC and each correlation estimator, the insurance premiums are computed using the zero utility principle and solving the following non linear equation modified to fine the the conditional cases:
\begin{equation}
\label{eq:premium_conditional}
    \mathbb{E}\left[u\left(w-Z^i\right)  \middle| Z^s\geq F^{-1}_{Z^S}(u),s \neq i \middle] = \mathbb{E}\middle[u\middle( w-P -Z^i+\sum_{n=1}^{N^i}\min(X_n^i,cw)\middle) \middle|  Z^s\geq F^{-1}_{Z^S}(u),s \neq i \right].
\end{equation}
We compute the conditional distribution for each case from the joint distribution generated by the algorithm in~\ref{algo:1}. Table~\ref{table:conditional_ce} shows the insurance premium computed on a quarterly basis, using the conditional equivalent principle of Equation~\ref{eq:premium_conditional}, for a representative company with 10 billion dollars of wealth. The responsiveness of each NAIC to losses greater than 75\% in the others business sectors can then be assessed in the subsequent results in Table \ref{table:conditional_ce}.

\begin{table}[ht!]
	\centering
	\caption[conditional Premiums]{This table shows conditional insurance premiums and their bootstrapped standard errors, for the 5 NAICs. Each premium is computed assuming losses greater than 75\% in the other NAICs, a company wealth of 10 billion dollars, and the relevant correlation structure. }
	\label{table:conditional_ce}
	\begin{tabular}{l|cccc}
		NAIC   &	Corr &	ssd	 & quadrant & 	mcd\\
		\hline
	    \multirow{2}*{NAIC52} & 7219.310 & 6607.928 & 6359.669 & 5820.162 \\
                              & (175.775)& (176.543)& (162.363)  &    (163.646)      \\
        \multirow{2}*{NAIC51} & 9521.399 & 9348.967 & 9202.159 & 8802.254 \\
                              &   (45.075)       &    (54.860)      &  (59.063)        &        (79.150)  \\
        \multirow{2}*{NAIC56} & 5017.681 & 5189.146 & 4768.389 & 4982.309 \\
                              &   (213.505)       &   (195.435)       &   (189.904)       &     (184.620)     \\
        \multirow{2}*{NAIC54} & 8692.571 & 8499.866 & 7889.639 & 7994.849 \\
                              &  (102.66)        &    (104.930)      &   (120.323)       &      (111.253)    \\
        \multirow{2}*{NAIC92} & 2151.384 & 1971.347 & 1655.238 & 1617.179 \\
                              & (211.624) & (183.206) & (154.234) & (147.890)\\

        \hline
		
	\end{tabular}
\end{table}

The insurance premiums in Table~\ref{table:conditional_ce} follow the same structure of the Hill estimates in Figure~\ref{Fig:hill_naics}, with NAIC51 reaching the highest values in all the considered dependence structures, suggesting that individual tail behaviour is still the main driver affecting insurance premiums calculation, even when dependence structure is considered in the modelling. Comparing the premium sizes with Table~\ref{table:net_ce} it can be seen that for each NAIC the highest conditional premium is achieved in the linear Pearson correlation case, while MCD and quadrant (sign) correlation return respecitively the lowest estimates. Moreover, except in the case of NAIC92 based on MCD and quadrant (sign) correlation, conditional premiums are greater than the net premiums, which is consistent with the increased risk of the positions analysed.

Insurance premiums from Tables~\ref{table:net_ce} and~\ref{table:conditional_ce} show the risk of a combination of NAICs and how pronounced is the effect of parameter uncertainty on the insurance pricing. Nonetheless, the two approaches can be used in different situations. The portfolio approach can be used in quantify how exposed a company is on the cyber risk front, and how variation in positions could improve the risk profile. In this context, parameter uncertainty \emph{externally} affects the enterprises under investigation, in the sense that among other things, the insurability of company cyber risk profile and stakeholder evaluation can be affected. The conditional calculation instead can be used \emph{internally} to evaluate strategically which risk management and mitigation strategies are best suited for the given company cyber risk profile. Here the effect of parameter uncertainty has the potential to be more catastrophic since, it could ultimately lead to sub-optimal or wrong decisions in the risk management department, or by the chief financial officer, where funds get misallocated to prevent or reduce the risk of catastrophic cyber events. Furthermore, there is a clear risk of mispricing insurance premiums associated with the misspecificed risk profile of the insurance portfolio, which could result in loss of competition, customers or even regulatory scrutiny and fines.

\subsection{Dependence Structures and Copula} \label{sec:background}
In this last section we will explore the copula dependence structure for pairs of the leading NAIC sectors. Fitting higher order copulas will be challenging due to the small sample sizes that arise from aggregating the loss data to a 3 monthly stratification. Recall, this period was selected to ensure reasonable sample sizes over time, so we have therefore intentionally restricted to 2-dimensional copula analysis as a result. Nevertheless this is still an insightful analysis to perform. Here we focus on the impact of selecting the \emph{right} dependence structure on insurance pricing and diversification measure. Given the quality and quantity of the data, instead of finding the best copula that fits the 5 NAICS jointly, we resort to variational approximation, where the true distribution is approximated by the combination of independent pair copulas that minimises the Kullback-Leibler divergence. We proceed with the following steps:
\begin{itemize}
    \item \emph{step 1}: we identify the best copula for each pair of NAIC according to an information criterion;
    \item \emph{step 2}: fit independence copula for each combination of pair copulas where marginals appear uniquely;
    \item \emph{step 3}: select the combination of independent pair copulas that minimises the Kullback-Leibler divergence.
\end{itemize}

Table~\ref{table:pair_copulas} show the results of the copula selection procedure according to the Akaike Information Criterion, the corresponding copula parameter estimated using maximum likelihood, and the Kendall's $\tau$. As show in in the table, there appears to be not much variation in terms of the selected copula, copula parameter and Kendall's $\tau$, as since the Joe copula is systematically selected as the best choice for each pair, and the parameters do not vary much between this model for each pair. This seems to suggest that when taking into account tail dependence, all the 5 NAICs have very similar behaviour, showing a positive tail dependence.

\begin{table}[]
    \centering
    \begin{tabular}{l|ccccc}
                                & NAIC52& NAIC51& NAIC56& NAIC54& NAIC92 \\
                                \hline
          \multirow{2}*{NAIC52} &    -  &  Survival Joe     & Survival Joe & Survival Joe & Survival Joe\\
                                &   -   & $\widehat{\theta} = 4.19$, $\widehat{\tau}= 0.63$  & $\widehat{\theta} = 2.96$, $\widehat{\tau}= 0.51$  & $\widehat{\theta} =4.09 $, $\widehat{\tau}= 0.62$ & $\widehat{\theta} =3.33 $, $\widehat{\tau}= 0.55$ \\
          \multirow{2}*{NAIC51} &    -  &  -  & Survival Joe & Survival Joe & Survival Joe\\
                                &   -   & -   &  $\widehat{\theta} = 3.85 , \widehat{\tau} = 0.6 $            &   $\widehat{\theta} = 4.7, \widehat{\tau} = 0.66 $  &   $\widehat{\theta} =3.58 , \widehat{\tau} = 0.58 $ \\
          \multirow{2}*{NAIC56} &   -   &   -    & - & Survival Joe & Survival Joe\\
                                &    -  &    -   &  -&  $\widehat{\theta} = 3.47, \widehat{\tau} =0.57 $    & $\widehat{\theta} =3.11 , \widehat{\tau} = 0.53$    \\
          \multirow{2}*{NAIC54} &  -    &   -    & - & - & Survival Joe\\
                                &   -   & -      & - &  -& $\widehat{\theta} = 3.17 , \widehat{\tau} = 0.54$\\
          \multirow{2}*{NAIC92} & -& -&- &-& -\\
                            & -& -&- &-& -\\

    \end{tabular}
    \caption{Best copula, and corresponding estimated parameter $\theta$ and Kendall's $\tau$ selected by the package VineCopula using Akaike information criterion}
    \label{table:pair_copulas}
\end{table}

\begin{figure}
    \centering
    \includegraphics[scale=.5]{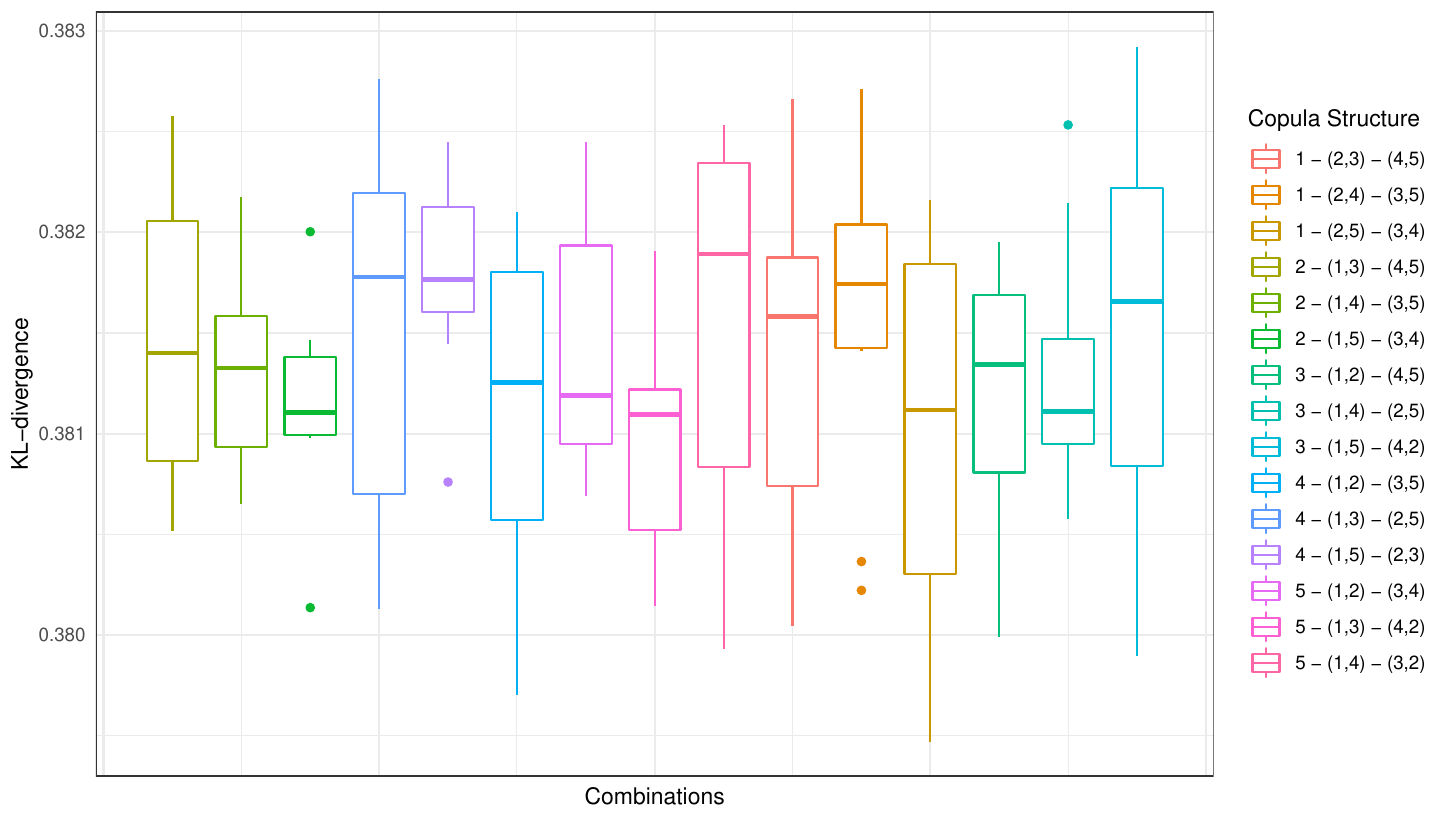}

    \caption{KL divergence of different copula structure ordered by tail parameter of the independent component for different seed. The values of KL divergence remains similar for different copula structure combinations. The legend reads as follows: 1:NAIC52, 2: NAIC51, 3:NAIC56, 4:NAIC54, 5:NAIC92. }
\label{Fig:kl_div}
\end{figure}

Given that our analysis focused on an odd number of NAICs, we form the combined five dimensional model as comprised of a product of 2-dimensional copulas in the variational approximation keeping one component independent, while allowing pair copulas for the other 4 NAICs. Figure~\ref{Fig:kl_div} shows the Kullback-Leibler divergence between all the possible combinations of pair copulas and independent component, for different seed in the random number generator, ordered according to the tail index estimates of the independent component. As it can be seen the relatively flat structure in Table~\ref{table:pair_copulas} affects also the KL divergence results, where the values are so close to each other that even the small and almost negligible variation due to the random number generation could affect the results. 

\begin{figure}[h] 
\subfloat[Insurance premiums for the copula resulting from the variational approximation and bootstrapped confidence intervals]{\includegraphics[scale=0.45]{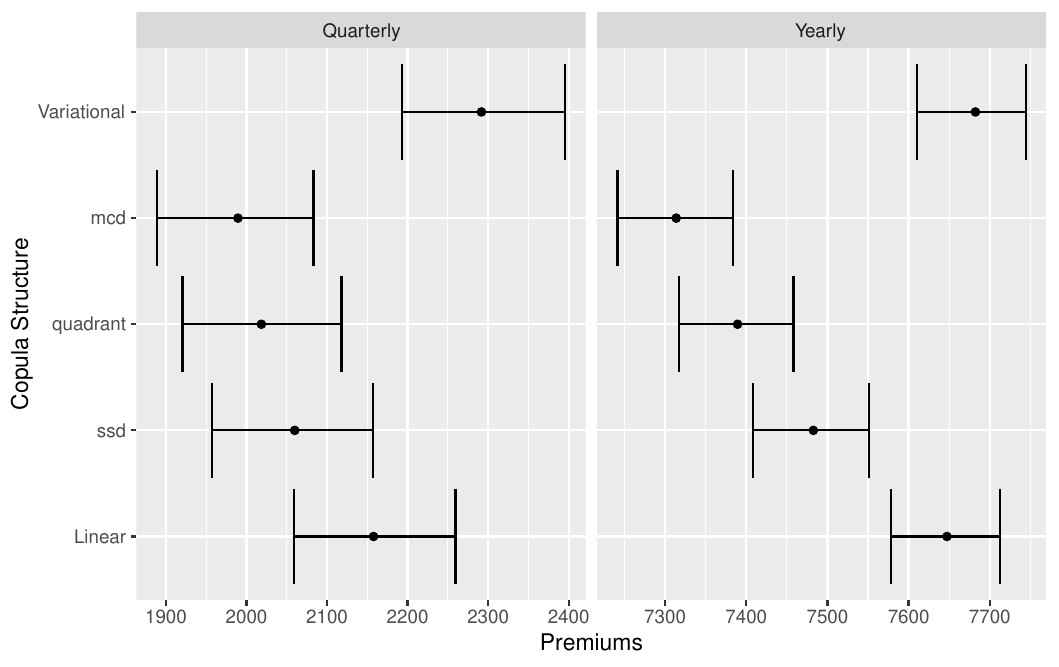}}\quad
\subfloat[Diversification coefficient for the copula resulting from the variational approximation and bootstrapped confidence intervals]{\includegraphics[scale=0.45]{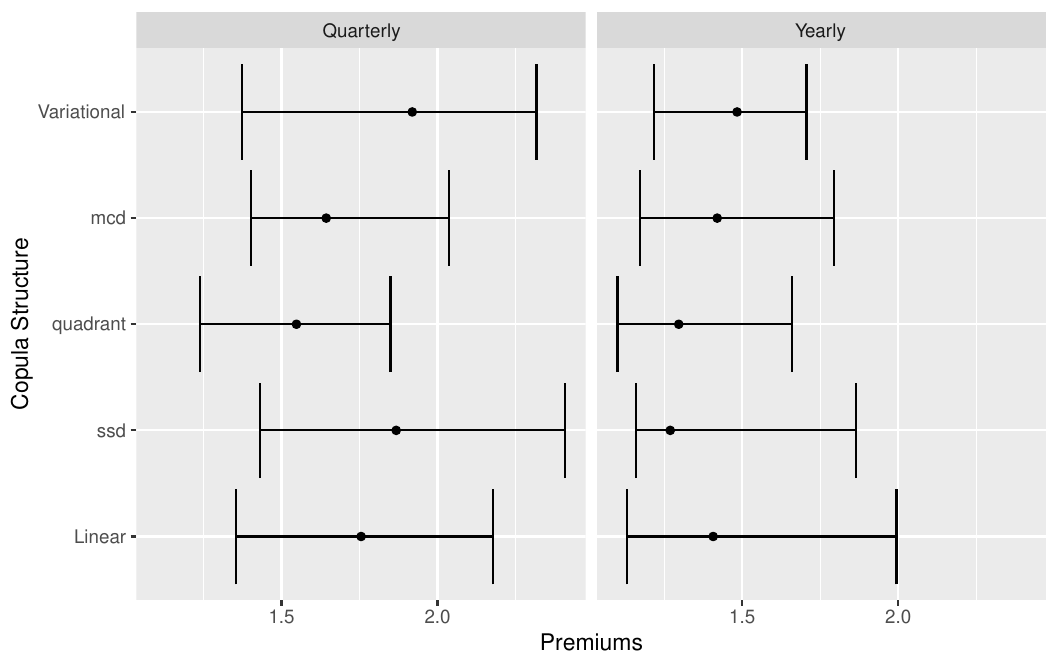}}\quad

\centering
\caption{Insurance premiums based on log utility and diversification measure of an equally weighted portfolio of 5 NAICs, with bootstrapped 95\% confidence intervals.}
\label{Fig:variational_premiums}

\end{figure}

Since, selecting a best performing approximated copula structure is not possible, we present the results for the case where NAIC51 is left as an independent component, NAIC52 and NAIC54 are fitted on Joe copula with parameter $\theta = 4.09$, and NAIC56 and NAIC92 are fitted on a Joe copula with parameter $\theta = 3.11$. Similar results can be obtained with other combinations of independent component and pair copulas. Then we use the selected copula structure as basis for the simulation in the algorithm \ref{algo:1} and compute the insurance premium of Equation \ref{eq:premium_copulas} in the case of an equally weighted portfolio of NAICs, and the corresponding diversification measure. Figure~\ref{Fig:variational_premiums} compares the results with those previously obtained using the Gaussian settings and the different robust correlation estimators, reporting also the bootstrapped 95\% confidence intervals. Insurance premiums computed using the approximated copula are statistically different from those computed using the Gaussian copula with the robust correlation estimators, both on quarterly and yearly basis. This indicates that not only parameter uncertainty affects insurance premium calculation in the case of cyber risk, but model risk does as well. Nevertheless,  premiums based on the linear correlation estimator are not statistically different from the one computed using the approximated copula. This can be explained due to the presence of two conflicting biases. On the one hand, the assumption of Gaussian copula as joint dependence structure seem to increase the values of insurance premiums. One the other hand, robust correlation estimators reduce the premium values, resulting in a premium not statistically different than the one computed using the approximated copula structure. Looking at the diversification measure results, there appear not to be any statistically significant differences between the considered underling dependence structures: diversification measure fails to be bounded in the interval $[0,1]$, due to the lack of subadditivity, and seems to have a more skewed bootstrapped distribution with respect to the insurance premiums, having the mean not centred in the confidence intervals. Finally, it can be notice how the confidence intervals for diversification measure seem to be less affected by time aggregation than the premiums counterparts. This can be explained by the lack of subadditivity due to the heavy tails of the considered risks: in the bootstraps procedure it's more likely that extreme scenarios, violating the subadditivity are generated more often. 

Figure~\ref{Fig:variational_premiums} provides statistical evidence that model risk and parameter uncertainty in cyber risk translate into insurance premiums and could affect the decision making process. 

\FloatBarrier

\section{Conclusions} 
\label{sec:Conclusions}
The paper explored the relationship between model risk and parameter uncertainty in insurance pricing in the setting of cyber risk. In particular the paper sought to explore whether the perspective previously held in the literature, that cyber risk losses are heavy tailed, was consistently found in the largest industry standard loss data based, obtained from Advisen. In this context the paper showed, that ones perspective on the tail behaviour of cyber risk loss processes is heavily dependent on the ability to rely upon the properties of the data obtained for calibration. Given that in the industry leading database there was evidence of some of the largest losses being incompletely reported, rounded, approximated and never settled or realised, we decided to assess what impact this may have on the actuaries perspective of the tail behaviour of such cyber risk loss processes. This is particularly compounded by the fact that necessarily so, the total cyber loss per event in Advisen dataset is a composition of many direct and indirect loss components, where direct losses are from the event itself and indirect losses are from the consequences of the event.  When the extremes of the data are contaminated, the classical Hill type estimators lead to inaccuracy in utility based cyber insurance premium calculations.  Furthermore, it poses a challenge to assess insurability of cyber risk losses. 

It was demonstrated that if robust estimators were adopted rather than the standard tail index estimators that used all the data as equally weighted and applied no trimming. To improve the robustness of tail index estimator reducing the effect of such observations (i.e. extreme outliers), we used the trimmed Hill estimator on aggregated 5 NAICs cyber related losses. We noticed that while it is highly sensitive to the choice of the trimming parameters for each NAIC, the uncertainty of the trimmed Hill estimator affects ultimately premium mispricing. Consequently, model risk makes it difficult to assess insurability of cyber risk losses. This led to the conclusion that significant model risk and parameter uncertainty may be present in the analysis depending on ones perspective on assessing the quality of the real data. Furthermore, we showed that once this was translated into insurance pricing, this led to significant mispricing potential in charged premiums.

We also investigated how uncertainty of the dependence structure of cyber event severity between 5 NAICs impacted on utility based conditional/unconditional cyber insurance premium pricing and diversification benefit measures.  Dependence between 5 NAICs was studied via robust dependence estimation methods, copula estimation methods, and a little known Monte Carlo based simulation method and the VaR ratio was used as diversification measure.

We provided statistical evidence that cyber premium mispricing and misleading diversification benefit measuring can arise from dependence model structure uncertainty as well as relevant parameter uncertainty.  We hope that what we have presented in this paper will provide practitioners with sensible approaches to quantify/assess univariate/multivariate cyber losses over a range of different industry sector dealing with model risk.

\section{Acknowledgement}
This research has been conducted within the Optus Macquarie University Cyber Security Hub and funded by its Risk Management, Governance and Control Program.

\section{Competing Interests}
The authors declare none

\section{Contributions}
GWP designed the research study, performed implementation of model analysis and data analysis in R for statistical estimation of models, and wrote the first paper draft. MM performed additional data analysis in implementing the R examples for insurance premium calculations and diversification analysis and edited the draft of the manuscript. PVS, GS, ST, JJ all provided ongoing consultation on Advisen data analysis and assisted in commenting the drafts of the manuscript.

\bibliographystyle{IEEEtran}
\bibliography{references}

\begin{thebibliography}{10}
\providecommand{\url}[1]{#1}
\csname url@samestyle\endcsname
\providecommand{\newblock}{\relax}
\providecommand{\bibinfo}[2]{#2}
\providecommand{\BIBentrySTDinterwordspacing}{\spaceskip=0pt\relax}
\providecommand{\BIBentryALTinterwordstretchfactor}{4}
\providecommand{\BIBentryALTinterwordspacing}{\spaceskip=\fontdimen2\font plus
\BIBentryALTinterwordstretchfactor\fontdimen3\font minus
  \fontdimen4\font\relax}
\providecommand{\BIBforeignlanguage}[2]{{%
\expandafter\ifx\csname l@#1\endcsname\relax
\typeout{** WARNING: IEEEtran.bst: No hyphenation pattern has been}%
\typeout{** loaded for the language `#1'. Using the pattern for}%
\typeout{** the default language instead.}%
\else
\language=\csname l@#1\endcsname
\fi
#2}}
\providecommand{\BIBdecl}{\relax}
\BIBdecl

\bibitem{eling2020cyber}
M.~Eling, ``Cyber risk research in business and actuarial science,''
  \emph{European Actuarial Journal}, pp. 1--31, 2020.

\bibitem{eling2016we}
M.~Eling and W.~Schnell, ``What do we know about cyber risk and cyber risk
  insurance?'' \emph{The Journal of Risk Finance}, 2016.

\bibitem{peters2018understanding}
G.~W. Peters, P.~V. Shevchenko, and R.~D. Cohen, ``Understanding cyber-risk and
  cyber-insurance,'' in \emph{Fintech: Growth and Deregulation}.\hskip 1em plus
  0.5em minus 0.4em\relax Risk Books, 2018, pp. 303--330.

\bibitem{romanosky2017content}
S.~Romanosky, L.~Ablon, A.~Kuehn, and T.~Jones, ``Content analysis of cyber
  insurance policies: How do carriers write policies and price cyber risk?''
  \emph{Available at SSRN 2929137}, 2017.

\bibitem{biener2015insurability}
C.~Biener, M.~Eling, and J.~H. Wirfs, ``Insurability of cyber risk: An
  empirical analysis,'' \emph{The Geneva Papers on Risk and Insurance-Issues
  and Practice}, vol.~40, no.~1, pp. 131--158, 2015.

\bibitem{edwards2016hype}
B.~Edwards, S.~Hofmeyr, and S.~Forrest, ``Hype and heavy tails: A closer look
  at data breaches,'' \emph{Journal of Cybersecurity}, vol.~2, no.~1, pp.
  3--14, 2016.

\bibitem{shevchenko2021quantification}
P.~V. Shevchenko, J.~Jang, M.~Malavasi, G.~W. Peters, G.~Sofronov, and
  S.~Tr{\"u}ck, ``Quantification of cyber risk--risk categories and business
  sectors,'' \emph{Available at SSRN 3858608}, 2021.

\bibitem{peters2018statistical}
G.~W. Peters, P.~V. Shevchenko, R.~D. Cohen, and D.~R. Maurice, ``Statistical
  machine learning analysis of cyber risk data: event case studies,'' in
  \emph{Fintech: Growth and Deregulation}.\hskip 1em plus 0.5em minus
  0.4em\relax Risk Books, 2018, pp. 75--99.

\bibitem{cruz2015}
M.~G. Cruz, G.~W. Peters, and P.~V. Shevchenko, \emph{Fundamental aspects of
  operational risk and insurance analytics: A handbook of operational
  risk}.\hskip 1em plus 0.5em minus 0.4em\relax John Wiley \& Sons, 2015.

\bibitem{ceross2017use}
A.~Ceross and A.~Simpson, ``The use of data protection regulatory actions as a
  data source for privacy economics,'' in \emph{International Conference on
  Computer Safety, Reliability, and Security}.\hskip 1em plus 0.5em minus
  0.4em\relax Springer, 2017, pp. 350--360.

\bibitem{eling2018cyber}
M.~Eling, ``Cyber risk and cyber risk insurance: Status quo and future
  research,'' \emph{The Geneva papers on risk and insurance-issues and
  practice}, vol.~43, no.~2, pp. 175--179, 2018.

\bibitem{xie2020cyber}
X.~Xie, C.~Lee, and M.~Eling, ``Cyber insurance offering and performance: an
  analysis of the us cyber insurance market,'' \emph{The Geneva Papers on Risk
  and Insurance-Issues and Practice}, vol.~45, no.~4, pp. 690--736, 2020.

\bibitem{eling2017}
M.~Eling and N.~Loperfido, ``Data breaches: Goodness of fit, pricing, and risk
  measurement,'' \emph{Insurance: mathematics and economics}, vol.~75, pp.
  126--136, 2017.

\bibitem{eling2019}
M.~Eling and J.~Wirfs, ``What are the actual costs of cyber risk events?''
  \emph{European Journal of Operational Research}, vol. 272, no.~3, pp.
  1109--1119, 2019.

\bibitem{edwards2016}
B.~Edwards, S.~Hofmeyr, and S.~Forrest, ``Hype and heavy tails: A closer look
  at data breaches,'' \emph{Journal of Cybersecurity}, vol.~2, no.~1, pp.
  3--14, 2016.

\bibitem{jung2021extreme}
K.~Jung, ``Extreme data breach losses: An alternative approach to estimating
  probable maximum loss for data breach risk,'' \emph{North American Actuarial
  Journal}, vol.~25, no.~4, pp. 580--603, 2021.

\bibitem{frohlich2018}
A.~Fr{\"o}hlich and A.~Weng, ``Parameter uncertainty and reserve risk under
  solvency ii,'' \emph{Insurance: Mathematics and Economics}, vol.~81, pp.
  130--141, 2018.

\bibitem{brazauskas2000robust}
V.~Brazauskas and R.~Serfling, ``Robust estimation of tail parameters for
  two-parameter pareto and exponential models via generalized quantile
  statistics,'' \emph{Extremes}, vol.~3, no.~3, pp. 231--249, 2000.

\bibitem{zou2020extreme}
J.~Zou, R.~A. Davis, and G.~Samorodnitsky, ``Extreme value analysis without the
  largest values: what can be done?'' \emph{Probability in the Engineering and
  Informational Sciences}, vol.~34, no.~2, pp. 200--220, 2020.

\bibitem{goegebeur2014robust}
Y.~Goegebeur, A.~Guillou, and A.~Verster, ``Robust and asymptotically unbiased
  estimation of extreme quantiles for heavy tailed distributions,''
  \emph{Statistics \& Probability Letters}, vol.~87, pp. 108--114, 2014.

\bibitem{peng2001robust}
L.~Peng and A.~Welsh, ``Robust estimation of the generalized pareto
  distribution,'' \emph{Extremes}, vol.~4, no.~1, pp. 53--65, 2001.

\bibitem{bhattacharya2017trimming}
S.~Bhattacharya, M.~Kallitsis, and S.~Stoev, ``Trimming the hill estimator:
  robustness, optimality and adaptivity,'' \emph{arXiv preprint
  arXiv:1705.03088}, 2017.

\bibitem{rea2017systematic}
A.~Rea-Guaman, T.~San~Feliu, J.~Calvo-Manzano, and I.~D.
  S{\'a}nchez-Garc{\'\i}a, ``Systematic review: Cybersecurity risk taxonomy,''
  in \emph{International Conference on Software Process Improvement}.\hskip 1em
  plus 0.5em minus 0.4em\relax Springer, 2017, pp. 137--146.

\bibitem{elnagdy2016understanding}
S.~A. Elnagdy, M.~Qiu, and K.~Gai, ``Understanding taxonomy of cyber risks for
  cybersecurity insurance of financial industry in cloud computing,'' in
  \emph{2016 IEEE 3rd International Conference on Cyber Security and Cloud
  Computing (CSCloud)}.\hskip 1em plus 0.5em minus 0.4em\relax IEEE, 2016, pp.
  295--300.

\bibitem{romanosky2016}
S.~Romanosky, ``Examining the costs and causes of cyber incidents,''
  \emph{Journal of Cybersecurity}, vol.~2, no.~2, pp. 121--135, 2016.

\bibitem{cyentia2020}
Cyentia, ``A clearer vision for assessing the risk of cyber incidents,''
  Cyentia Institute, Tech. Rep., 2020.

\bibitem{malavasi2021cyber}
M.~Malavasi, G.~W. Peters, P.~V. Shevchenko, S.~Tr{\"u}ck, J.~Jang, and
  G.~Sofronov, ``Cyber risk frequency, severity and insurance viability,''
  \emph{arXiv preprint arXiv:2111.03366}, 2021.

\bibitem{resnick1997smoothing}
S.~Resnick and C.~St{\u{a}}ric{\u{a}}, ``Smoothing the hill estimator,''
  \emph{Advances in Applied Probability}, vol.~29, no.~1, pp. 271--293, 1997.

\bibitem{davis2009extreme}
R.~A. Davis and T.~Mikosch, ``Extreme value theory for garch processes,'' in
  \emph{Handbook of financial time series}.\hskip 1em plus 0.5em minus
  0.4em\relax Springer, 2009, pp. 187--200.

\bibitem{peters2015advances}
G.~W. Peters and P.~V. Shevchenko, \emph{Advances in heavy tailed risk
  modeling: A handbook of operational risk}.\hskip 1em plus 0.5em minus
  0.4em\relax John Wiley \& Sons, 2015.

\bibitem{foss2011introduction}
S.~Foss, D.~Korshunov, S.~Zachary \emph{et~al.}, \emph{An introduction to
  heavy-tailed and subexponential distributions}.\hskip 1em plus 0.5em minus
  0.4em\relax Springer, 2011, vol.~6.

\bibitem{pitman1968behaviour}
E.~Pitman, ``On the behaviour of the characteristic function of a probability
  distribution in the neighbourhood of the origin,'' \emph{Journal of the
  Australian Mathematical Society}, vol.~8, no.~3, pp. 423--443, 1968.

\bibitem{hii1975simple}
B.~HiI, ``A simple genera/approach to inference about the tail of a
  distribution,'' \emph{Ann. Statist}, vol.~3, pp. 1163--1174, 1975.

\bibitem{hall1982some}
P.~Hall, ``On some simple estimates of an exponent of regular variation,''
  \emph{Journal of the Royal Statistical Society: Series B (Methodological)},
  vol.~44, no.~1, pp. 37--42, 1982.

\bibitem{fedotenkov2020review}
I.~Fedotenkov, ``A review of more than one hundred pareto-tail index
  estimators,'' \emph{Statistica}, vol.~80, no.~3, pp. 245--299, 2020.

\bibitem{welsh1986use}
A.~Welsh, ``On the use of the empirical distribution and characteristic
  function to estimate parameters of regular variation,'' \emph{Australian
  Journal of Statistics}, vol.~28, no.~2, pp. 173--181, 1986.

\bibitem{jia2014heavy}
M.~Jia, ``Heavy-tailed phenomena and tail index inference,'' Ph.D.
  dissertation, University of Trento, 2014.

\bibitem{hill1975simple}
B.~M. Hill, ``A simple general approach to inference about the tail of a
  distribution,'' \emph{The annals of statistics}, pp. 1163--1174, 1975.

\bibitem{pickands1975statistical}
J.~Pickands~III, ``Statistical inference using extreme order statistics,''
  \emph{the Annals of Statistics}, pp. 119--131, 1975.

\bibitem{hall1982limit}
P.~Hall, ``Limit theorems for estimators based on inverses of spacings of order
  statistics,'' \emph{The Annals of Probability}, pp. 992--1003, 1982.

\bibitem{embrechts2013modelling}
P.~Embrechts, C.~Kl{\"u}ppelberg, and T.~Mikosch, \emph{Modelling extremal
  events: for insurance and finance}.\hskip 1em plus 0.5em minus 0.4em\relax
  Springer Science \& Business Media, 2013, vol.~33.

\bibitem{munasinghetail}
R.~Munasinghe, P.~Kossinna, D.~Jayasinghe, and D.~Wijeratne, ``Tail index
  estimation for power law distributions in {R}.''

\bibitem{hubert2013detecting}
M.~Hubert, G.~Dierckx, and D.~Vanpaemel, ``Detecting influential data points
  for the hill estimator in pareto-type distributions,'' \emph{Computational
  Statistics \& Data Analysis}, vol.~65, pp. 13--28, 2013.

\bibitem{vandewalle2007robust}
B.~Vandewalle, J.~Beirlant, A.~Christmann, and M.~Hubert, ``A robust estimator
  for the tail index of pareto-type distributions,'' \emph{Computational
  Statistics \& Data Analysis}, vol.~51, no.~12, pp. 6252--6268, 2007.

\bibitem{newman2005power}
M.~E. Newman, ``Power laws, pareto distributions and zipf's law,''
  \emph{Contemporary physics}, vol.~46, no.~5, pp. 323--351, 2005.

\bibitem{rizzo2009new}
M.~L. Rizzo, ``New goodness-of-fit tests for pareto distributions,''
  \emph{ASTIN Bulletin: The Journal of the IAA}, vol.~39, no.~2, pp. 691--715,
  2009.

\bibitem{nair2013fundamentals}
J.~Nair, A.~Wierman, and B.~Zwart, ``The fundamentals of heavy-tails:
  Properties, emergence, and identification,'' in \emph{Proceedings of the ACM
  SIGMETRICS/international conference on Measurement and modeling of computer
  systems}, 2013, pp. 387--388.

\bibitem{bhatti2018efficient}
S.~H. Bhatti, S.~Hussain, T.~Ahmad, M.~Aslam, M.~Aftab, and M.~A. Raza,
  ``Efficient estimation of pareto model: Some modified percentile
  estimators,'' \emph{PloS one}, vol.~13, no.~5, p. e0196456, 2018.

\bibitem{aban2004generalized}
I.~B. Aban and M.~M. Meerschaert, ``Generalized least-squares estimators for
  the thickness of heavy tails,'' \emph{Journal of Statistical Planning and
  Inference}, vol. 119, no.~2, pp. 341--352, 2004.

\bibitem{beirlant2016tail}
J.~Beirlant, I.~F. Alves, and I.~Gomes, ``Tail fitting for truncated and
  non-truncated pareto-type distributions,'' \emph{Extremes}, vol.~19, no.~3,
  pp. 429--462, 2016.

\bibitem{beirlant2004statistics}
J.~Beirlant, Y.~Goegebeur, J.~Segers, and J.~L. Teugels, \emph{Statistics of
  extremes: theory and applications}.\hskip 1em plus 0.5em minus 0.4em\relax
  John Wiley \& Sons, 2004, vol. 558.

\bibitem{shevlyakov2011robust}
G.~Shevlyakov and P.~Smirnov, ``Robust estimation of the correlation
  coefficient: An attempt of survey,'' \emph{Austrian Journal of Statistics},
  vol.~40, no. 1\&2, pp. 147--156, 2011.

\bibitem{falk1998note}
M.~Falk, ``A note on the comedian for elliptical distributions,'' \emph{Journal
  of Multivariate Analysis}, vol.~67, no.~2, pp. 306--317, 1998.

\bibitem{blomqvist1950measure}
N.~Blomqvist, ``On a measure of dependence between two random variables,''
  \emph{The Annals of Mathematical Statistics}, pp. 593--600, 1950.

\bibitem{lopuhaa1991breakdown}
H.~P. Lopuhaa and P.~J. Rousseeuw, ``Breakdown points of affine equivariant
  estimators of multivariate location and covariance matrices,'' \emph{The
  Annals of Statistics}, pp. 229--248, 1991.

\bibitem{wang1998}
S.~Wang and J.~Dhaene, ``Comonotonicity, correlation order and premium
  principles,'' \emph{Insurance: Mathematics and Economics}, vol.~22, no.~3,
  pp. 235--242, 1998.

\bibitem{dahen2010}
H.~Dahen and G.~Dionne, ``Scaling models for the severity and frequency of
  external operational loss data,'' \emph{Journal of Banking \& Finance},
  vol.~34, no.~7, pp. 1484--1496, 2010.

\bibitem{ibragimov2011}
R.~Ibragimov, D.~Jaffee, and J.~Walden, ``Diversification disasters,''
  \emph{Journal of Financial Economics}, vol.~99, no.~2, pp. 333--348, 2011.

\bibitem{nevslehova2006}
J.~Ne{\v{s}}lehov{\'a}, P.~Embrechts, and V.~Chavez-Demoulin, ``Infinite mean
  models and the {LDA} for operational risk,'' \emph{Journal of Operational
  Risk}, vol.~1, no.~1, pp. 3--25, 2006.

\end{thebibliography}

\FloatBarrier

\section{Appendix 1}
\vspace{-0.25cm}

\begin{figure}[h]
    \centering
    \subfloat[]{\includegraphics[width = 0.5\textwidth, height = 0.25\textwidth]{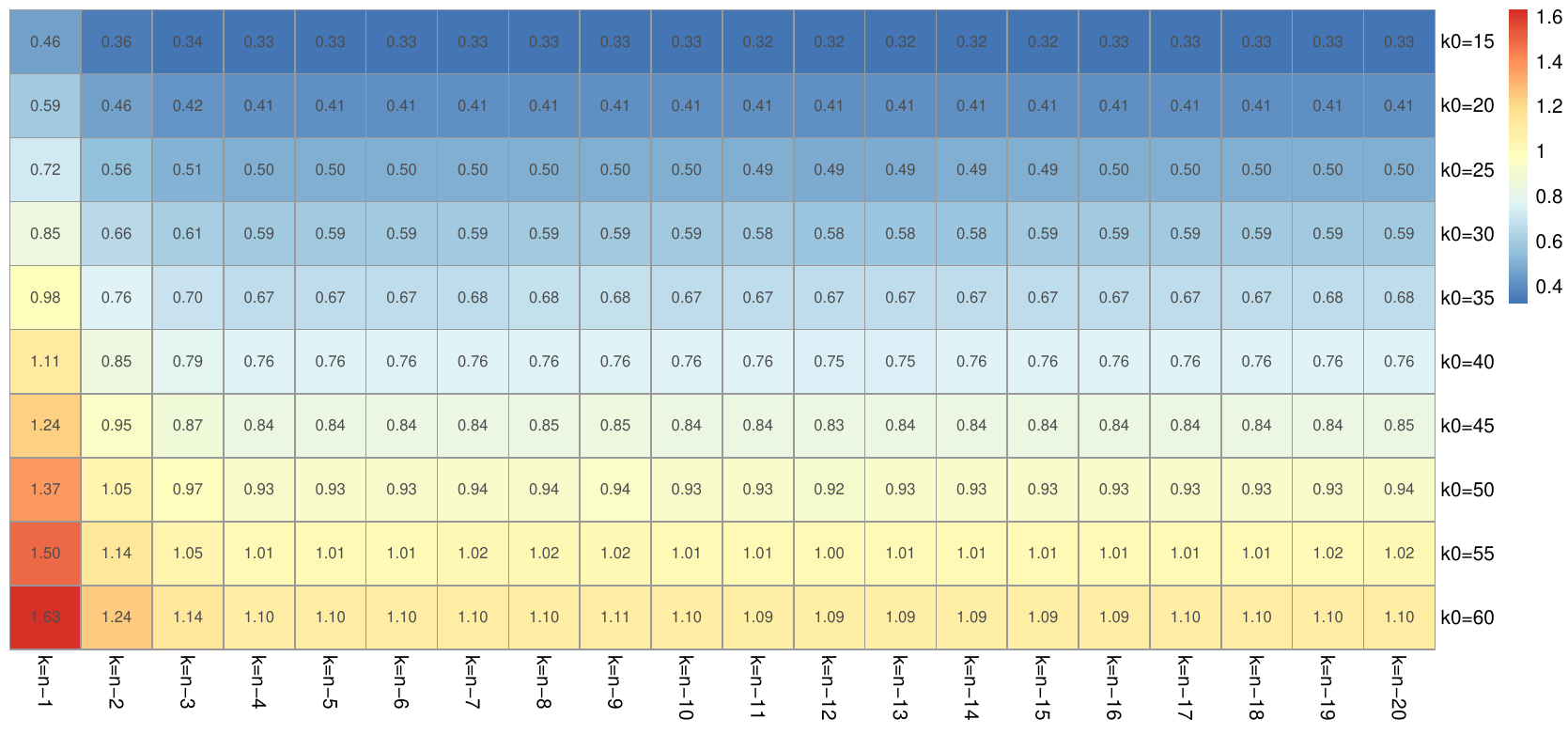}}
    \subfloat[]{\includegraphics[width = 0.5\textwidth, height = 0.25\textwidth]{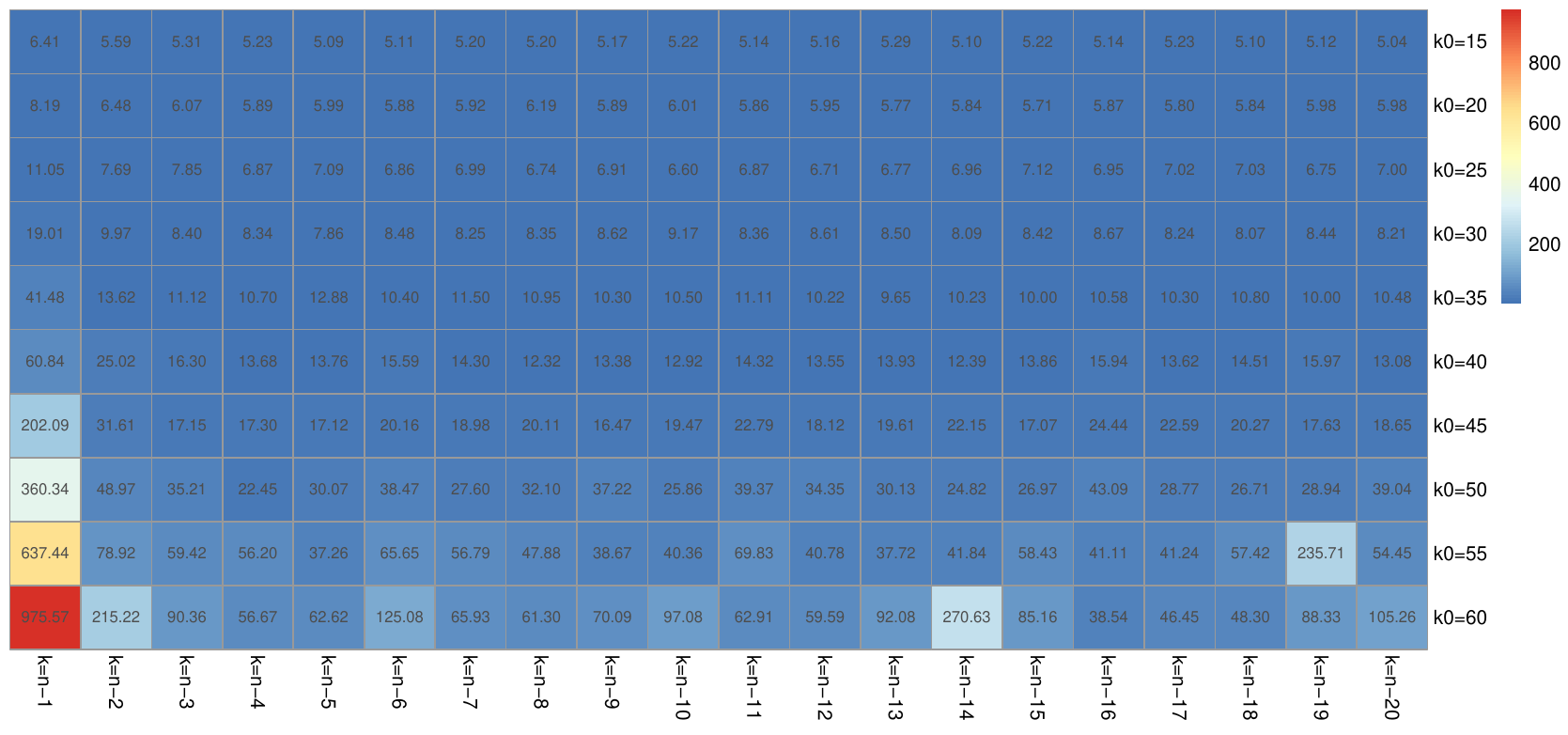}}\\
    \subfloat[]{\includegraphics[width = 0.5\textwidth, height = 0.25\textwidth]{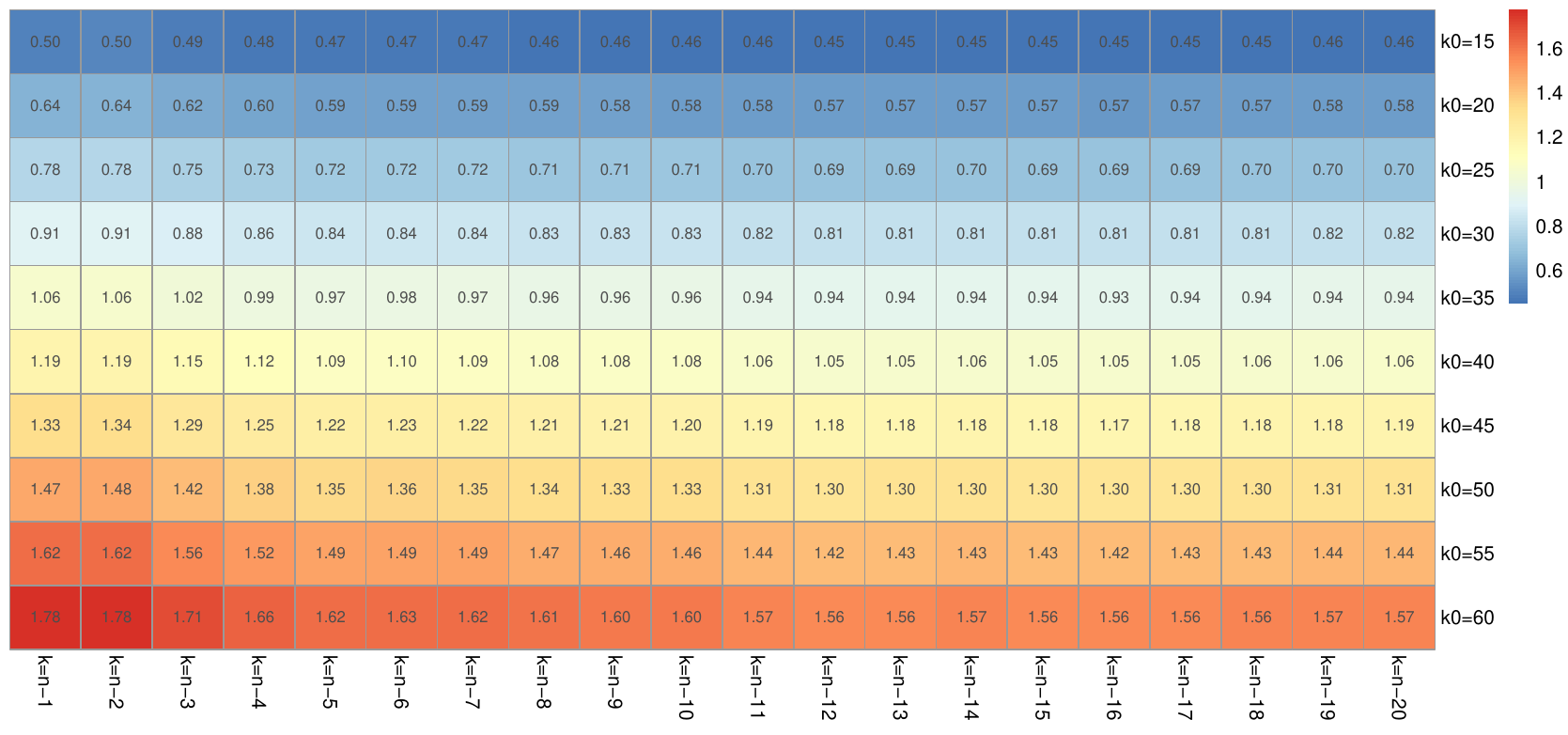}}
    \subfloat[]{\includegraphics[width = 0.5\textwidth, height = 0.25\textwidth]{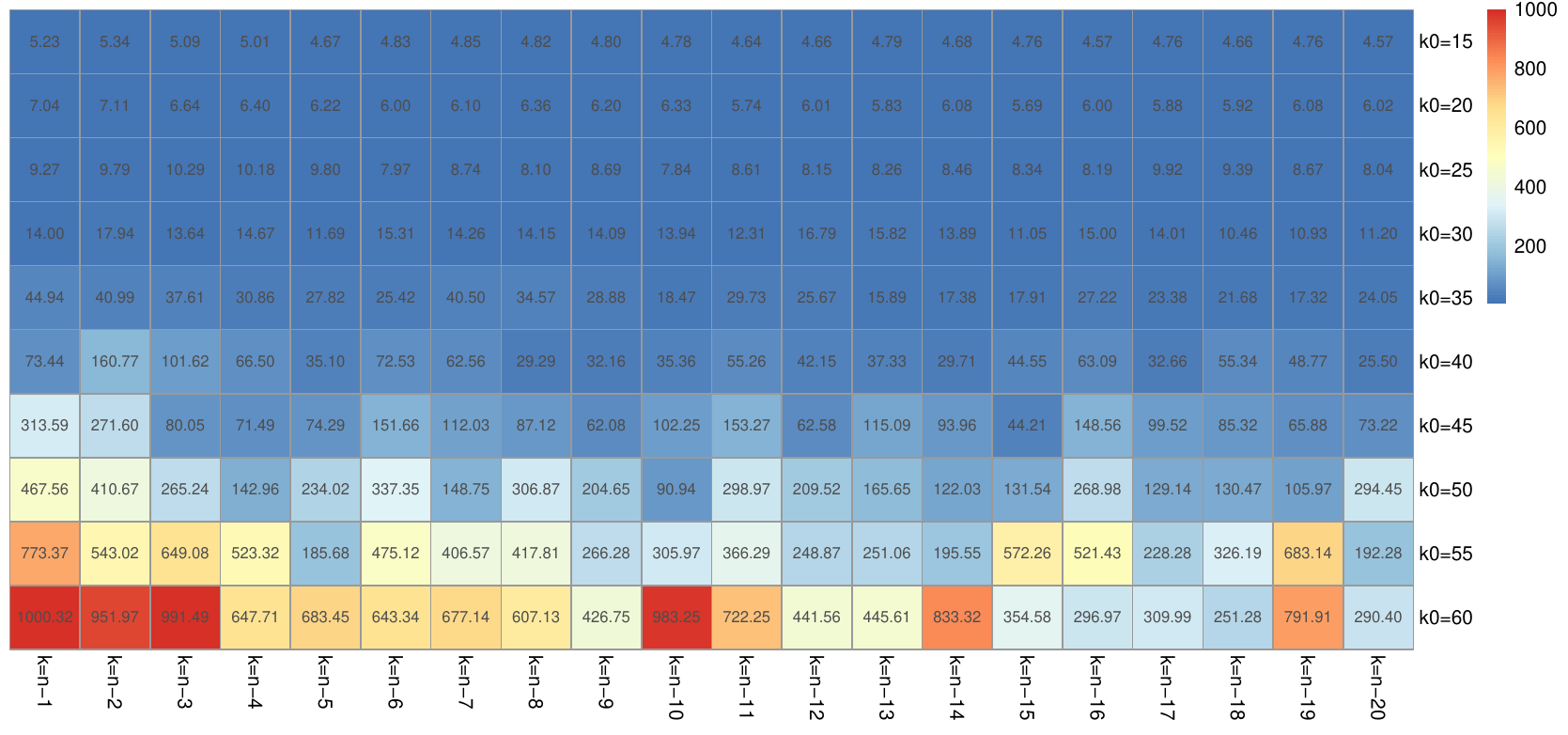}}\\
    \subfloat[]{\includegraphics[width = 0.5\textwidth, height = 0.25\textwidth]{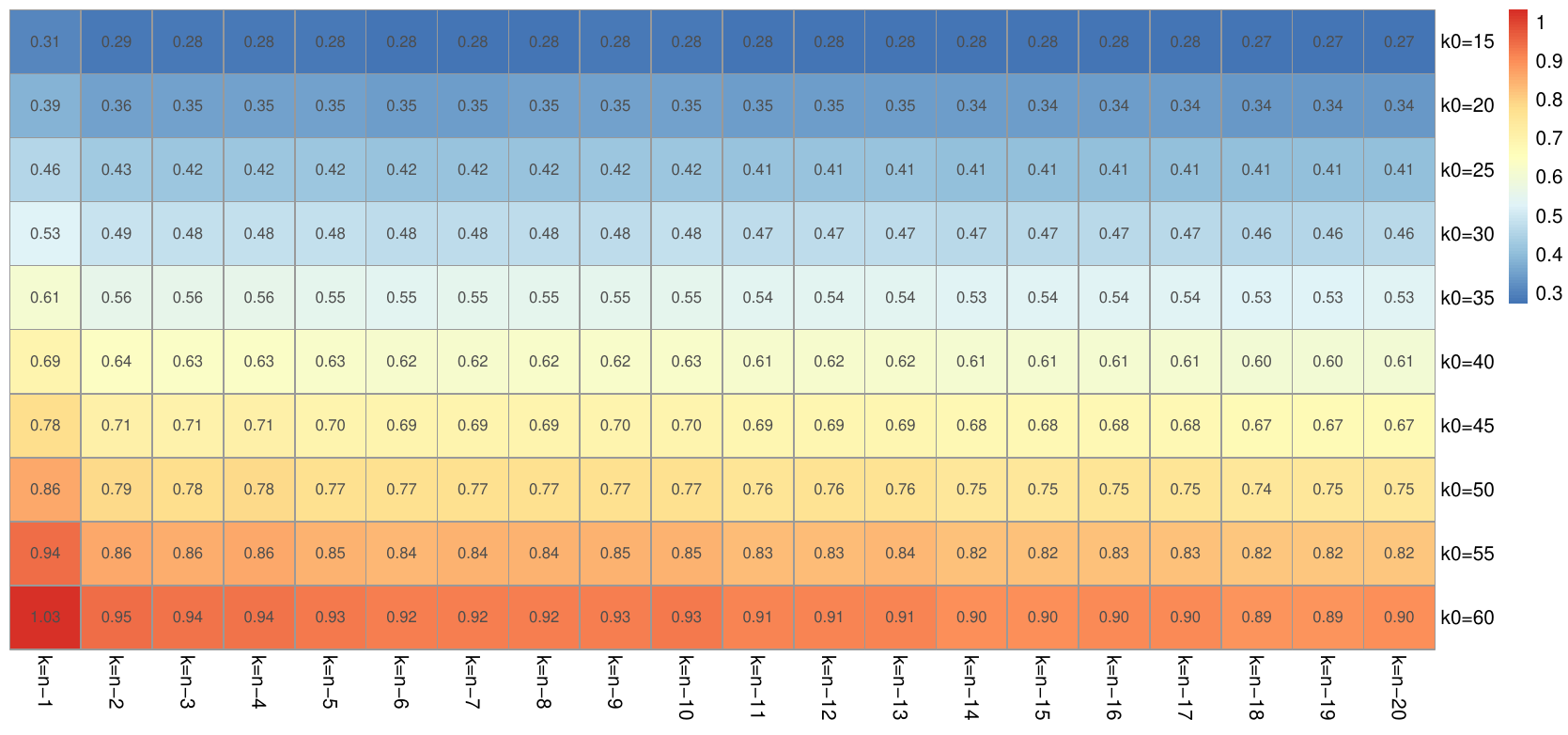}}
    \subfloat[]{\includegraphics[width = 0.5\textwidth, height = 0.25\textwidth]{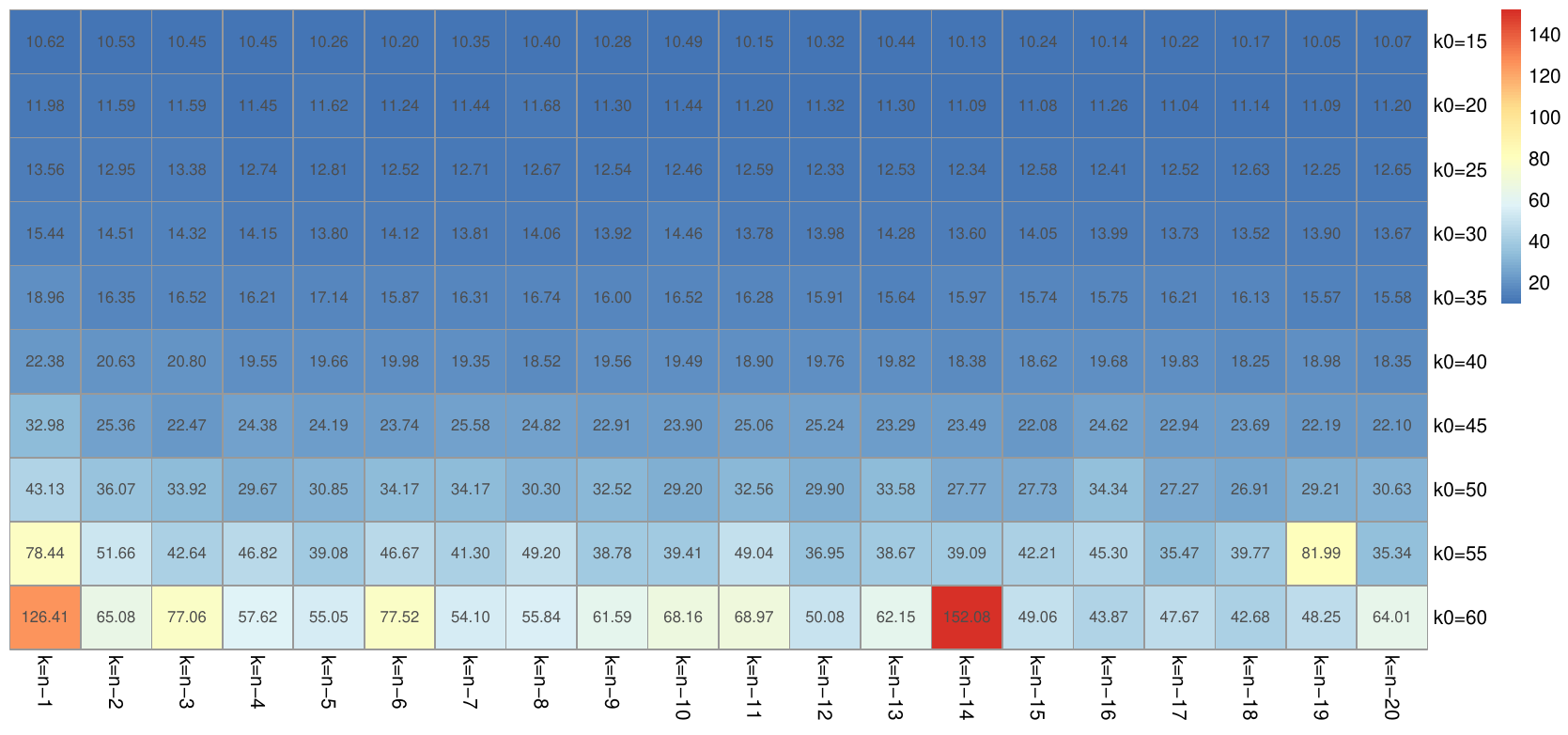} }\\
    \subfloat[]{\includegraphics[width = 0.5\textwidth, height = 0.25\textwidth]{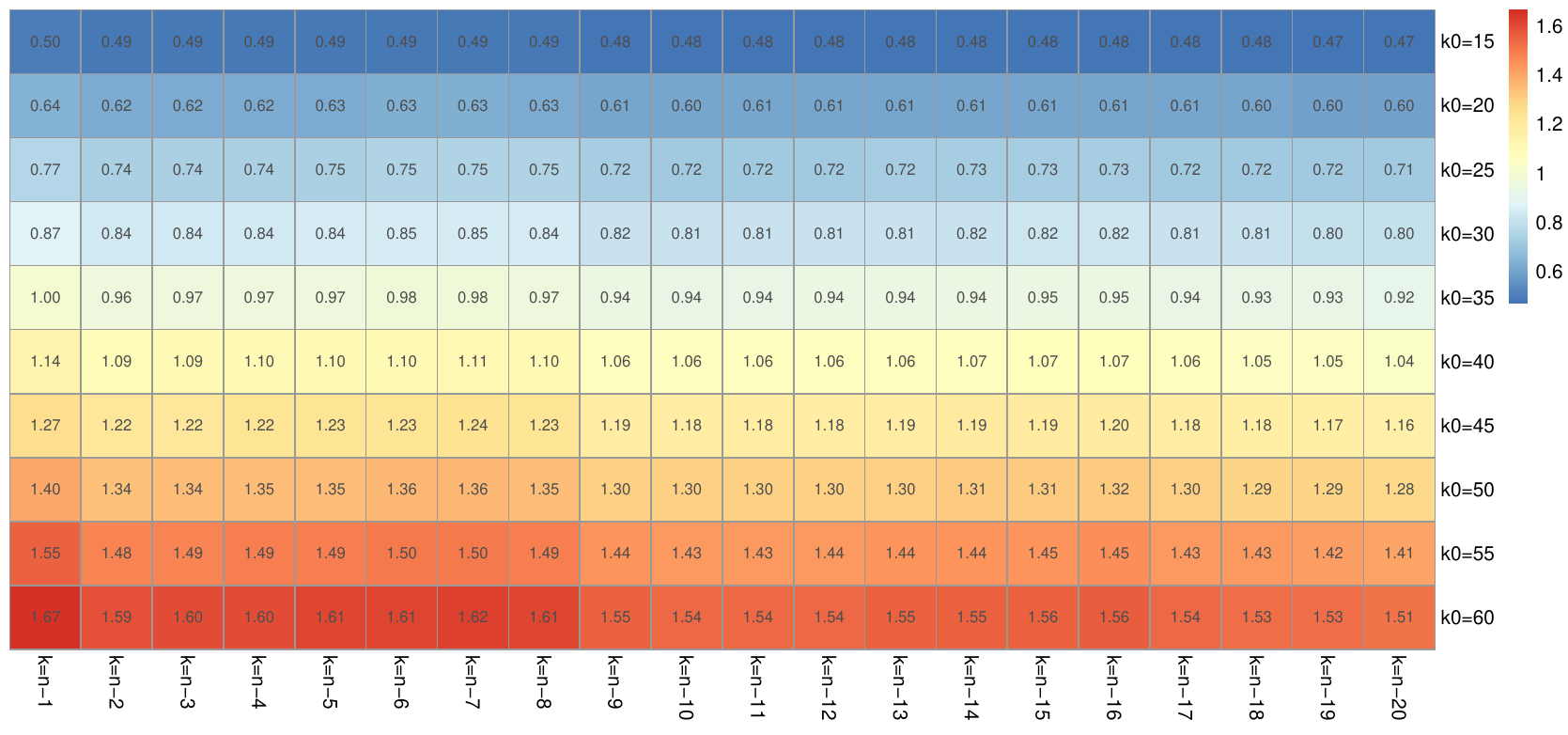}}
    \subfloat[]{\includegraphics[width = 0.5\textwidth, height = 0.25\textwidth]{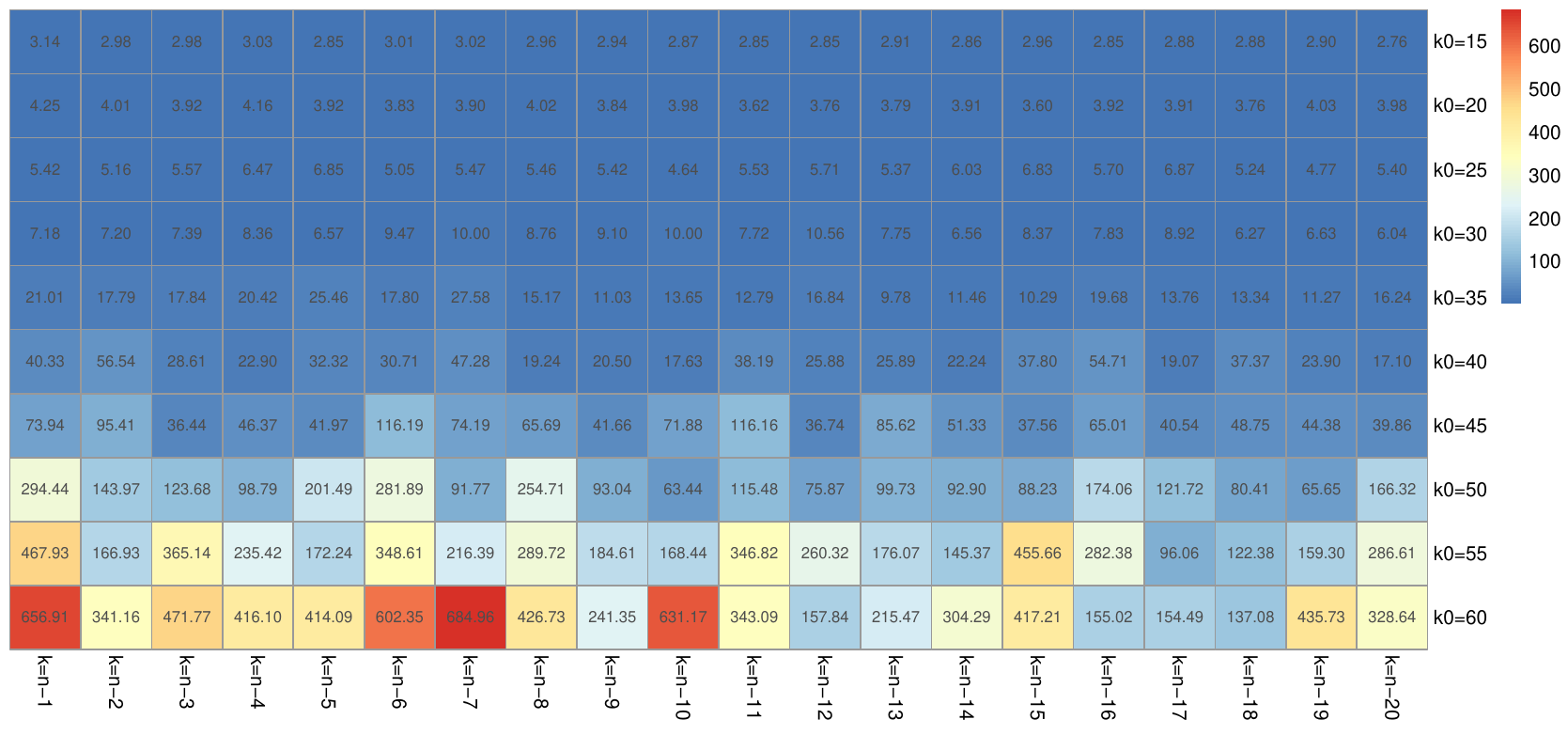}}  
    \caption{Trimmed inverse tail index Hill estimators (on the left column) and corresponding insurance premiums (on the right column) for varying trimming parameters $(k_0,k)$ for (a) NAIC51, (b) NAIC54, (c) NAIC56, (d) NAIC92. Premiums are computed on quarterly basis using 1000,000 Monte Carlo draws. The variation in the trimmed Hill estimates translates into premium calculations.}
    \label{fig:trimmed_NAIC51}
\end{figure}

\end{document}